\documentclass[11pt,a4paper]{article}
\usepackage{booktabs} 
\usepackage{enumitem} 

\usepackage{natbib}
\defcitealias{camsdata}{INERIS et.~al, 2022}
\defcitealias{nasadataset}{CIESIN et.~al, 2018}
\defcitealias{meteoseasons}{NCEI, 2023}
\defcitealias{cds}{CDS, 2020}
\defcitealias{cds2}{CDS, 2021}

\usepackage[pdftex,colorlinks,citecolor=blue,urlcolor=blue,linkcolor=red]{hyperref} 

\usepackage{doi}

\usepackage{graphicx} 
\usepackage[small,bf,hang]{caption}	
\usepackage{subcaption}
\usepackage{nicefrac}
\usepackage{tabularx}
\usepackage[export]{adjustbox}
\usepackage{afterpage}

\renewcommand*{\thesubfigure}{\alph{subfigure}}
\newcommand\nsubcap[1]{\phantomcaption%
       \caption*{\textbf{(\thesubfigure}) #1}}
              
\usepackage{amssymb} 
\usepackage{amsmath} 
\usepackage{accents}
\usepackage{bbm} 
\usepackage{amsbsy} 
\usepackage{array}
\usepackage{multirow} 
\usepackage[dvipsnames]{xcolor} 
\usepackage{mathtools}
\usepackage{eurosym}

\usepackage{floatrow}
\newfloatcommand{capbtabbox}{table}[][\FBwidth]

\usepackage[ruled]{algorithm2e} 
\SetAlCapFnt{\small} 
\SetKw{KwTo}{in}

\usepackage{authblk} 
\usepackage{footnote}

\DeclareMathOperator*{\argmax}{arg\,max}
\DeclareMathOperator*{\argmin}{arg\,min}
\bibpunct[, ]{(}{)}{;}{a}{,}{,}

\usepackage{tikz,pgfplots}
\usetikzlibrary{fit, positioning, chains,shapes.arrows,fit}
\usetikzlibrary{shapes, arrows, arrows.meta}
\usetikzlibrary{decorations.pathreplacing}
\usetikzlibrary{calc} 
\usetikzlibrary{matrix} 
\usepackage{pgf}
\usetikzlibrary{decorations.text}
\tikzset{cross/.style={cross out, draw=black, minimum size=2*(#1-\pgflinewidth), inner sep=0pt, outer sep=0pt},
cross/.default={1pt}}
\usetikzlibrary{shapes.misc}
\usepackage{environ}
\makeatletter
\newsavebox{\measure@tikzpicture}
\NewEnviron{scaletikzpicturetowidth}[1]{%
  \def\tikz@width{#1}%
  \begin{lrbox}{\measure@tikzpicture}%
  \BODY
  \end{lrbox}%
  \pgfmathparse{#1/\wd\measure@tikzpicture}%
  \BODY
}
\makeatother
\tikzdeclarecoordinatesystem{page}{
    \parsecomma#1\endparsecomma
    \pgfpointanchor{current page}{north east}
    \pgf@xc=\pgf@x%
    \pgf@yc=\pgf@y%
    \pgfpointanchor{current page}{south west}
    \pgf@xb=\pgf@x%
    \pgf@yb=\pgf@y%
    \pgfmathparse{(\pgf@xc-\pgf@xb)/2.*\page@x+(\pgf@xc+\pgf@xb)/2.}
    \expandafter\pgf@x\expandafter=\pgfmathresult pt
    \pgfmathparse{(\pgf@yc-\pgf@yb)/2.*\page@y+(\pgf@yc+\pgf@yb)/2.}
    \expandafter\pgf@y\expandafter=\pgfmathresult pt
}
\makeatother
\usetikzlibrary{arrows,automata}
\usepackage{ragged2e}

\tikzset{
    max width/.style args={#1}{
        execute at begin node={\begin{varwidth}{#1}},
        execute at end node={\end{varwidth}}
    }
}
\definecolor{newblue}{RGB}{86, 180, 233}
\definecolor{newred}{RGB}{248, 118, 109}
\definecolor{newgreen}{RGB}{163, 213, 150}

\colorlet{textcolor}{white}
\colorlet{bordercolor}{white}

\pgfdeclarelayer{background}
\pgfsetlayers{background,main}

\definecolor{airforceblue}{rgb}{0.36, 0.54, 0.66}
\definecolor{forestgreen}{rgb}{0.13, 0.55, 0.13}\definecolor{fulvous}{rgb}{0.86, 0.52, 0.0}
\definecolor{gray}{rgb}{0.5, 0.5, 0.5}
\definecolor{bistre}{rgb}{0.24, 0.17, 0.12}\definecolor{bostonuniversityred}{rgb}{0.8, 0.0, 0.0}
\definecolor{purpleheart}{rgb}{0.41, 0.21, 0.61}
\definecolor{lightsalmonpink}{rgb}{1.0, 0.6, 0.6}\definecolor{arrowcolor}{rgb}{0.92, 0.92, 0.92}
\tikzset{
inner/.style={
  on chain,
  circle,
  inner sep=4pt,
  fill=circlecolor,
  line width=1.5pt,
  draw=bordercolor,
  text width=1.2em,
  align=center,
  text height=1.25ex,
  text depth=0ex
},
on grid
}

\usepackage{regexpatch}

\usepackage{pbox}
\newcommand\drawarrow{

\node[on chain] (f) {};
\begin{pgfonlayer}{background}
\node[
  inner sep=10pt,
  single arrow,
  single arrow head extend=0.6cm,
  draw=none,
  fill=arrowcolor,
  fit= (c1) (f)
] (arrow) {};
\fill[white] 
  (arrow.before tail) -- (c1|-arrow.west) -- (arrow.after tail) -- cycle;
\end{pgfonlayer}
}

\numberwithin{equation}{section}

\usepackage{relsize}

\usepackage{setspace}
\def\spacingset#1{\renewcommand{\baselinestretch}%
{#1}\small\normalsize} \spacingset{1}

\usepackage{xr}
\makeatletter
\let\ref\@refstar
\makeatother

\begin{document}
\def\spacingset#1{\renewcommand{\baselinestretch}%
{#1}\small\normalsize} \spacingset{1}

\title{\bf{Granular mortality modeling with temperature and epidemic shocks: a three-state regime-switching approach}}

\author[1,2,*]{Jens Robben}
\author[3]{Karim Barigou}
\author[1,2]{Torsten Kleinow}
\affil[1]{Faculty of Economics and Business, University of Amsterdam, The Netherlands.}
\affil[2]{RCLR, Research Centre for Longevity Risk, University of Amsterdam, The Netherlands.}
\affil[3]{Institute of Statistics, Biostatistics and Actuarial Science (ISBA), Louvain Institute of Data Analysis and Modeling (LIDAM), UCLouvain, Louvain-la-Neuve, Belgium }

\affil[*]{Corresponding author: \href{mailto:j.robben@uva.nl}{j.robben@uva.nl}}
\date{\today} 
\maketitle
\thispagestyle{empty} 

\begin{abstract} \noindent
This paper develops a granular regime-switching framework to model mortality deviations from seasonal baseline trends driven by temperature and epidemic shocks. The framework features three states: (1) a baseline state that captures observed seasonal mortality patterns, (2) an environmental shock state for heat waves, and (3) a respiratory shock state that addresses mortality deviations caused by strong outbreaks of respiratory diseases due to influenza and COVID-19. Transition probabilities between states are modeled using covariate-dependent multinomial logit functions. These functions incorporate, among others, lagged temperature and influenza incidence rates as predictors, allowing dynamic adjustments to evolving shocks. Calibrated on weekly mortality data across 21 French regions and six age groups, the regime-switching framework accounts for spatial and demographic heterogeneity. Under various projection scenarios for temperature and influenza, we quantify uncertainty in mortality forecasts through prediction intervals constructed using an extensive bootstrap approach. These projections can guide healthcare providers and hospitals in managing risks and planning resources for potential future shocks.
\end{abstract}

\noindent%
\vfill

\newpage

\section{Introduction}
Mortality rates typically exhibit seasonal variation throughout the year, where the highest rates occur during winter \citep{marti2014seasonal, gemmell2000seasonal,mackenbach1992seasonal}. External events, including increased respiratory infections, such as influenza outbreaks, and extreme weather conditions, such as heat waves, have historically led to substantial deviations from these expected seasonal trends, leading to what is known as excess mortality \citep{iuliano2018estimates, nielsen2011excess, basu2002relation}. Since mortality is an important public health indicator, accurately estimating and forecasting mortality rates in the short term is essential for effective health care and public health actions \citep{hajat2010health,kok2010behavioural}. In this paper, we analyze and aim to better understand the impact of factors such as increased temperatures or heightened respiratory infections on weekly, regional mortality rates using a novel regime-switching framework.

The proposed framework consists of two key components. The first component captures seasonal, age-specific mortality trends within each region, which we call the baseline model. The second component is a three-state regime-switching model, where mortality rates in the first regime follow the seasonal baseline trend. The second and third regimes amplify the baseline based on the current and past week's temperature and respiratory characteristics. This framework is particularly useful, as it enables the generation of future mortality scenarios under different weather and respiratory conditions. In addition, we provide guidance on the uncertainty surrounding these scenario projections. Such insights help to anticipate and respond to potential mortality spikes due to extreme weather events or respiratory outbreaks. Importantly, this paper focuses on short-term mortality projections. Long-term projections, which would require accounting for future improvements in heat resilience and public health interventions, are beyond the scope of this study.

Related work has been done in the EuroMOMO and FluMOMO projects, established to operate a coordinated and timely monitoring of mortality rates within 28 participating European countries. Similarly to our setting, they use the Serfling model as a baseline, incorporating seasonality through Fourier terms \citep{serfling1963methods}. \cite{vestergaard2020excess} and \cite{norgaard2021real} use the EuroMOMO framework to estimate excess mortality during the first and second waves of the COVID-19 pandemic. \cite{nielsen2018influenza} use the FluMoMo model to calculate influenza-attributable mortality during the winter seasons 2010/2011 - 2016/2017 in Denmark, whereas \cite{nielsen2019european} use it to analyze the influenza-attributable mortality in Europe during the winter season 2017/2018.

The epidemiological and medical literature has extensively investigated the associations between temperature, epidemic variables, and mortality. \cite{gasparrini2015mortality} conduct a systematic assessment of the impact of temperature on mortality across multiple countries worldwide. Their findings reveal that, on average, 7.71$\%$ of total daily death counts could be attributed to cold and heat. \cite{song2017impact} provide an overview of systematic reviews to summarize evidence regarding the impact of ambient temperature on morbidity and mortality. \cite{iuliano2018estimates} estimate global influenza-associated respiratory deaths for 1999–2015. Their results reveal a higher annual mortality burden than previously reported and find significant variation across regions and age groups. \cite{wang2022estimating} conduct a systematic analysis to estimate COVID-19 excess mortality.

Statistical methods used to analyze mortality rates are manifold. Examples include \cite{lee1992modeling} who model age-specific death rates as a linear function of an unobserved period-specific intensity index, with parameters depending on age. \cite{berger2023modeling} use a novel survival modeling approach to analyze postoperative mortality in older patients and incorporate risk factors measured at hospital admission. \cite{berger2001bayesian} use Bayesian random-effects logit models to model hospital mortality rates for a heart attack or acute myocardial infarction.

Our paper contributes to the statistical literature in three ways. First, we gather daily gridded temperature data from the Copernicus Climate Data Store and outline how to convert these data into weekly, region-specific population-weighted temperature levels. Furthermore, we use epidemic data from the French Sentinelles network and hospital admission records from Santé Publique France to capture influenza outbreaks and COVID-19 excess mortality. Second, we introduce a novel mortality modeling framework that integrates a granular age- and region-specific baseline with a regime-switching model that captures mortality shocks driven by environmental and epidemic factors. Using covariate-driven transition probabilities and spatial dependencies, this framework offers a comprehensive and interpretable approach to understanding how these shocks impact mortality patterns across various regions and age groups. While previous studies have primarily focused on the effects of temperature or epidemic shocks on mortality, this paper is among the first to integrate both factors within a unified mortality modeling framework. Third, we quantify uncertainty in the model's estimates and forecasts and construct prediction intervals for the death counts under various weather and respiratory scenarios.

The paper is organized as follows: Section~\ref{subsec:data} discusses the notation and the sources of mortality, environmental and epidemic data considered. Section~\ref{sec:granmomors} specifies the model, consisting of the baseline model in Section~\ref{modelspec:baseline} and the regime-switching model in Section~\ref{subsec:modelspec2}. We outline the calibration procedure in Section~\ref{sec:modelcal}, and the uncertainty quantification in in- and out-of-sample predictions in Section~\ref{sec:quantify.uncertainty}. We outline the results of the practical application to 21 French regions and six older age groups in Section~\ref{sec:casestudy}. Section~\ref{sec:conclusion} concludes.

\section{Notations and data} \label{subsec:data}
\paragraph{Notations.} Let $d_{x,t}^{(r)}$ denote the number of deaths in region $r$ for age group $x$ in week $t$, where $t$ represents the number of ISO weeks elapsed since a selected start date. We adhere to the ISO 8601 standard, where each ISO week is 7 days long, and ISO week 1 starts on the Monday of the week that contains the first Thursday of the year. We define the set of regions as $\mathcal{R}$, the set of age groups as $\mathcal{X}$, and the range of weeks under consideration as $\mathcal{T} = \{1,2,...,T\}$.  Although the methodologies presented in this paper apply to both men and women, our analysis uses combined data for both sexes. Furthermore, let $E_{x,t}^{(r)}$ be the exposure-to-risk in region $r \in \mathcal{R}$ at age group $x$ in week $t$. The observed weekly death rate $\smash{m_{x,t}^{(r)}}$ then equals $\nicefrac{d_{x,t}^{(r)}}{E_{x,t}^{(r)}}$.

\begin{figure}[ht!]
\centering
\begin{subfigure}{0.42\textwidth}
\centering
\includegraphics[width = \textwidth]{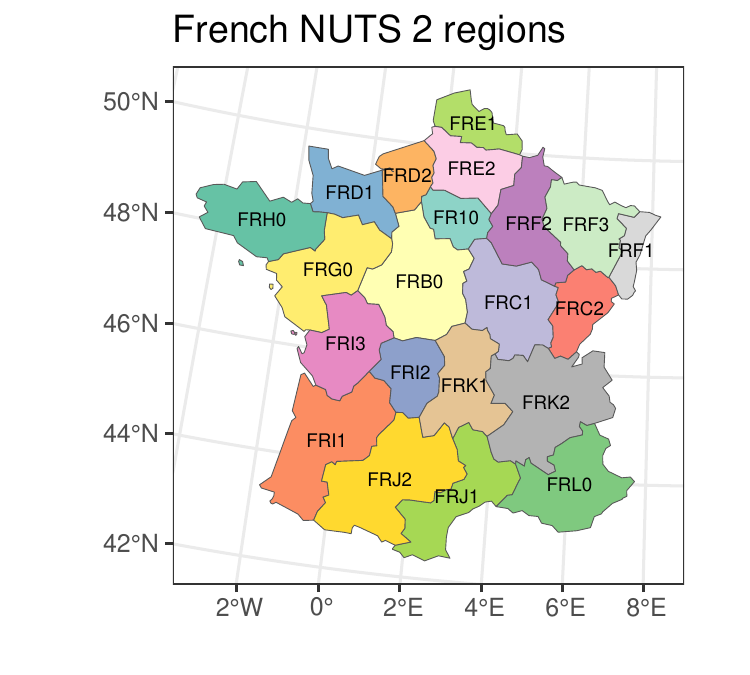}
\end{subfigure}
\hspace{0.1cm}
\begin{subfigure}{0.42\textwidth}
\centering
\includegraphics[width = \textwidth]{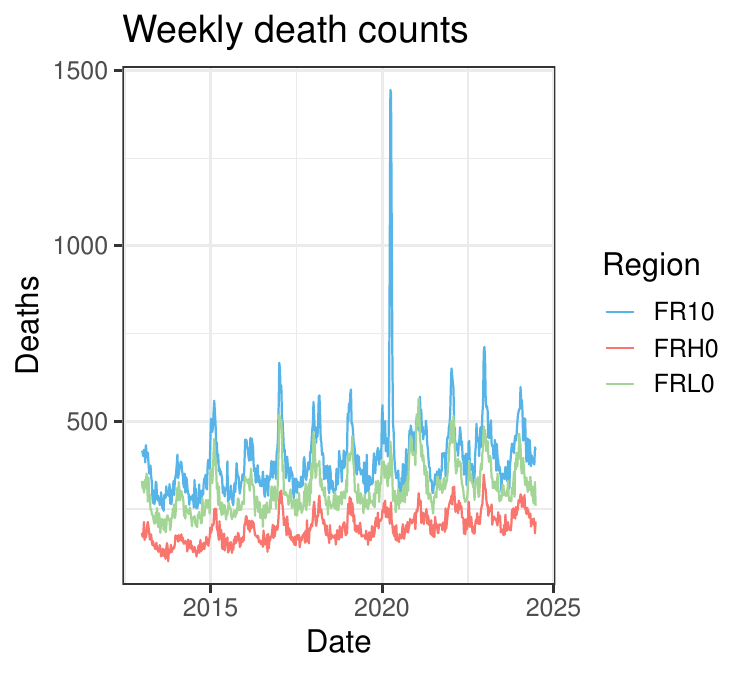}
\end{subfigure}
\caption{Left panel: the French NUTS 2 regions. Right panel: weekly death counts for the 90+ age group in Ile-de-France (FR10), Provence-Alpes-Côte d'Azur (FRL0), and Bretagne (FRH0) from 2013-2024. \label{fig:nutsdeaths}}
\end{figure}

\paragraph{Mortality data.} We retrieve weekly death counts in Metropolitan France from the first ISO week of 2013 to the 26th ISO week of 2024 ($T = 600$) from Eurostat. Deaths are categorized by sex, 5-year age groups, and NUTS 2 regions \citep{Eurostat_DEMO_R_MWEEK2}. NUTS (Nomenclature of Territorial Units for Statistics) classifies European administrative areas into hierarchical levels (NUTS 1, 2, 3) for statistical purposes. The NUTS 2 regions in Metropolitan France correspond to the 22 former regions. We exclude Corsica due to the unavailability of weekly death counts. Figure~\ref{fig:nutsdeaths} (left panel) displays the considered French NUTS 2 regions for which Table~\ref{tab:nuts2overview} in Suppl.~Mat.~\ref{app:overview} lists the corresponding names. 

This paper focuses on the elderly population, given their heightened vulnerability to environmental factors and illnesses such as influenza. Hereto, we consider 5-year age groups starting from the age of 65, i.e., $\mathcal{X} = \{65$-$69, 70$-$74, 75$-$79, 80$-$84, 85$-$89, 90$+$\}$. Figure~\ref{fig:nutsdeaths} (right panel) presents the weekly death counts from 2013 to 2024 for individuals aged 90 and above in Ile-de-France (FR10), Provence-Alpes-Côte d'Azur (FRL0), and Bretagne (FRH0). The data reveal a pronounced seasonal variation, where death counts increase during winter and decrease during summer. Furthermore, we observe a clear COVID-19 peak in Ile-de-France, where the death counts exceeded the usual level by more than three times. Figures~\ref{figA:overviewDE} and~\ref{figA:overviewAge} in Suppl.~Mat.~\ref{app:overview} provide a more detailed overview of the structure of the death count data. In order to calculate weekly mortality rates, it is necessary to construct a weekly exposure measure, Suppl.~Mat.~\ref{app:overview} also details this construction.

\begin{figure}[!htb]
\centering
\begin{subfigure}{0.42\textwidth}
\centering
\includegraphics[width = \textwidth]{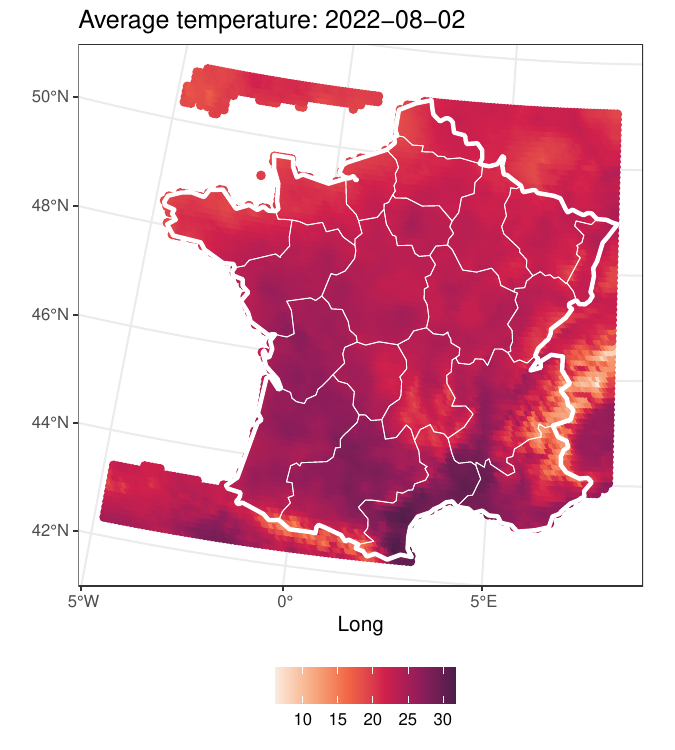}
\end{subfigure}
\hspace{0.1cm}
\begin{subfigure}{0.42\textwidth}
\centering
\includegraphics[width = \textwidth]{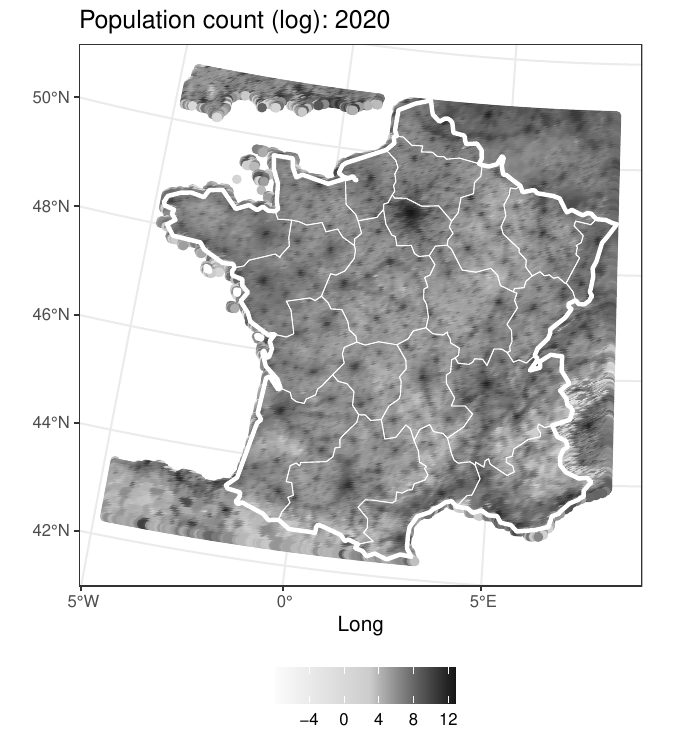}
\end{subfigure}
\caption{Left panel: the daily average temperature in a spatial grid covering Metropolitan France on August 2, 2022. Right panel: population count (log-scale) in 2020 on a spatial grid covering Metropolitan France. We highlight the boundaries of the French NUTS 2 regions in white. \label{fig:eobssedac}}
\end{figure}

\paragraph{Temperature data.} We use E-OBS gridded meteorological data from the Copernicus Climate Data Store (CDS) to extract daily average temperature levels in Metropolitan France at a 0.10$^\circ$ resolution \citepalias{cds}. To align with the NUTS 2-level mortality data, we aggregate daily temperatures into population-weighted averages within each NUTS 2 region using NASA’s gridded population data \citepalias{nasadataset}. This ensures that the temperature averages reflect areas where more people reside. Figure~\ref{fig:eobssedac} illustrates a snapshot of the temperature and population data on a randomly selected date and year.

\begin{figure}[!htb]
\centering
\begin{subfigure}{0.42\textwidth}
\centering
\includegraphics[width = \textwidth]{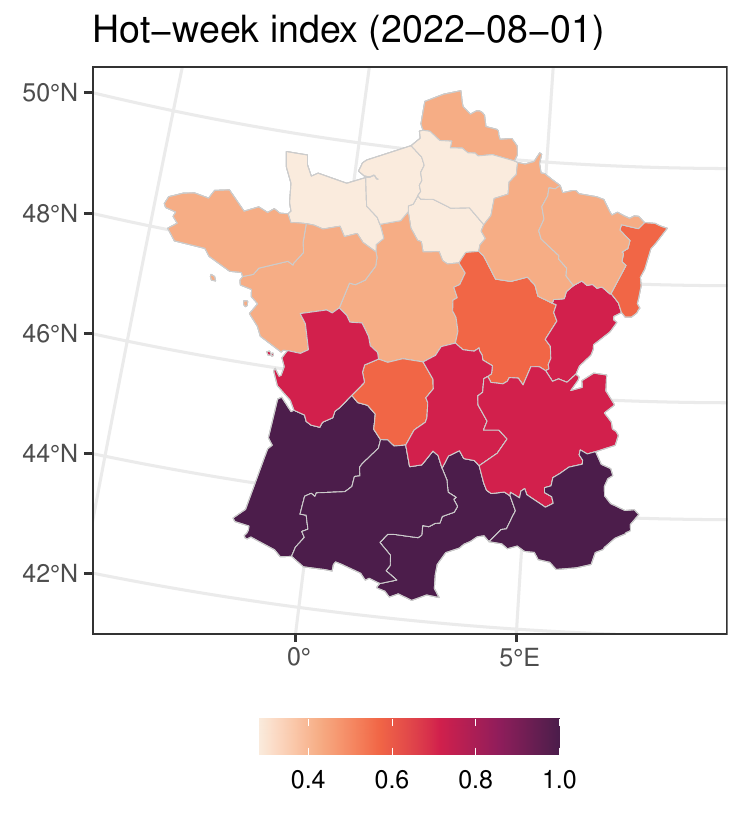}
\begin{minipage}{1.1\textwidth}
\vspace{-0.25cm}
\nsubcap{\label{fig:hwi}}
\end{minipage}
\end{subfigure}
\hspace{0.1cm}
\begin{subfigure}{0.42\textwidth}
\centering
\includegraphics[width = \textwidth]{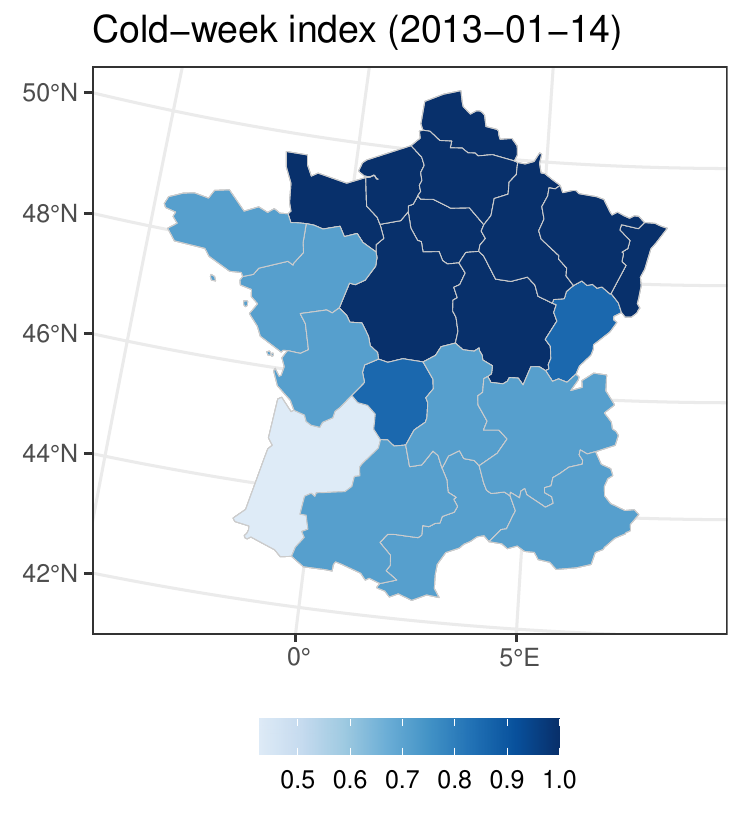}
\begin{minipage}{1.1\textwidth}
\vspace{-0.25cm}
\nsubcap{\label{fig:cwi}}
\end{minipage}
\end{subfigure}
\caption{Left panel: the weekly average of the hot-day indicator at the ISO week starting August 1, 2022. Right panel: the weekly average of the cold-day indicator at the ISO week starting January 14, 2013. We highlight the boundaries of the French NUTS 2 regions in grey. \label{fig:hwicwi}}
\end{figure}

We construct population-weighted daily temperature averages and compare them to region-specific high (95$\%$) and low ($5\%$) thresholds based on 2013–2024 data. We then average daily indicators for exceeding these thresholds weekly to create hot- and cold-week indices, see Figure~\ref{fig:hwicwi} for an example. We also compute weekly averages of daily temperature anomalies by comparing observed temperatures to a region-specific daily baseline. While the hot-week index measures the frequency of hot days in a week, temperature anomalies reflect their severity. A similar interpretation applies to the cold-week index. We refer to \cite{robben2024association} for further details on this feature construction step. 

\begin{figure}[!h]
\centering
\begin{subfigure}{0.44\textwidth}
\centering
\includegraphics[width = \textwidth]{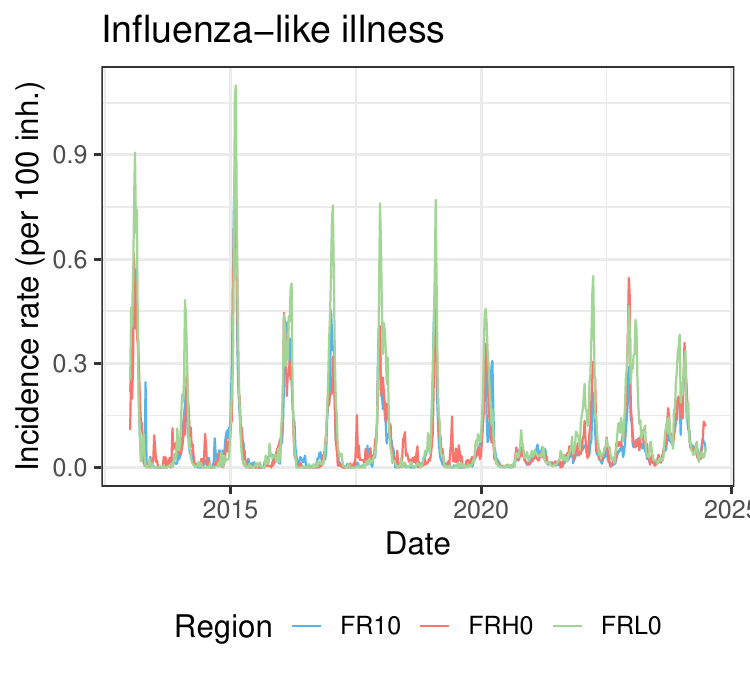}
\end{subfigure}
\hspace{0.1cm}
\begin{subfigure}{0.44\textwidth}
\centering
\includegraphics[width = \textwidth]{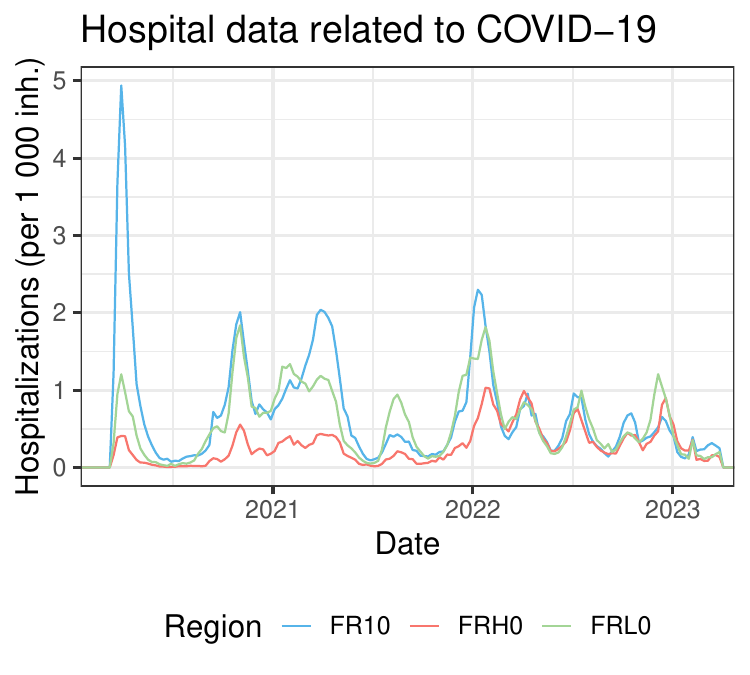}
\end{subfigure}
\caption{We show the weekly number of new influenza-like incidences per 100 inhabitants (left) and the COVID-19 hospitalizations per 1 000 inhabitants (right) from 2013-2014 and 2020-2023, respectively, in Ile-de-France (FR10), Provence-Alpes-Côte d'Azur (FRL0), and Bretagne (FRH0). \label{fig:ariili}}
\end{figure}

\paragraph{Influenza data.} We consult the French Sentinelles network, established in 1984 by Inserm and Sorbonne University, to collect near real-time epidemic data from $\approx$ 1\ 300 general practitioners across France \citep{valleron1986computer}. For this study, we extract weekly influenza incidence rates per 100 inhabitants at the NUTS 2 level. These incidences are based on patient visits with fever, myalgia, and respiratory symptoms. Figure~\ref{fig:ariili} (left panel) shows influenza trends in Île-de-France (FR10), Provence-Alpes-Côte d'Azur (FRL0), and Bretagne (FRH0) from 2013 to 2024. The figure highlights peaks such as in the 2014/2015 winter season and a very low influenza activity in 2020/2021 due to the imposed COVID-19 lockdown measures. Since we aim to explain mortality deviations from a seasonal baseline, we work with so-called influenza anomalies. These anomalies indicate whether influenza activity in a given week was higher or lower than its seasonal norm.

\paragraph{COVID-19 hospitalizations.} We use daily data on COVID-19 hospitalizations per 1 000 inhabitants in French NUTS 2 regions, sourced from Santé Publique France, to capture the impact of COVID-19 on mortality. We aggregate this data, available from March 19, 2020, to March 31, 2023, weekly. Post-March 2023 hospitalization rates are set to zero. Figure~\ref{fig:ariili} shows weekly COVID-19 hospitalizations in Île-de-France (FR10), Provence-Alpes-Côte d'Azur (FRL0), and Bretagne (FRH0), and highlights clear regional variations in the timing and intensity of COVID-19 waves.

\paragraph{Feature overview.} Table~\ref{tab:features} summarizes the variables constructed from the environmental, influenza, and COVID-19 hospital admission data discussed in this section.
\begin{table}[h!]
    \centering
    \caption{Variables derived from the temperature, influenza, and COVID-19 hospitalization data. \label{tab:features}}
    \label{tab:features_summary}
    \begin{small}
    \begin{tabular}{p{1.4cm}p{13.1cm}}
        \toprule
        \textbf{Feature} & \textbf{Description} \\
        \midrule
        HI  & Hot-week index: weekly average of daily indicators for unusually high temperatures. \\
        CI  & Cold-week index: weekly average of daily indicators for unusually low temperatures. \\
        TA  & Temperature anomalies: weekly average of daily deviations from baseline temperature levels. \\
        IA  & Influenza activity index: influenza anomalies, deviations from expected seasonal influenza incidences (per 100 inhabitants). \\
        HA  & Hospital admission: weekly count of new hospitalizations for COVID-19 (per 1\ 000 inhabitants). \\
        \bottomrule
    \end{tabular}
    \end{small}
\end{table}

\section{Modeling mortality with regime-switching dynamics} \label{sec:granmomors}
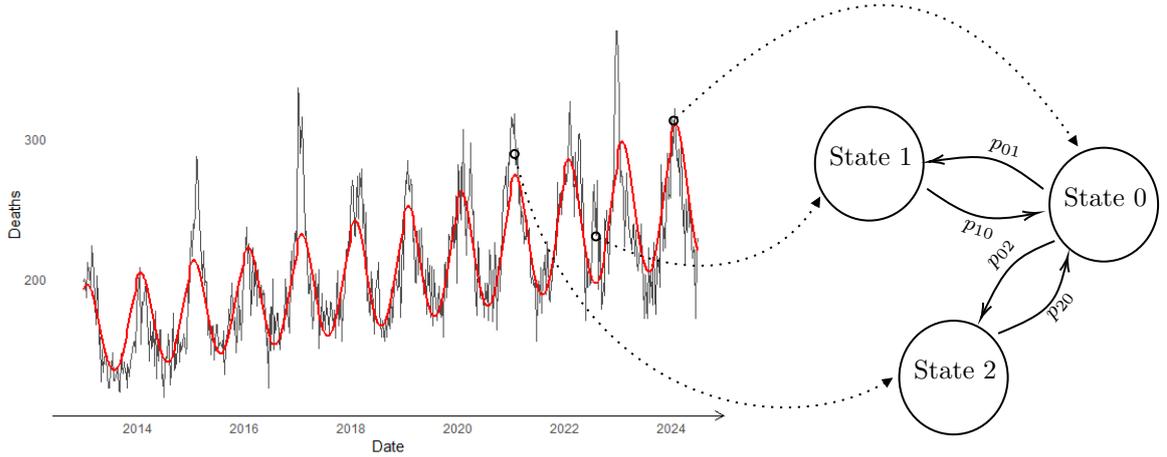
\begin{figure}[!ht]
\centering
\tikzset{every picture/.style={line width=0.75pt}} 

\begin{tikzpicture}[x=0.75pt,y=0.75pt,yscale=-1,xscale=1]

\draw (189.5,162.71) node  {\includegraphics[width=273.75pt,height=172.07pt]{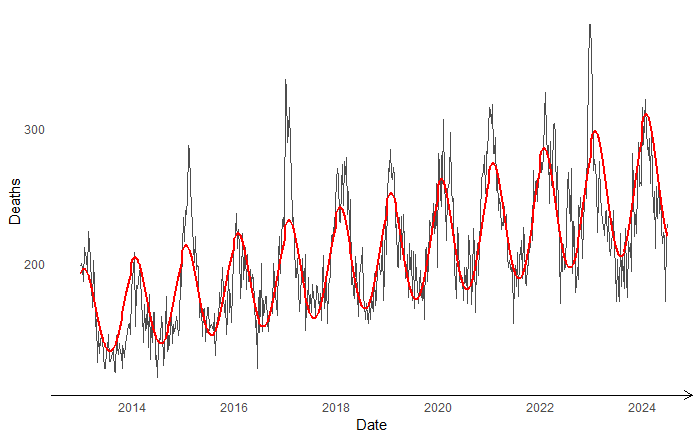}};
\draw   (530.8,148.06) .. controls (530.8,132.26) and (542.93,119.45) .. (557.9,119.45) .. controls (572.87,119.45) and (585,132.26) .. (585,148.06) .. controls (585,163.86) and (572.87,176.67) .. (557.9,176.67) .. controls (542.93,176.67) and (530.8,163.86) .. (530.8,148.06) -- cycle ;

\draw   (413.74,127.46) .. controls (413.74,111.66) and (425.87,98.85) .. (440.83,98.85) .. controls (455.8,98.85) and (467.93,111.66) .. (467.93,127.46) .. controls (467.93,143.26) and (455.8,156.07) .. (440.83,156.07) .. controls (425.87,156.07) and (413.74,143.26) .. (413.74,127.46) -- cycle ;

\draw   (455.79,235.04) .. controls (455.79,219.24) and (467.93,206.43) .. (482.89,206.43) .. controls (497.86,206.43) and (509.99,219.24) .. (509.99,235.04) .. controls (509.99,250.84) and (497.86,263.65) .. (482.89,263.65) .. controls (467.93,263.65) and (455.79,250.84) .. (455.79,235.04) -- cycle ;

\draw [line width=0.75]  [dash pattern={on 0.84pt off 2.51pt}]  (264.26,124.56) .. controls (295.62,252.78) and (401.98,267.03) .. (450.5,236.51) ;
\draw [shift={(452.67,235.09)}, rotate = 145.71] [fill={rgb, 255:red, 0; green, 0; blue, 0 }  ][line width=0.08]  [draw opacity=0] (5.36,-2.57) -- (0,0) -- (5.36,2.57) -- cycle    ;
\draw [shift={(263.78,122.59)}, rotate = 76.7] [color={rgb, 255:red, 0; green, 0; blue, 0 }  ][line width=0.75]      (0, 0) circle [x radius= 2.01, y radius= 2.01]   ;
\draw [color={rgb, 255:red, 0; green, 0; blue, 0 }  ,draw opacity=1 ] [dash pattern={on 0.84pt off 2.51pt}]  (305.92,164.88) .. controls (355.4,191.29) and (397.4,177.37) .. (415.11,146.35) ;
\draw [shift={(416.42,143.93)}, rotate = 119.24] [fill={rgb, 255:red, 0; green, 0; blue, 0 }  ,fill opacity=1 ][line width=0.08]  [draw opacity=0] (5.36,-2.57) -- (0,0) -- (5.36,2.57) -- cycle    ;
\draw [shift={(304.38,164.05)}, rotate = 27.41] [color={rgb, 255:red, 0; green, 0; blue, 0 }  ,draw opacity=1 ][line width=0.75]      (0, 0) circle [x radius= 2.01, y radius= 2.01]   ;
\draw    (505.22,212.84) .. controls (531.01,202.3) and (534.54,198.12) .. (539.52,175.86) ;
\draw [shift={(539.91,174.11)}, rotate = 102.33] [color={rgb, 255:red, 0; green, 0; blue, 0 }  ][line width=0.75]    (7.65,-2.3) .. controls (4.86,-0.97) and (2.31,-0.21) .. (0,0) .. controls (2.31,0.21) and (4.86,0.98) .. (7.65,2.3)   ;
\draw    (533.4,166.33) .. controls (513.43,174.4) and (508.87,178.52) .. (497.27,203.67) ;
\draw [shift={(496.55,205.24)}, rotate = 294.57] [color={rgb, 255:red, 0; green, 0; blue, 0 }  ][line width=0.75]    (7.65,-2.3) .. controls (4.86,-0.97) and (2.31,-0.21) .. (0,0) .. controls (2.31,0.21) and (4.86,0.98) .. (7.65,2.3)   ;
\draw    (527.77,140.23) .. controls (505.67,123.18) and (500.65,121.96) .. (472.3,126.23) ;
\draw [shift={(470.53,126.5)}, rotate = 351.3] [color={rgb, 255:red, 0; green, 0; blue, 0 }  ][line width=0.75]    (7.65,-2.3) .. controls (4.86,-0.97) and (2.31,-0.21) .. (0,0) .. controls (2.31,0.21) and (4.86,0.98) .. (7.65,2.3)   ;
\draw    (469.67,140.05) .. controls (491.23,155.4) and (499.19,157.53) .. (523.28,152.53) ;
\draw [shift={(525.17,152.13)}, rotate = 167.88] [color={rgb, 255:red, 0; green, 0; blue, 0 }  ][line width=0.75]    (7.65,-2.3) .. controls (4.86,-0.97) and (2.31,-0.21) .. (0,0) .. controls (2.31,0.21) and (4.86,0.98) .. (7.65,2.3)   ;
\draw  [dash pattern={on 0.84pt off 2.51pt}]  (344.44,104.59) .. controls (428.11,23.39) and (483.53,30.34) .. (544.11,117.49) ;
\draw [shift={(545.02,118.81)}, rotate = 235.42] [fill={rgb, 255:red, 0; green, 0; blue, 0 }  ][line width=0.08]  [draw opacity=0] (5.36,-2.57) -- (0,0) -- (5.36,2.57) -- cycle    ;
\draw [shift={(343.17,105.83)}, rotate = 315.64] [color={rgb, 255:red, 0; green, 0; blue, 0 }  ][line width=0.75]      (0, 0) circle [x radius= 2.01, y radius= 2.01]   ;

\draw (536.55,138.07) node [anchor=north west][inner sep=0.75pt]   [align=left] {{\small State 0}};
\draw (419.48,117.47) node [anchor=north west][inner sep=0.75pt]   [align=left] {{\small State 1}};
\draw (461.54,225.05) node [anchor=north west][inner sep=0.75pt]   [align=left] {{\small State 2}};
\draw (487.82,153.43) node [anchor=north west][inner sep=0.75pt]  [font=\scriptsize,rotate=-16.56] [align=left] {$\displaystyle p_{10}$};
\draw (501.11,112.84) node [anchor=north west][inner sep=0.75pt]  [font=\scriptsize,rotate=-14.95] [align=left] {$\displaystyle p_{01}$};
\draw (497.4,176.99) node [anchor=north west][inner sep=0.75pt]  [font=\scriptsize,rotate=-308.47] [align=left] {$\displaystyle p_{02}$};
\draw (526.89,203.54) node [anchor=north west][inner sep=0.75pt]  [font=\scriptsize,rotate=-308.47] [align=left] {$\displaystyle p_{20}$};
\end{tikzpicture}
\caption{Schematic representation of the proposed regime-switching mortality model. The left panel shows a robust seasonal baseline trend (red line) for each age group and region, alongside observed death counts (grey). The right panel depicts the regime-switching structure, transitioning between three states: baseline mortality (state 0), excess mortality from environmental shocks (state 1), and excess mortality from respiratory infections (state 2). Arrows indicate state transitions, with probabilities driven by environmental and epidemic covariates.\label{tikz:vizmod}}
\end{figure}

Figure~\ref{tikz:vizmod} illustrates the proposed model. On the left, the red line visualizes the estimation of a robust seasonal baseline mortality trend, tailored to each considered age category and region within France. However, the observed death counts (in gray) deviate from this baseline due to various potential factors. Hereto, we introduce a regime-switching model, illustrated on the right, which alternates between three states: 
\begin{enumerate}
\item State 0 (baseline state): applies when the baseline accurately reflects the true underlying death counts.  
\item State 1 (environmental shock state): captures deviations caused by environmental shocks, such as elevated summer temperatures.
\item State 2 (respiratory shock state): addresses deviations driven by an increase in respiratory infections, including influenza and COVID-19.
\end{enumerate}

The arrows in Figure~\ref{tikz:vizmod} represent possible state transitions and their corresponding probabilities. The model assumes the system can only be in one of the three states at any time, and direct transitions between the environmental shock state (state 1) and respiratory shock state (state 2) are not allowed. As described in Section~\ref{subsec:data}, transition probabilities depend on environmental and respiratory-related covariates. Incorporating such covariates into the transition probabilities helps to determine the current state. For instance, if very high temperatures are observed at time point $t$, the model will facilitate a transition to state 1. As such, regime states are not entirely hidden but influenced by observable conditions.

\subsection{Weekly baseline mortality model} \label{modelspec:baseline}
We design a region- and age-group-specific baseline to capture the overall seasonal mortality patterns by following the approach in \cite{serfling1963methods}. Hereto, we incorporate seasonality using Fourier terms and assume that the observed weekly death counts in region $r$ and at age group $x$ are realizations from a Poisson-distributed random variable, $D_{x,t}^{(r)}$:  
\begin{align} \label{eq:baselinemodelspec}
D_{x,t}^{(r)} \sim \text{Poisson}\left(E_{x,t}^{(r)}\: \mu_{x,t}^{(r)}\right),
\end{align}
where the structure of $\mu_{x,t}^{(r)}$ is given by \citep{serfling1963methods}:  
\begin{equation} \label{eq:baselinedeathsstructure}
\begin{aligned}
\log \mu_{x,t}^{(r)} = \: &\gamma_{x,0}^{(r)} + \gamma_{x,1}^{(r)} t + \gamma_{x,2}^{(r)} \sin\left(\frac{2\pi w(t)}{52.18}\right) + \gamma_{x,3}^{(r)} \cos\left(\frac{2\pi w(t)}{52.18}\right) + \\ 
&\:\:\:\gamma_{x,4}^{(r)} \sin\left(\frac{2\pi w(t)}{26.09}\right) + \gamma_{x,5}^{(r)} \cos\left(\frac{2\pi w(t)}{26.09}\right),
\end{aligned}
\end{equation}  
with $w(t)$ the ISO-week number in the year $y(t)$ of week $t$, and $52.18$ represents the average number of weeks in a year. This baseline structure is tailored to each region, age, and week. The parameters \(\gamma_{x,p}^{(r)}\) (\(p=0,1,\ldots,5\)) are region- and age-group-specific and account for the level, trend, and seasonal age-specific variations. We outline the estimation process of this baseline in Section~\ref{subsec:modelcal1} and denote the estimated expected weekly baseline death counts as:  
\begin{align} \label{eq:btwr}
\hat{b}^{(r)}_{x,t} := \hat{E}\left[D_{x,t}^{(r)}\right] =  E_{x,t}^{(r)} \: \hat{\mu}_{x,t}^{(r)},
\end{align}  
where \(\hat{\mu}_{x,t}^{(r)}\) represents the fitted baseline structure in~\eqref{eq:baselinedeathsstructure}. 

\subsection{Regime-switching mortality framework} \label{subsec:modelspec2}
Consider the estimated baseline deaths, $\smash{\hat{b}_{x,t}^{(r)}}$, as fixed. We assume that the distribution of $\smash{D_{x,t}^{(r)}}$ depends on the state of an underlying Markov chain $\smash{S_{t}^{(r)}}$, which varies by region and time. At each time point $t$, $\smash{S_{t}^{(r)}}$ can be in a baseline (state 0), environmental shock (state 1), or respiratory shock (state 2) state. For each state $i$, $\smash{D_{x,t}^{(r)}}$ follows a Poisson distribution:
\begin{equation} \label{eq:cond.dens}
\begin{aligned}
D_{x,t}^{(r)} \mid S_{t}^{(r)} = i \sim 
\text{POI}\left(\hat{b}_{x,t}^{(r)} \cdot \exp\left[\left(\boldsymbol{z}_{t}^{(r)}\right)^\top \boldsymbol{\alpha}_{i,x}\right]\right),
\end{aligned}
\end{equation}
where $\boldsymbol{\alpha}_{i,x}$ captures the effects of the covariate vector $\smash{\boldsymbol{z}_{t}^{(r)}}$ (e.g., temperature, influenza, COVID-19) on deviations from baseline mortality for age group $x$ in state $i$. This age dependency allows the model to account for how covariates, such as extreme temperatures, influence excess mortality differently across the considered age groups. 

Each regime state corresponds to a distinct mortality pattern. For $S_t^{(r)} = 0$, the baseline state, we assume mortality rates follow the seasonal baseline trend $\smash{\hat{b}_{x,t}^{(r)}}$, where no additional covariate effects influence mortality. We set $\boldsymbol{\alpha}_{0,x} = 0$. For $S_t^{(r)} = 1$, the environmental-shock state, only temperature covariates in $\smash{\boldsymbol{z}_t^{(r)}}$ play a role, and we restrict $\boldsymbol{\alpha}_{1,x}$ to have non-zero values only for these variables. Lastly, for $S_t^{(r)} = 2$, the respiratory shock state, we assume only epidemic covariates affect mortality, with non-zero values in $\boldsymbol{\alpha}_{2,x}$. 

To avoid overcomplicating the model, we assume the Markov chain $S_t^{(r)}$ is independent of the age group $x$. As a result, state transitions affect all ages simultaneously, but through age-dependent parameters in~\eqref{eq:cond.dens} we can still account for age-specific differences in the impact on mortality. The state transition probabilities are described by a time-varying, region-specific transition matrix $\smash{\boldsymbol{P}_{t}^{(r)}}$:
\begin{small}
\begin{center}
$\boldsymbol{P}_{t}^{(r)} = \left( \text{
\begin{tabular}{ccc}
$1 - p_{t}^{01} \left(\boldsymbol{z}_{t}^{(r)};\boldsymbol{\beta}, U_r\right)$  & 
\multirow{2}{*}{$p_{t}^{01} \left(\boldsymbol{z}_{t}^{(r)};\boldsymbol{\beta}, U_r\right)$} &
\multirow{2}{*}{$p_{t}^{02} \left(\boldsymbol{z}_{t}^{(r)};\boldsymbol{\beta}, U_r\right)$} \\
 $\hspace{1cm} -\: p_{t}^{02} \left(\boldsymbol{z}_{t}^{(r)};\boldsymbol{\beta}, U_r\right)$  &   &  \vspace{0.2cm}\\
$1-p_{t}^{11} \left(\boldsymbol{z}_{t}^{(r)};\boldsymbol{\beta}, U_r\right)$ & 
$p_{t}^{11} \left(\boldsymbol{z}_{t}^{(r)};\boldsymbol{\beta}, U_r\right)$ &
0  \vspace{0.1cm}\\
$1-p_{t}^{22} \left(\boldsymbol{z}_{t}^{(r)};\boldsymbol{\beta}, U_r\right)$ & 
0 &
$ p_{t}^{22} \left(\boldsymbol{z}_{t}^{(r)};\boldsymbol{\beta},U_r\right)$ 
\end{tabular}}\right)$,
\end{center}
\end{small}
with $\boldsymbol{\beta}$ the parameter vector, and $U_r$ a spatial effect to capture regional disparities. This spatial effect ensures that neighbouring regions are more likely to have similar transition probabilities. In this paper, we apply an Intrinsic Conditional Auto-Regressive (ICAR) model to the spatial effect vector \citep{besag1995conditional}:
\begin{align} \label{eq:icardens}
\boldsymbol{U} = (U_1, U_2, \ldots, U_R) \sim \mathcal{N}\left( \boldsymbol{0}, \left[\tau \cdot (\boldsymbol{D}-\boldsymbol{W}) \right]^{-1}\right),
\end{align}
where $R := |\mathcal{R}|$ is the region count, $\boldsymbol{D}$ a diagonal matrix with entries equal to the number of neighbours for each region, and $\boldsymbol{W}$ the adjacency matrix. The entry $W_{rr'}$ equals $1$ if the regions $r$ and $r'$ are neighbors and $0$ otherwise. The diagonal entries of $\boldsymbol{W}$ are zero. The precision parameter $\tau$ controls the level of spatial autocorrelation among the regions. 

We model the probability of transitioning from state $i$ to state $j$ using a multinomial logit specification. We therefore define $\mathcal{J}_i$ as the set of possible states, excluding the baseline state, to which the system can transition starting from state $i$. Specifically, $\mathcal{J}_0 = \{1,2\}$, $\mathcal{J}_1 = \{1\}$, and $\mathcal{J}_2 = \{2\}$.  For $ij \notin \{12,21\}$, the probability of transitioning from $i$ to $j$ is:
\begin{equation} \label{eq:trans.prob}
\begin{aligned}
p_{t}^{ij}\left(\boldsymbol{z}_{t}^{(r)};\boldsymbol{\beta}, u_r \right) = \begin{cases} \dfrac{\exp\left( \left(\boldsymbol{z}_{t}^{(r)}\right)^\top \boldsymbol{\beta}_{ij} + U_r\right)}{1 + \displaystyle \sum_{j' \in \mathcal{J}_i} \exp\left(\left(\boldsymbol{z}_{t}^{(r)}\right)^\top \boldsymbol{\beta}_{ij'} + U_r\right)} \quad  \quad j \neq  0\\
\dfrac{1}{1 + \displaystyle \sum_{j' \in \mathcal{J}_i} \exp\left(\left(\boldsymbol{z}_{t}^{(r)}\right)^\top \boldsymbol{\beta}_{ij'} + U_r \right)}  \quad  \quad j = 0, \end{cases}
\end{aligned}
\end{equation}
with $\boldsymbol{\beta} = (\boldsymbol{\beta}_{ij})_{ij}$. For transitions toward (or persistence in) the environmental shock state, only temperature-related covariates are included. For transitions toward (or persistence in) the respiratory shock state, only epidemic and COVID-19 hosptizalization variables are considered. Other coefficients are set to zero.

\section{Model calibration} \label{sec:modelcal}
\subsection{Weekly baseline mortality model} \label{subsec:modelcal1}
We fit the weekly baseline mortality model in Section~\ref{modelspec:baseline} using a Poisson generalized linear model (GLM) to capture region- and age-group-specific seasonal patterns in the death counts. To ensure smooth parameter estimates across neighboring regions, see \eqref{eq:baselinedeathsstructure}, we add a penalty term to the objective function \citep{robben2024association, li2024boosting}: 
\begin{align*}
\boldsymbol{\gamma}^\ast = \argmin_{\boldsymbol{\gamma}} \left(-\ell_P(\boldsymbol{\gamma}) + \sum_{p=0}^5 \lambda_p \left(\sum_{x\in\mathcal{X}} \boldsymbol{\gamma}_{x,p}^\top \left(\boldsymbol{D} - \boldsymbol{W}\right) \boldsymbol{\gamma}_{x,p}\right)\right),
\end{align*}
where $\boldsymbol{\gamma} = (\boldsymbol{\gamma}_{x,p})_{x,p}$, with $\boldsymbol{\gamma}_{x,p} = (\gamma_{x,p}^{(r)})_{r\in\mathcal{R}}$, for each age group $x \in \mathcal{X}$ and $p=0,1,...,5$. Furthermore, $\ell_P(\boldsymbol{\gamma})$ is the Poisson log-likelihood, and the matrices $\boldsymbol{D}$ and $\boldsymbol{W}$ are defined in~\eqref{eq:icardens}. The smoothness parameters $\lambda_p$ determine the amount of smoothing applied, with larger values imposing similar parameter estimates across regions. We select optimal $\lambda_p$-values using Un-Biased Risk Estimation scores. We calculate the expected baseline death counts for region $r$, week period $t$, and age group $x$ in~\eqref{eq:baselinedeathsstructure} using the optimal parameter vector $\boldsymbol{\gamma}^\ast$, and denote the result as $\hat{b}_{x,t}^{(r)}$.

\subsection{Regime-switching mortality framework} \label{subsec:modelcal2}
\subsubsection{Notation} \label{subsubsec:notation}
Let $\mathcal{D}_T$ denote the collection of age-specific death counts observed throughout the calibration period $\mathcal{T} = \{1,2,...,T\}$, i.e., $\mathcal{D}_{T} = \{D_{x,t}^{(r)} \mid r \in \mathcal{R}, x \in \mathcal{X}, t \in \mathcal{T} \}.$ We introduce a similar notation for the collection of all states, $\mathcal{S}_T$, and covariate vectors, $\mathcal{Z}_T$. Lastly, we denote the parameter vector for the regime-switching process as:
\begin{align} \label{eq:RSpvec}
\boldsymbol{\theta} = \left\{\boldsymbol{\alpha}_1, \boldsymbol{\alpha}_2, \boldsymbol{\beta}_{01}, \boldsymbol{\beta}_{02}, \boldsymbol{\beta}_{11}, \boldsymbol{\beta}_{22} \right\},
\end{align}
where $\boldsymbol{\alpha}_i = (\boldsymbol{\alpha}_{i,x})_{x \in \mathcal{X}}$  for $i=1,2$. The vector $\boldsymbol{\theta}$ consists of the parameters in~\eqref{eq:cond.dens} and~\eqref{eq:trans.prob}. 

\subsubsection{From the complete-data to the incomplete-data log-likelihood}
The complete-data likelihood assumes that the observed death counts $\mathcal{D}_T$ and latent regime states $\mathcal{S}_T$ are fully known. However, this likelihood depends on the spatial random effect $\boldsymbol{U}$. To account for this, we integrate out $\boldsymbol{U}$:
\begin{equation}\label{eq:likelihood.cond}
\begin{aligned}
\mathcal{L}\left(\boldsymbol{\theta},\tau \mid \mathcal{D}_T, \mathcal{S}_T, \mathcal{Z}_T\right) &= \displaystyle \int_{\boldsymbol{u}} \mathbb{P}\left(\mathcal{D}_T, \mathcal{S}_T, \boldsymbol{u} \mid \mathcal{Z}_T \:; \boldsymbol{\theta}, \tau \right) d\boldsymbol{u} \\
&= \int_{\boldsymbol{u}} \mathbb{P}\left(\mathcal{D}_T, \mathcal{S}_T\mid \boldsymbol{u},\mathcal{Z}_T; \boldsymbol{\theta}\right) f(\boldsymbol{u} \:; \tau) d\boldsymbol{u},
\end{aligned}
\end{equation}
where $\mathbb{P}\left(\mathcal{D}_T, \mathcal{S}_T\mid \boldsymbol{u},\mathcal{Z}_T; \boldsymbol{\theta}\right)$ is the complete-data likelihood conditionally on the spatial effect $\boldsymbol{u}$, and $f(\boldsymbol{u} \:; \tau)$ is the prior distribution of $\boldsymbol{u}$, which follows a multivariate normal distribution with precision parameter $\tau$, see~\eqref{eq:icardens}: 
\begin{align*}
f(\boldsymbol{u} \:; \tau) &= \dfrac{1}{(2\pi)^{R/2} (\det(\boldsymbol{Q}(\tau)^{-1}))^{1/2}} e^{-\frac{1}{2} \boldsymbol{u}^\top \boldsymbol{Q}(\tau)\boldsymbol{u}},
\end{align*}
where $\boldsymbol{Q}(\tau) := \tau [\boldsymbol{D} - \boldsymbol{W}]$ is called the precision matrix that governs the degree of spatial autocorrelation across the regions. 

It is computationally challenging for regime-switching models to evaluate the integral in~\eqref{eq:likelihood.cond}. To simplify this, we apply Laplace's approximation method to avoid a direct calculation of the integral expression \citep{laplace1774memoire, tierney1986accurate}:
\begin{equation} \label{eq:objL}
\begin{aligned} 
&\log \mathcal{L}(\boldsymbol{\theta}, \tau\mid \mathcal{D}_T, \mathcal{S}_T, \mathcal{Z}_T) 
\\ &\hspace{1cm}= \log \mathbb{P}\left(\mathcal{D}_T, \mathcal{S}_T \mid \boldsymbol{u}^\ast, \mathcal{Z}_T \:; \boldsymbol{\theta} \right) + \log f(\boldsymbol{u}^\ast\:;\tau) + \frac{R}{2} \log (2\pi) - \frac{1}{2} \log(\det(\boldsymbol{S}^{-1})),
\end{aligned}
\end{equation}
where $\boldsymbol{u}^\ast$ is the mode of the expected log joint probability distribution, and  $\boldsymbol{S}^{-1}$ is the matrix of second-order derivatives evaluated at $\boldsymbol{u}^\ast$. Suppl.~Mat.~\ref{app:condloglik} and~\ref{app:hessian} provide details on the calculation of the Laplace-approximated log-likelihood and the involved Hessian.

We are not able to directly optimize the complete-data log-likelihood in~\eqref{eq:objL} because the regime states $S_t^{(r)}$ are unobserved. Instead, we focus on maximizing the incomplete-data log-likelihood, which aggregates over all possible regime state configurations:
\begin{align*}
&\log \tilde{\mathcal{L}}(\boldsymbol{\theta}, \tau \mid \mathcal{D}_T, \mathcal{Z}_T) = \:\:\:\:\: \displaystyle \smashoperator{\sum_{\mathcal{S}_T \in \{0,1,..,N\}^{R\cdot T}}} \: \log \mathcal{L}(\boldsymbol{\theta}, \tau \mid \mathcal{D}_T, \mathcal{S}_T, \mathcal{Z}_T).
\end{align*}
However, it is practically impossible to directly evaluate this expression, as it would require us to sum over a very large number of possible state combinations. Therefore, we apply the Expectation-Maximization (EM) algorithm to optimize the incomplete-data log-likelihood.

\subsubsection{Optimizing the incomplete-data log-likelihood using the EM-algorithm} \label{subsec:modelestimationEM}
Since the values of the spatial effect $\boldsymbol{u}^\ast$ depend on the parameter vector $\boldsymbol{\theta}$, we incorporate Laplace's approximation of the complete-data log-likelihood into the EM-algorithm to iteratively update both the parameter vector $\boldsymbol{\theta}$ and the spatial effect $\boldsymbol{u}^\ast(\boldsymbol{\theta})$. Below is a summary and explanation of the key steps of which Suppl.~Mat.~\ref{app:modelestimationEM} provides the details:
\begin{enumerate}
\item Initialization: Start with initial estimates $\boldsymbol{\theta}^\ast_0$ and $\boldsymbol{u}^\ast_0$.
\item EM-algorithm (iteration $k+1$):
\begin{enumerate} 
\item E-step: Take the expected value of the complete-data log-likelihood with respect to the regime state process, given the observed data, covariate information, and the current best-estimates of the parameter vector, $\boldsymbol{\theta}^\ast_k$, and spatial effect, $\boldsymbol{u}^\ast_k$.
\item M-step: Update the parameter vector $\boldsymbol{\theta}$ by maximizing the expected complete-data log-likelihood and obtain $\boldsymbol{\theta}_{k+1}^\ast$.
\item Update spatial effect: Update the spatial effect, $\boldsymbol{u}^\ast$, under the current optimal parameter vector, $\boldsymbol{\theta}^\ast_{k+1}$, and obtain $\boldsymbol{u}_{k+1}^\ast$.
\end{enumerate}
\item Convergence check: Repeat steps (a), (b), and (c) until we reach an absolute convergence of $\epsilon$ (a predefined small threshold) in the expected complete-data log-likelihood.
\end{enumerate}
The final step is to estimate the precision parameter $\tau$ of the ICAR prior. We do this maximizing the profile log-likelihood over a grid of candidate values $\mathcal{G}$. For each $\tau \in \mathcal{G}$, we apply the EM algorithm to obtain the optimal parameter vector $\boldsymbol{\theta}^\ast(\tau)$. We then select the optimal $\tau^\ast$ as the value that maximizes the profile log-likelihood, see Suppl.~Mat.~\ref{subsubsec:precisionparam}.

We can then assess the uncertainty in the estimated parameters of the regime-switching process by deriving the so-called Fisher-Information matrix (FIM). Existing methods, such as the ones proposed by \cite{louis1982finding}, \cite{meng1991using} and \cite{oakes1999direct}, often involve computational challenges. Therefore, we follow the approach of \cite{meng2017efficient}, who propose a Monte Carlo-based method using simultaneous perturbation stochastic approximation to estimate the Hessian matrix of the incomplete data log-likelihood \citep{spall2000adaptive}. Suppl.~Mat.~\ref{app:paramuncertainty} outlines the estimation procedure.

\section{Multi-layered uncertainty in mortality predictions} \label{sec:quantify.uncertainty}
When projecting weekly death counts with the proposed model, we account for parameter, spatial, regime-switching state, and Poisson uncertainty. We account for these uncertainties to improve the reliability of the mortality predictions. We do this by generating $B$ bootstrap samples according to the below outlined prediction process. For each sample $\iota$, we proceed as follows:

\begin{itemize}
    \item[1.] \textbf{Parameter uncertainty.} We account for parameter uncertainty in the model's predictions by drawing randomly from the following multivariate normal distributions:
    \begin{align}
        \boldsymbol{\gamma}_{\iota} &\sim \mathcal{N}\left(\boldsymbol{\gamma}^\ast, \boldsymbol{\Sigma}_1(\boldsymbol{\gamma}^\ast) \right), \hspace{0.75cm}
        \boldsymbol{\theta}_{\iota} \sim \mathcal{N}\left( \boldsymbol{\theta}^\ast, \boldsymbol{\Sigma}_2(\boldsymbol{\theta}^\ast)\right),
    \end{align}
    where $\boldsymbol{\gamma}^\ast$ and $\boldsymbol{\theta}^\ast$ are the optimal parameter vectors in the baseline and regime-switching model, respectively. We construct the variance-covariance matrices $\boldsymbol{\Sigma}_1(\boldsymbol{\gamma}^\ast)$ and $\boldsymbol{\Sigma}_2(\boldsymbol{\theta}^\ast)$ from the Fisher information of their corresponding log-likelihood functions. The construction of $\boldsymbol{\Sigma}_2(\boldsymbol{\theta}^\ast)$ is more intricate and detailed in Suppl.~Mat.~\ref{app:paramuncertainty}.
    

    \item[2.] \textbf{Spatial uncertainty.} We account for regional disparities in the transition probabilities by drawing a random sample from the ICAR model, specified in~\eqref{eq:icardens}:
    \begin{align*}
        \boldsymbol{u}_{\iota} = \left(u_{1,\iota}, u_{2,\iota}, \ldots, u_{R,\iota}\right) \sim \mathcal{N}\left( \boldsymbol{0}, \left[\tau \cdot (\boldsymbol{D}-\boldsymbol{W}) \right]^{-1}\right).
    \end{align*}
    
    \item[3.] \textbf{State uncertainty.}  The state uncertainty reflects the uncertainty about the true values of the latent regime states when making predictions. To account for this, we generate regime state trajectories using the estimated parameters in the transition probabilities, see~\eqref{eq:trans.prob}. 
    
    We start by selecting an initial state $i$. This is either the most probable state at the end of the calibration period (for forecasting) or the most probable initial state (for in-sample prediction). The next state is then generated by sampling from a discrete distribution with possible values $j\in\{0,1,2\}$, weighted by transition probabilities $p_t^{ij}(\boldsymbol{z}_t^{(r)}; \boldsymbol{\beta}_{\iota},u_{r,\iota})$, where $\boldsymbol{z}_t^{(r)}$ is the covariate vector, $u_{r,\iota}$ is the sampled spatial effect, and $\boldsymbol{\beta}_{\iota}$ is the sampled parameter vector contained in $\boldsymbol{\theta}_{\iota}$. We repeat this iterative procedure for all $t$ in the prediction period $\mathcal{T}_{\text{pred}}$. We denote the resulting regime state trajectory as $s_{t,\iota}^{(r)}$ for $t \in \mathcal{T}_{\text{pred}}$.

    \item[4.] \textbf{Poisson uncertainty.} Using the sampled states $s_{t,\iota}^{(r)}$, we sample death counts from the state-dependent Poisson distributions in~\eqref{eq:cond.dens} with the parameter vector $\boldsymbol{\theta}_\iota$. We calculate the baseline death counts using~\eqref{eq:baselinemodelspec} and \eqref{eq:baselinedeathsstructure}. This corresponds to either the calibrated baseline (for in-sample prediction) or a forecasted mean from the Poisson GLM (for out-of-sample forecasting). In the latter case, we need to use scenarios for the covariate vectors $\boldsymbol{z}_t^{(r)}$.
\end{itemize}

Next, we construct confidence intervals for the predicted death counts by calculating the $2.5\%$, $50\%$, and $97.5\%$ percentiles of the $B$ bootstrap samples. As a result, we obtain a range of outcomes in which the predicted death counts may fall with a $95\%$ confidence rather than a point estimate. Lastly, note that we can switch any source of uncertainty on or off, independently of each other. This allows us to control for the factors contributing to the uncertainty and to detect the sources of uncertainty in the predicted death counts.

\section{Case study on 21 French NUTS 2 regions} \label{sec:casestudy}

\subsection{Baseline and regime-switching model: specification and calibration}
\subsubsection{Baseline mortality model}
\begin{figure}[!ht]
\centering
\includegraphics[width=0.85\textwidth]{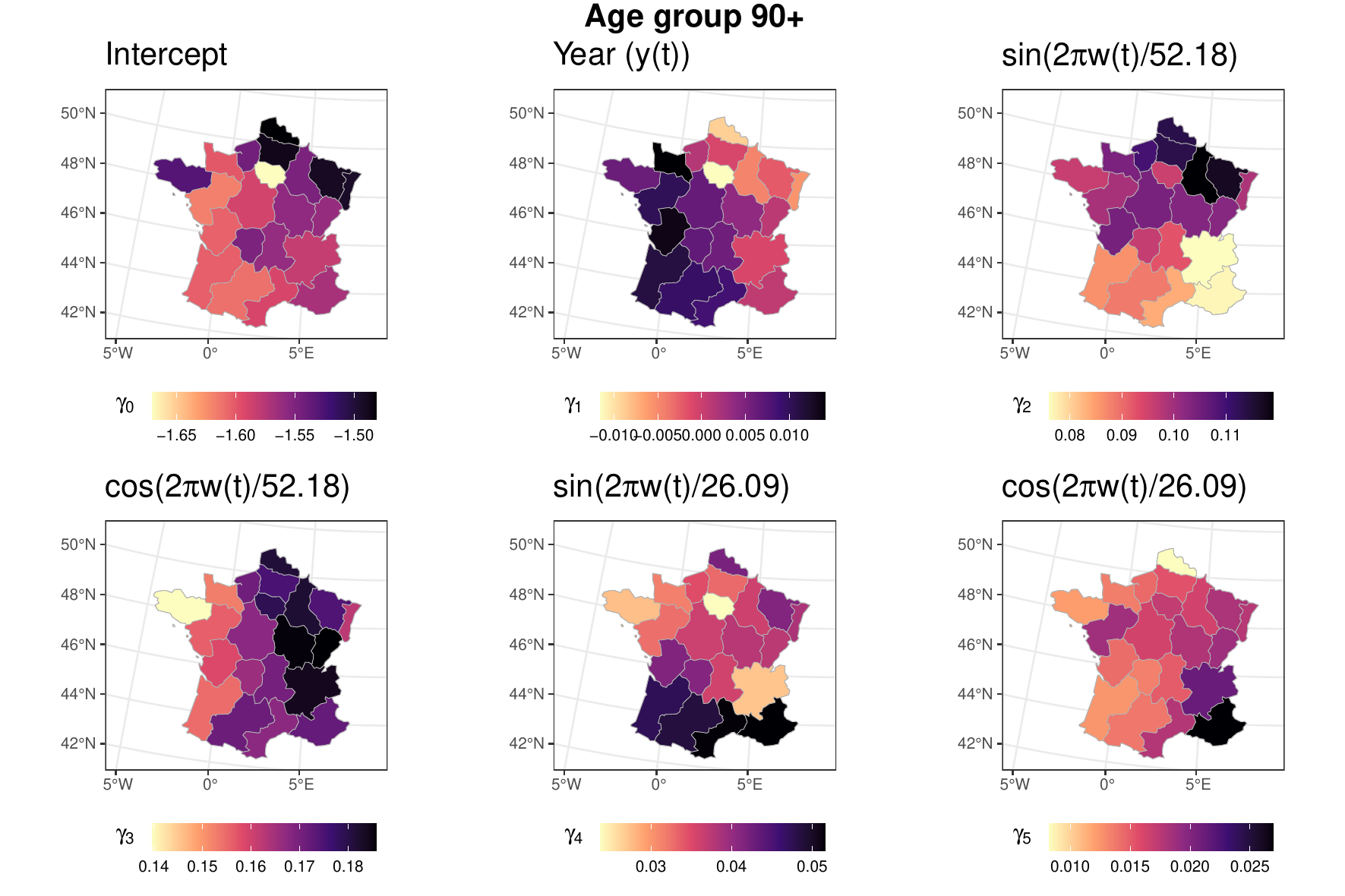}
\caption{The estimated parameters in the baseline mortality model, i.e.,  $\gamma_{x,p}^{(r)}$, for the age group 90+ in the French NUTS 2 regions.\label{fig:bparamage90}}
\end{figure}

We calibrate the baseline parameters $\gamma_{x,p}^{(r)}$, using the strategy in Section~\ref{subsec:modelcal1}, on the set $\mathcal{R}$ of French NUTS 2 regions, age groups $\mathcal{X} = \{65$-$69, 70$-$74, 75$-$79, 80$-$84, 85$-$89, 90$+$\}$, and weeks $\mathcal{T} = \{1,2,.., 600\}$, where $t=1$ corresponds to the first ISO week in 2013 and $t = 600$ to the 26th ISO week in 2024. We omit the weeks related to the first wave of the COVID-19 pandemic (weeks 12-16), as they heavily influence the overall performance of the baseline trend. Figure~\ref{fig:bparamage90} shows the estimated parameters for the age group 90+. The top left and middle panels show that Île-de-France had the lowest overall death rate in 2013 ($\gamma_0$) and experienced the largest mortality decline. Eastern French regions also saw a decline in mortality, while the Western regions experienced a slight increase ($\gamma_1$). The seasonal variation was most prominent in Northern(-Eastern) regions. Section A of the Online Appendix contains the parameter estimates for the five remaining age groups.

\subsubsection{Regime-switching mortality model}

\paragraph{State-specific model specifications.} We use the variables constructed from the temperature, influenza, and hospitalization data in Table~\ref{tab:features} as features in the state-dependent Poisson models in~\eqref{eq:cond.dens}. Since their impact on mortality differs by age, we use age-dependent parameters. We further assume shared parameters among two consecutive age groups, and focus, as such, on the reduced set of age groups: $\mathcal{X}_{\text{red}} = \{$65–74, 75–84, 85+$\}$.

In state 1, we model heatwave-related shocks by incorporating weekly averages of daily temperature anomalies and extreme temperature indices with up to two weeks' lag, based on evidence from \cite{armstrong2006models} and \cite{gasparrini2011impact}. These features provide a good measure of the heat frequency and intensity within a given week. We model the expected death counts in regime state 1 as:
\begin{equation} \label{eq:denshot1}
\begin{aligned}
\log \mathbb{E}\left[D_{x,t}^{(r)} \mid S_t^{(r)} = 1 \right] = \log \hat{b}_{x,t}^{(r)} +  &\displaystyle \sum_{a \in \mathcal{X}_{\text{red}}} \Bigg(\alpha_{1,1}^{(a)} \text{TA}_{t}^{(r)} +  \alpha_{1,2}^{(a)} \text{TA}_{t-1}^{(r)} + \alpha_{1,3}^{(a)} \text{TA}_{t-2}^{(r)}  + \\
&\hspace{1.4cm}\alpha_{1,4}^{(a)} \text{HI}_{t}^{(r)} +  \alpha_{1,5}^{(a)} \text{HI}_{t-1}^{(r)} + \alpha_{1,6}^{(a)} \text{HI}_{t-2}^{(r)}\Bigg) \scalebox{1.2}{$\mathbbm{1}$} \left\{ x \subset a\right\},
\end{aligned}
\end{equation}
where the $\alpha_{1,j}^{(a)}$'s are age-specific parameters, $\hat{b}_{x,t}^{(r)}$ refers to the baseline death count estimated as in Section~\ref{subsec:modelcal1}, and $\mathbbm{1}\{x\subset a\}$ is an indicator function which equals one whenever the smaller age group $x\in \mathcal{X}$ is a subset of the bigger age group $a \in \mathcal{X}_{\text{red}}$. We refer to Table~\ref{tab:features_summary} for the notation of the covariates, where, for instance, $\text{TA}_{t-1}^{(r)}$ refers to the weekly average of the daily temperature anomalies at week $t-1$ in region $r$ (lag 1).

In state 2, we model mortality shocks from influenza activity and COVID-19 hospitalizations, and incorporate a cold indicator due to its additional impact on mortality. Since respiratory-related mortality effects are often delayed, we use a maximum lag of 3 weeks, based on evidence from \cite{wong2004influenza} and \cite{lytras2019mortality}. Furthermore, we group the lags in short-term (weeks 0–1) and mid-term (weeks 2–3) intervals to reduce multicollinearity issues. Long-term impacts are not considered in this paper. We model the expected death counts in state 2 as:
\begin{equation} \label{eq:denshot2}
\begin{aligned}
\log \mathbb{E}\left[D_{x,t}^{(r)} \mid S_t^{(r)} = 2 \right] &= \log \hat{b}_{x,t}^{(r)} +  \displaystyle \sum_{a \in \mathcal{X}_{\text{red}}} \Bigg( \alpha_{2,1}^{(a)} \text{IA}_{t,t-1}^{(r)} + \alpha_{2,2}^{(a)} \text{IA}_{t-2,t-3}^{(r)} + \alpha_{2,3}^{(a)} \text{CI}_{t,t-1}^{(r)}\\
&\hspace{1.75cm}\alpha_{2,4}^{(a)} \text{CI}_{t-2,t-3} + \alpha_{2,5}^{(a)} \text{HA}_{t,t-1}^{(r)} + \alpha_{2,6}^{(a)} \text{HA}_{t-2,t-3}^{(r)}\Bigg)  \scalebox{1.2}{$\mathbbm{1}$} \left\{ x \subset a\right\},
\end{aligned}
\end{equation}
where $\text{IA}^{(r)}_{t,t-1}$ and $\smash{\text{IA}^{(r)}_{t-2,t-3}}$ refer to the influenza activity index at week $t$ averaged over lags 0-1 and 2-3, respectively. Analogous notation is introduced for the cold indicator (CI) and the COVID-19 hospitalization index (HA). For the respiratory-related features, we use exceedances over their 75$\%$ quantile since extremer values, rather than moderate fluctuations, are more likely to contribute to significant mortality shocks.

\paragraph{Modeling regime transition probabilities.} We construct multinomial logistic models for the transition probabilities, as defined in~\eqref{eq:trans.prob}. Following Section~\ref{subsec:modelspec2}, state transition probabilities are independent of age group $x$. To capture regional variations, we introduce a spatial effect $U_r$, which we model using an ICAR process. Specifically, we define:
\begin{equation}\label{eq:transprobcasest}
\begin{aligned} 
    \text{logit} \: p_t^{01}\left(\boldsymbol{z}_t^{(r)};\: \boldsymbol{\beta}, u_r\right) &= \beta_{01,0} + \beta_{01,1} \text{HI}_t + U_r \\
    \text{logit} \: p_t^{02}\left(\boldsymbol{z}_t^{(r)};\: \boldsymbol{\beta}, u_r\right) &= \beta_{02,0} + \beta_{02,1} \text{IA}_{t,t-1}^{(r)} + \beta_{02,2} \text{HA}_{t,t-1}^{(r)} + U_r\\ 
    \text{logit} \: p_t^{11}\left(\boldsymbol{z}_t^{(r)};\: \boldsymbol{\beta}, u_r\right) &= \beta_{11,0} + \beta_{11,1} \text{HI}_t + \beta_{11,2} \text{HI}_{t-1} + \beta_{11,3} \text{HI}_{t-2} + U_r\\    
    \text{logit} \: p_t^{22}\left(\boldsymbol{z}_t^{(r)};\: \boldsymbol{\beta}, u_r\right) &= \beta_{22,0} + \beta_{22,1} \text{IA}_{t,t-1}^{(r)}  +  \beta_{22,2} \text{IA}_{t-2,t-3}^{(r)} +  \\ 
    &\hspace{3cm}\beta_{22,3}\text{HA}_{t,t-1}^{(r)}   +  \beta_{22,4} \text{HA}_{t-2,t-3}^{(r)} + U_r.
\end{aligned}
\end{equation}
Here, we focus solely on features capturing short-term mortality impacts to model the transition probabilities from the baseline to one of the shock states. This facilitates a direct transition to the shock state whenever we observe increased temperature levels or respiratory infections. Conversely, to determine whether to remain in the current state, we incorporate lagged features to account for state persistence and delayed effects.

\paragraph{Parameter estimates.} We follow Section~\ref{subsec:modelestimationEM} and apply the EM-algorithm to estimate the regime-switching parameter vector $\boldsymbol{\theta}$. Following Suppl.~Mat.~\ref{subsubsec:precisionparam}, we apply profile likelihood maximization to select the precision parameter $\tau$. We use the grid $\mathcal{G} = $ $\{0.001,$ $0.01,0.1,1,10,100,1000\}$ as candidate values and select $\tau^\ast = 10$.

\begin{figure}[!ht]
    \centering
    \includegraphics[width=0.85\linewidth]{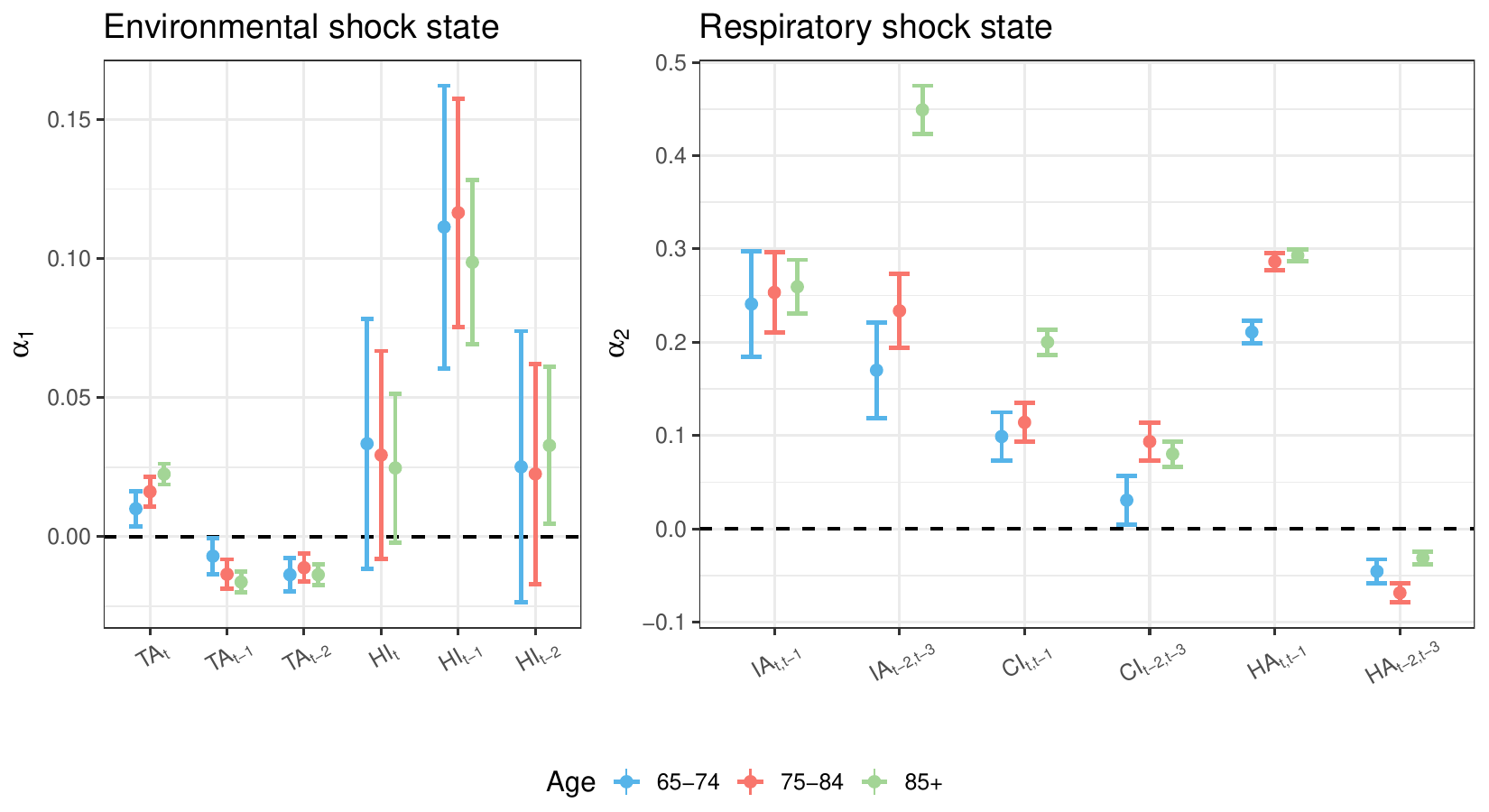}
    \caption{Parameter estimates for $\boldsymbol{\alpha}_1$ (environmental shock state) and $\boldsymbol{\alpha}_2$ (respiratory shock state) in the Poisson models of the regime-switching model, along with their 95$\%$ confidence intervals based on 25\ 000 bootstrap samples.}
    \label{fig:PAsev}
\end{figure}

Figure~\ref{fig:PAsev} (left panel) shows the parameter estimates for $\boldsymbol{\alpha}_1$. Following Suppl.~Mat.~\ref{app:paramuncertainty}, we construct 95$\%$ confidence intervals using the Fisher Information matrix based on 25 000 bootstrap samples. Elevated temperature anomalies in the current week are strongly associated to excess mortality, especially for those aged 85+. At lag 1, effects differ depending on the interaction between heat frequency (hot-week index) and severity (temperature anomalies). At lag 2, a harvesting effect is likely to occur across all age groups for severely hot weeks. However, the wide confidence intervals for the hot-week index, potentially due to collinearity with temperature anomalies, suggest caution in interpreting this effect.

The right panel of Figure~\ref{fig:PAsev} shows the estimates and 95$\%$ confidence intervals for the parameter vector $\boldsymbol{\alpha}_2$, which captures the effects of increased influenza activity or COVID-19 hospitalizations on mortality. Results reveal significant excess mortality from high influenza activity in both the short- and mid-term, especially for ages 85+. Cold temperatures also significantly increase mortality in both periods and COVID-19 hospitalizations show a significant short-term impact on excess mortality, increasing with age. We observe a modest harvesting effect at lags 2–3 for the COVID-19 hospitalizations.

\begin{figure}[!ht]
    \centering
    \includegraphics[width=0.85\linewidth]{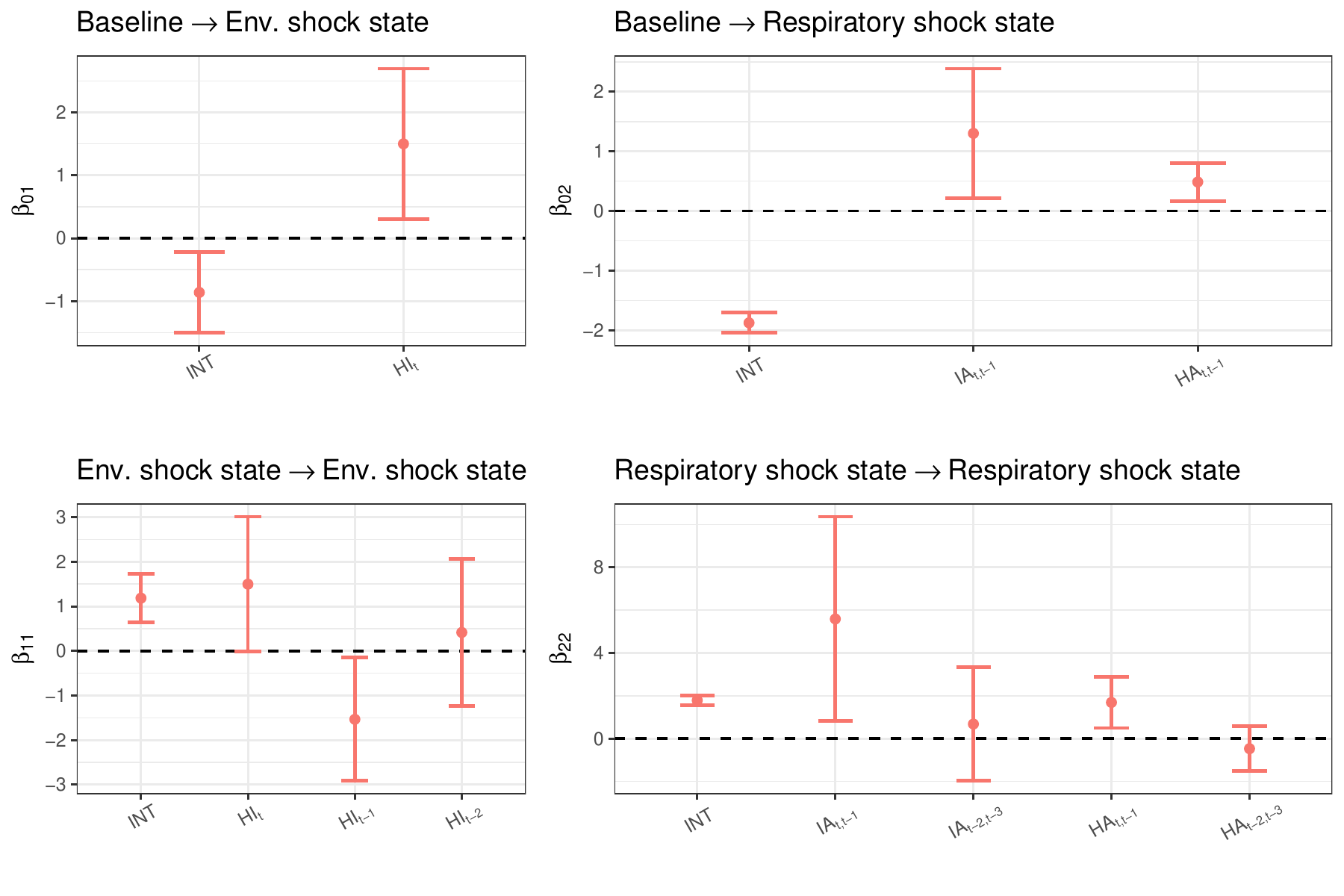}
    \caption{Estimates of the parameter vectors in the transitional probabilities and their 95$\%$ confidence intervals based on 25\ 000 bootstrap samples: $\boldsymbol{\beta}_{01}$ (top left), $\boldsymbol{\beta}_{02}$ (top right), $\boldsymbol{\beta}_{11}$ (bottom left), and $\boldsymbol{\beta}_{22}$ (bottom right). The label `INT' refers to the intercept }
    \label{fig:PAtrans}
\end{figure}

Figure~\ref{fig:PAtrans} displays the estimates and 95$\%$ confidence intervals of the parameter vectors $\boldsymbol{\beta}_{01}$, $\boldsymbol{\beta}_{02}$, $\boldsymbol{\beta}_{11}$, and $\boldsymbol{\beta}_{22}$ in the transition probabilities of~\eqref{eq:transprobcasest}. Transitions from the baseline to the environmental shock state are more likely with a high current-week hot-week index. Transitions to the respiratory shock state increase with higher short-term influenza activity and COVID-19 hospitalizations. Once in the environmental shock state, the probability of remaining in this state increases with the current-week hot-week index but decreases with the previous week's index. In the respiratory shock state, the likelihood of staying increases with higher short-term influenza activity or hospital admissions. The long-term effects (lags 2–3) are insignificant.

\begin{figure}[!ht]
    \centering
    \includegraphics[width=0.4\linewidth]{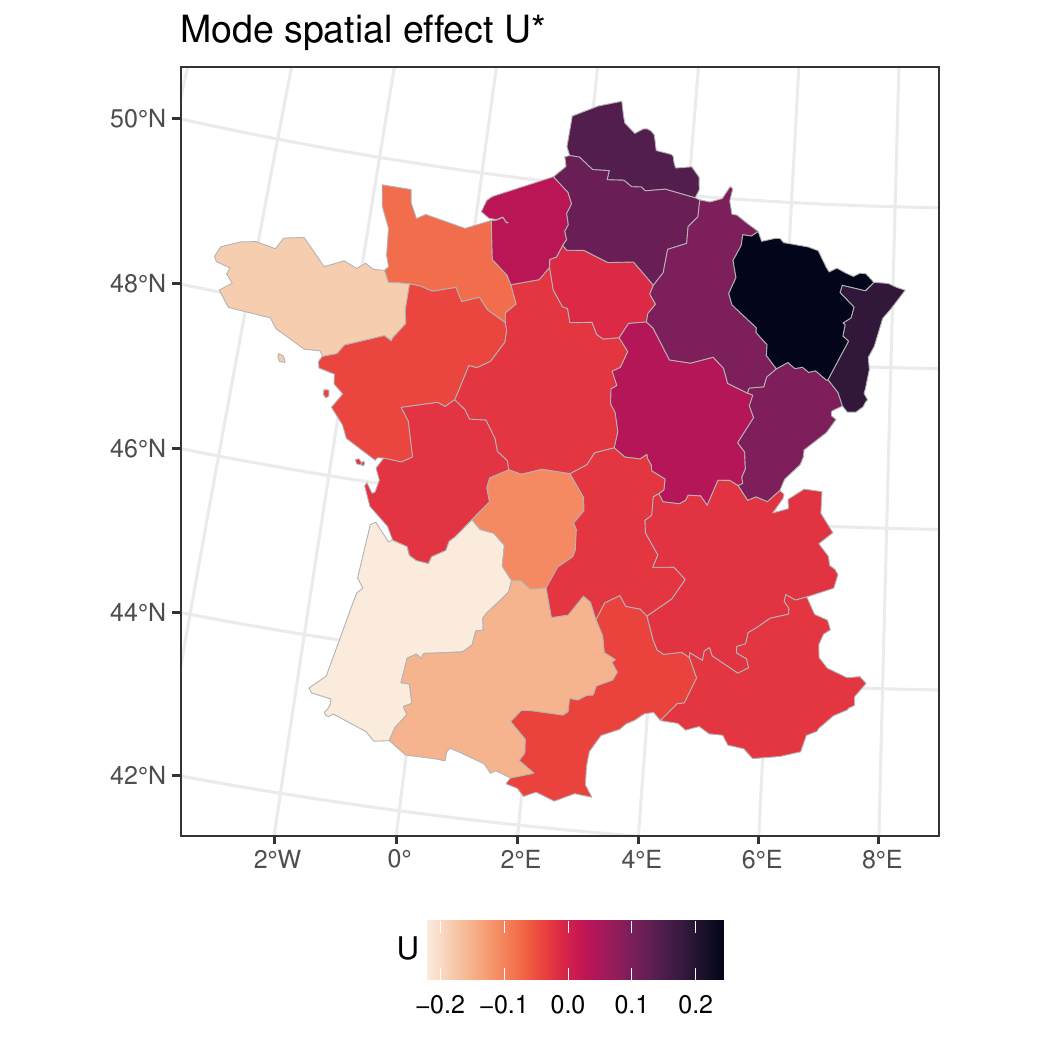}
    \caption{The mode of the spatial effect $\boldsymbol{u}^\ast$ in the expected log joint density,.}
    \label{fig:modeU}
\end{figure}

\paragraph{Spatial effect.} Figure~\ref{fig:modeU} visualizes the estimated mode of the expected log joint density, i.e., the spatial effect vector $\boldsymbol{u}^\ast$ (see Suppl.~Mat.~\ref{app:modelestimationEM}). The mode is the largest in North-East France, implying that regime state transitions occur more frequently in these regions.

\subsection{Quantifying uncertainty in in-sample model predictions}

\subsubsection{Best-estimate in-sample fit}
We use the estimated baseline mortality model and construct a BE in-sample fit for the regime-switching framework. We do this by using the estimated marginal state probabilities. For any $t \in \mathcal{T}$, we define the BE regime-switching trajectory as:
\begin{align} \label{eq:bestate}
     s_{\text{BE},t}^{(r)} = \argmax_{j\in\{0,1,2\}} \hat{P}\left(S_t^{(r)} = j \mid \mathcal{D}_t^{(r)}, \mathcal{Z}_t^{(r)}; \boldsymbol{\theta}^{\ast}, \boldsymbol{u}^\ast \right),
\end{align}
where $\boldsymbol{\theta}^\ast$ and $\boldsymbol{u}^\ast$ are the optimal regime-switching parameter vector and the mode of the spatial effect. The sets $\mathcal{D}_t^{(r)}$ and $\mathcal{Z}_t^{(r)}$ denote the collection of death counts and covariate vectors up to time $t$ in region $r$. Suppl.~Mat.~\ref{app:modelestimationEM} outlines the calculation of the marginal state probabilities. We define the BE prediction for the weekly death counts, see~\eqref{eq:cond.dens}, as:
\begin{align*}
    \hat{d}^{(r)}_{\text{BE},x,t} = \hat{\mathbb{E}}\left[D_{x,t}^{(r)} \mid S_t^{(r)} = s_{\text{BE},t}^{(r)}\right].
\end{align*}

\begin{figure}[!ht]
    \centering
    \includegraphics[width=0.85\linewidth]{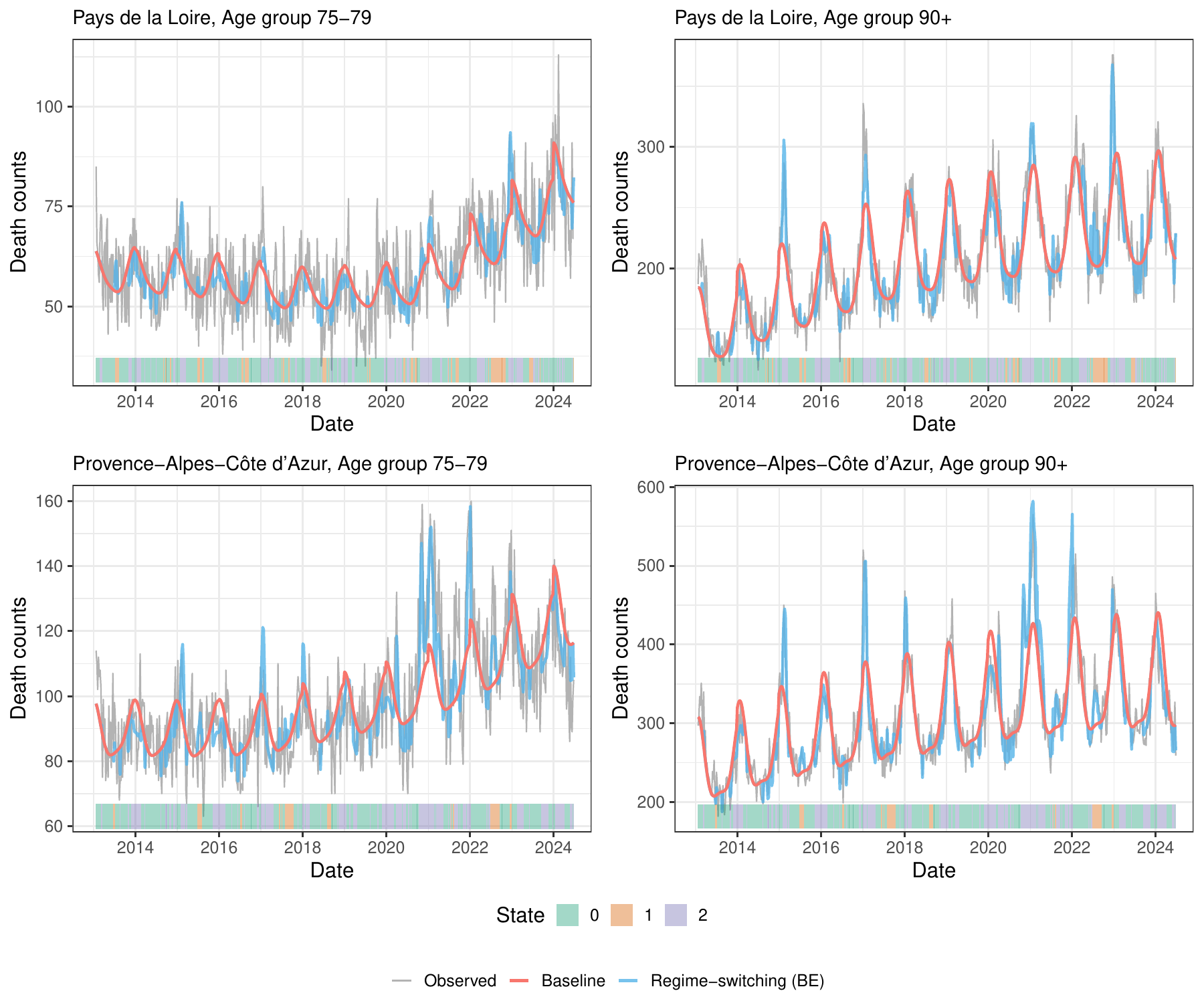}
    \caption{The grey line represents the observed death counts from the first ISO week of 2013 to the 26th ISO week of 2024. The red line visualizes the estimated baseline number of deaths, while the blue line visualizes the best-estimate (BE) prediction of the death counts resulting from the regime-switching model. At the bottom, we show the BE regime-switching trajectory. We show the results for the age groups 75-79 (left) and 90+ (right) in Pays de la Loire (top) and Provence--Alpes-Côte d'Azur (bottom).}
    \label{fig:bedtxr}
\end{figure}

Figure~\ref{fig:bedtxr} presents the BE in-sample predicted weekly death counts in Pays de la Loire and Provence-Alpes-Côte d'Azur for the age groups 75–79 and 90+. Section B of the Online Appendix contains the figures for all French NUTS 2 regions and age groups. The coloured bars at the bottom of each panel indicate the BE regime-switching trajectory. The figure shows that our model captures well the observed spikes in the weekly mortality patterns. The inclusion of the (lagged) hospital admissions index is useful to account for the impact of the COVID-19 pandemic on mortality. Furthermore, the model successfully captures excess deaths linked with increased influenza activity during the winter seasons of 2014/2015 and 2016/2017, and those related to heatwaves in the summer seasons of 2015 and 2022.

\subsubsection{Uncertainty in in-sample predictions}
\begin{figure}[!ht]
    \centering
    \includegraphics[width=0.88\linewidth]{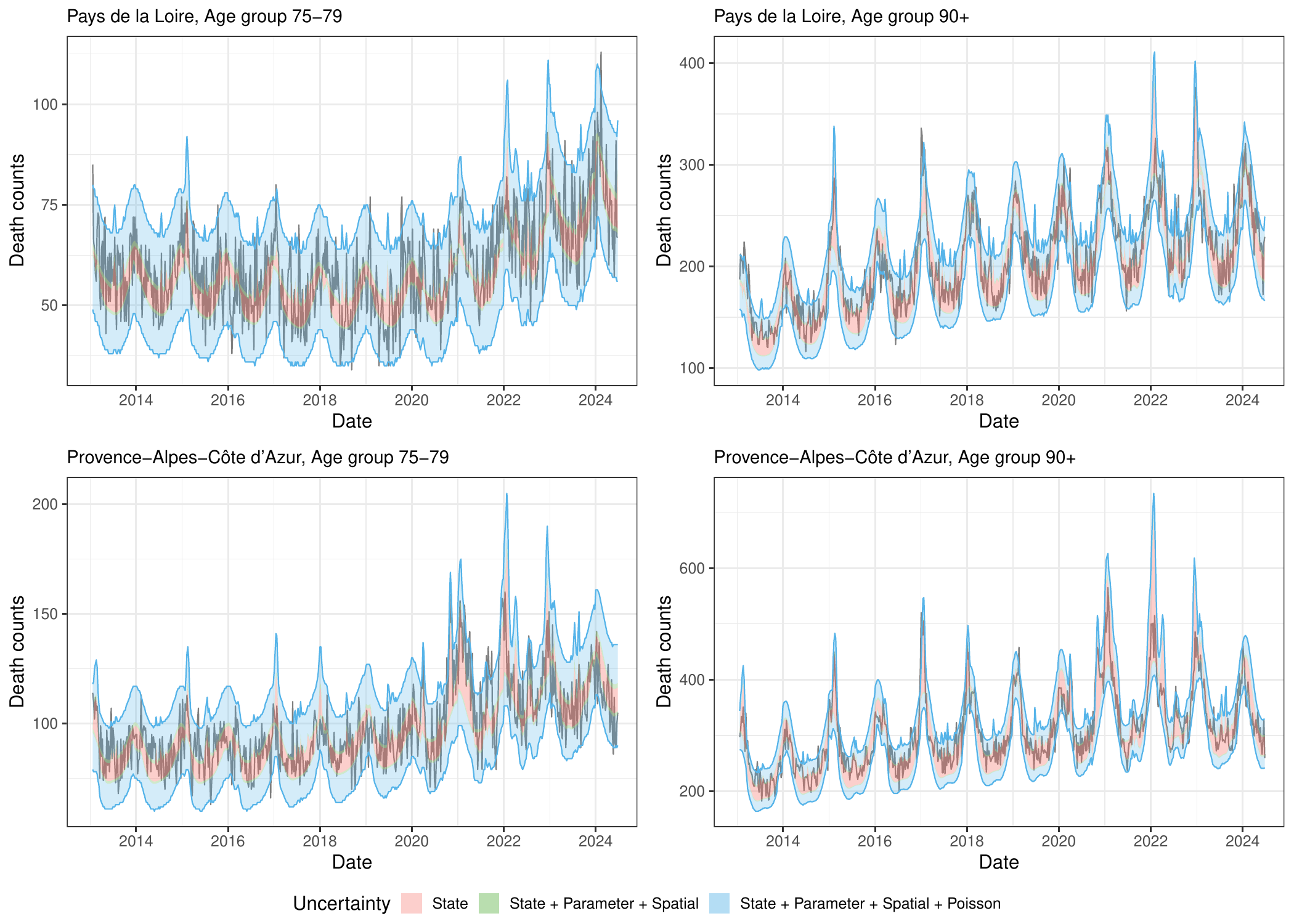}
    \caption{We present $95\%$ confidence intervals for the weekly death counts in Pays de la Loire (top) and Provence-Alpes-Côte d'Azur (bottom) for the age groups 75–79 (left), and 90+ (right) from the first ISO week of 2013 till the 26th ISO week of 2024. The intervals represent different sources of uncertainty: state uncertainty only (red), state, parameter, and spatial uncertainty (green), and state, parameter, spatial, and Poisson uncertainty (blue).}
    \label{fig:uncdtxr}
\end{figure}

We follow the methodology in Section~\ref{sec:quantify.uncertainty} and quantify the uncertainty in the in-sample predicted weekly death counts of the regime-switching modeling framework. Using a bootstrap sampling approach with 25\ 000 samples, we construct 95$\%$ prediction intervals. Figure~\ref{fig:uncdtxr} provides an example of how state, parameter, spatial, and Poisson uncertainty contribute to the intervals for weekly death counts in the age groups 75-79 and 90+ within the Pays de la Loire and Provence-Alpes-Côte d'Azur NUTS 2 region. Section C of the Online Appendix contains the in-sample prediction intervals for all age groups and NUTS 2 regions considered in this paper. First, state uncertainty effectively captures spikes in weekly mortality patterns whenever they correspond to external factors such as heatwaves, heightened influenza activity, or increases in COVID-19 hospitalization rates. Second, Poisson uncertainty is especially prominent in the younger old-age groups. The underlying reason is the relatively small number of weekly death counts, and, consequently, the more random fluctuations. In contrast, older age groups report higher weekly death counts, leading to substantially less Poisson uncertainty. Third, the relatively limited parameter and spatial uncertainty is more pronounced in the younger old-age groups. 

\subsection{Out-of-sample back-testing}
We conduct out-of-sample back-testing to evaluate the model's performance on unseen data. Specifically, we limit the calibration period to the first ISO week of 2013 until the 26th ISO week of 2022, denoted as $\tilde{\mathcal{T}}$. The out-of-sample period ranges from the 27th week of 2022 to the 26th week of 2024, covering two years, and we refer to it as $\mathcal{T}_{\text{pred}}$.

We first reconstruct the features using the smaller calibration period $\tilde{\mathcal{T}}$. For instance, we recompute the temperature anomalies based on a temperature baseline calibrated on $\tilde{\mathcal{T}}$. Second, we recalibrate the baseline mortality model in Section~\ref{subsec:modelcal1}, and the three-state regime-switching model in Section~\ref{subsec:modelspec2}, on $\tilde{\mathcal{T}}$. Third, we generate projections for the death counts in $\mathcal{T}_{\text{pred}}$ while accounting for parameter, spatial, state, and Poisson uncertainty, as described in Section~\ref{sec:quantify.uncertainty}. In the back-testing, we explore a hypothetical scenario that uses observed values for the temperature anomalies, influenza activity, and COVID-19 hospitalizations for the two-year prediction period. This allows us to gain insights into how well the model captures excess deaths conditionally on the observed evolution of these covariates.

\begin{figure}[!htb]
    \centering
    \includegraphics[width=0.85\linewidth]{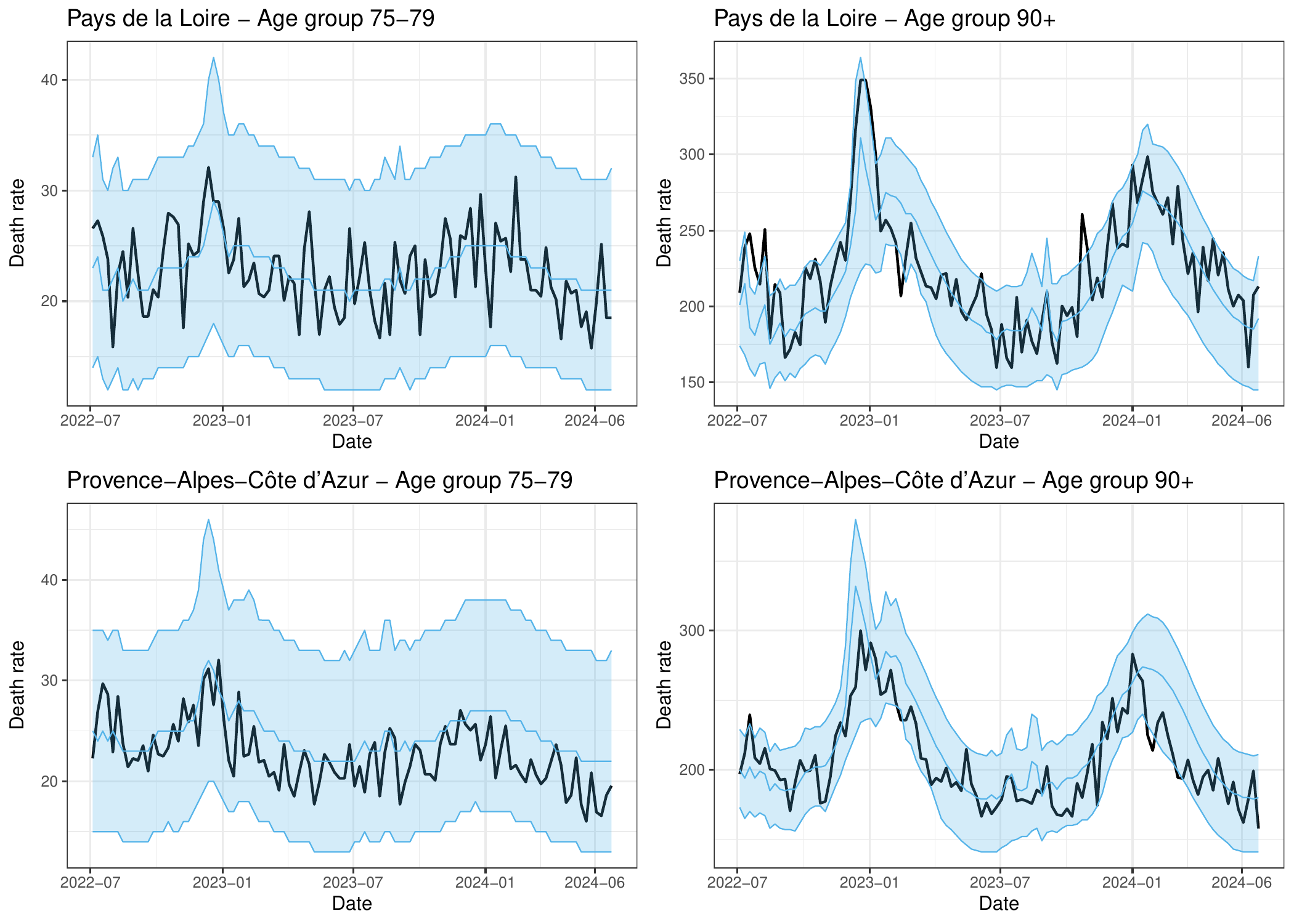}
    \caption{We present the 95$\%$ prediction intervals for the weekly death rates per 1 000 person-years in Pays de la Loire (top) and Provence-Alpes-Côte d'Azur (bottom) for the age groups 75-79 (left) and 90+ (right), from the 27th ISO week of 2022 to the 26th ISO week of 2024. The intervals are based on 25\ 000 bootstrap samples and consider parameter, spatial, state, and Poisson uncertainty. The blue lines show the 2.5$\%$, 50$\%$, and 97.5$\%$ quantile, while the black line visualizes the observed death rates per 1\ 000 person-years.}
    \label{fig:scenpred1}
\end{figure}

To avoid constructing a point forecast for the exposure-to-risk, we focus on the death rates per 1\ 000 person-years. This approach facilitates making comparisons across different regions. Figure~\ref{fig:scenpred1} illustrates the 95$\%$ prediction intervals for the weekly death rates in Pays de la Loire and Provence-Alpes-Côte d'Azur for the age groups 75-79 and 90+. The remaining illustrations can be found in Section D of the Online Appendix. These intervals incorporate parameter, spatial, state, and Poisson uncertainty, as described in Section~\ref{sec:quantify.uncertainty}. As Figure~\ref{fig:scenpred1} indicates, the prediction intervals effectively capture the uncertainty inherent in the observed death counts. The model captures the excess deaths observed in the 2022/2023 winter season for the age group 90+. Furthermore, the prediction bounds are narrower for the winter of 2023/2024, aligning with the observed death rates. 

\subsection{Short-term mortality forecasting with scenario-based approaches} \label{subsec:shorttermmortforecasting}
To forecast death rates in the short-term, we use temperature scenarios based on the RCP 2.6, RCP 4.5, and RCP 8.5 pathways from the Copernicus Climate Change Service \citep{cds2}. For influenza, we generate moderate, high, and severe incidence scenarios using a Bayesian stochastic SIRS model \citep{bucyibaruta2023discrete}, which is able to generate forecasts for weekly new influenza cases based on posterior predictive samples. COVID-19 hospitalizations are set to zero, but alternative scenarios can be explored. We include all technical details in Suppl.~Mat.~\ref{apps:scenariooooos}. Figure shows the results of the constructed scenarios for Pays de la Loire and Provences-Alpes-Côte d'Azur. Section E of the Online Appendix contains the results for the other regions.

\begin{figure}[!htb]
    \centering
    \includegraphics[width=0.85\linewidth]{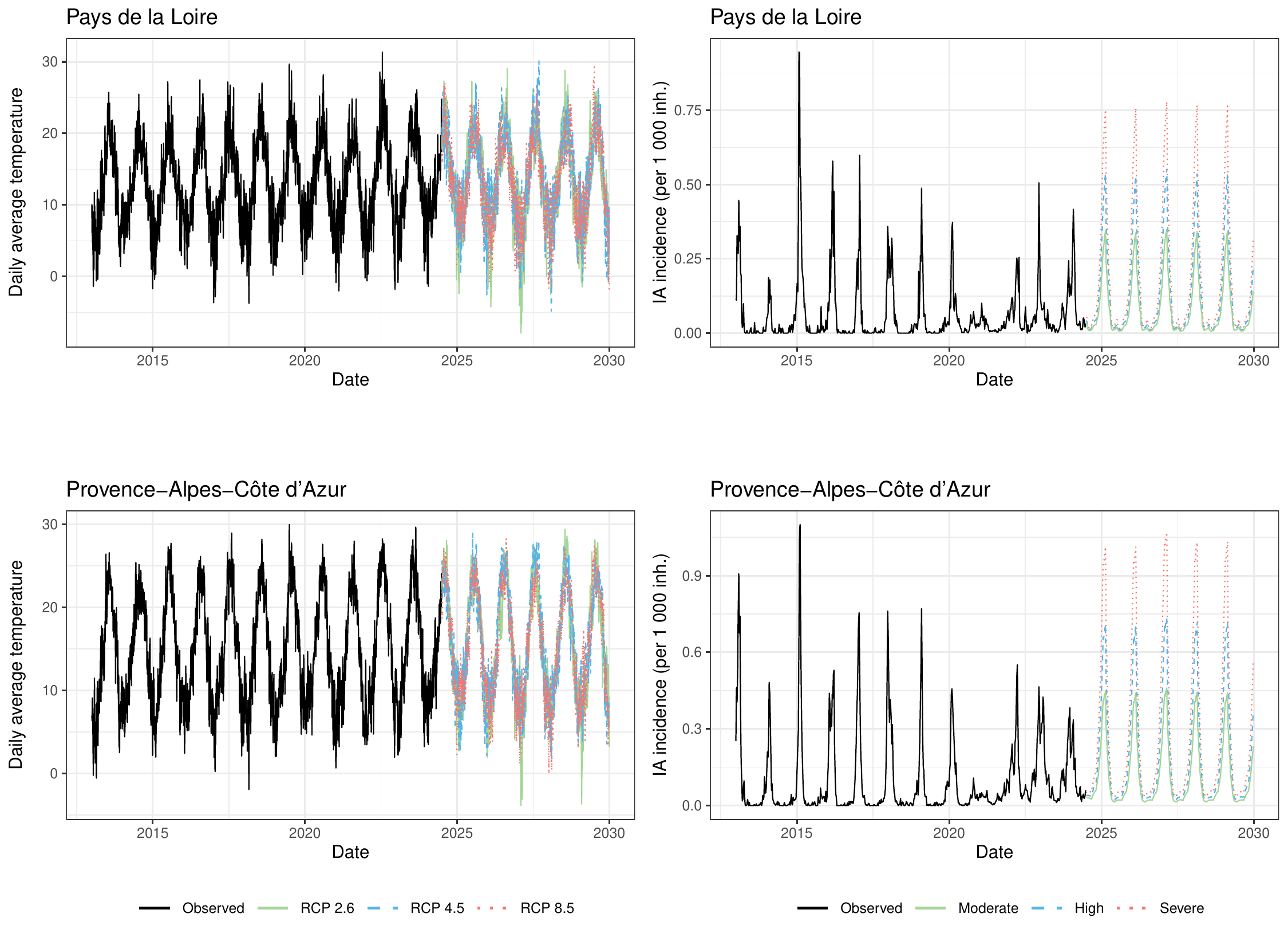}
    \caption{Projections for the daily average temperature based on the RCP 2.6, RCP 4.5, and RCP 8.5 scenarios (left panel), and for the weekly influenza activity per 100 inhabitants based on the proposed Bayesian stochastic SIRS model (right panel). We provide a moderate, high, and severe incidence scenario for influenza activity using the point-wise $50\%$, $80\%$, and $95\%$ quantiles of the posterior predictive samples, respectively. The top panels show the result for Pays de la Loire, while the bottom panels depict the results for Provences-Alpes-Côte d'Azur.}
    \label{fig:rcpsirs}
\end{figure}

\begin{figure}[!ht]
    \centering
    \includegraphics[width=0.85\linewidth]{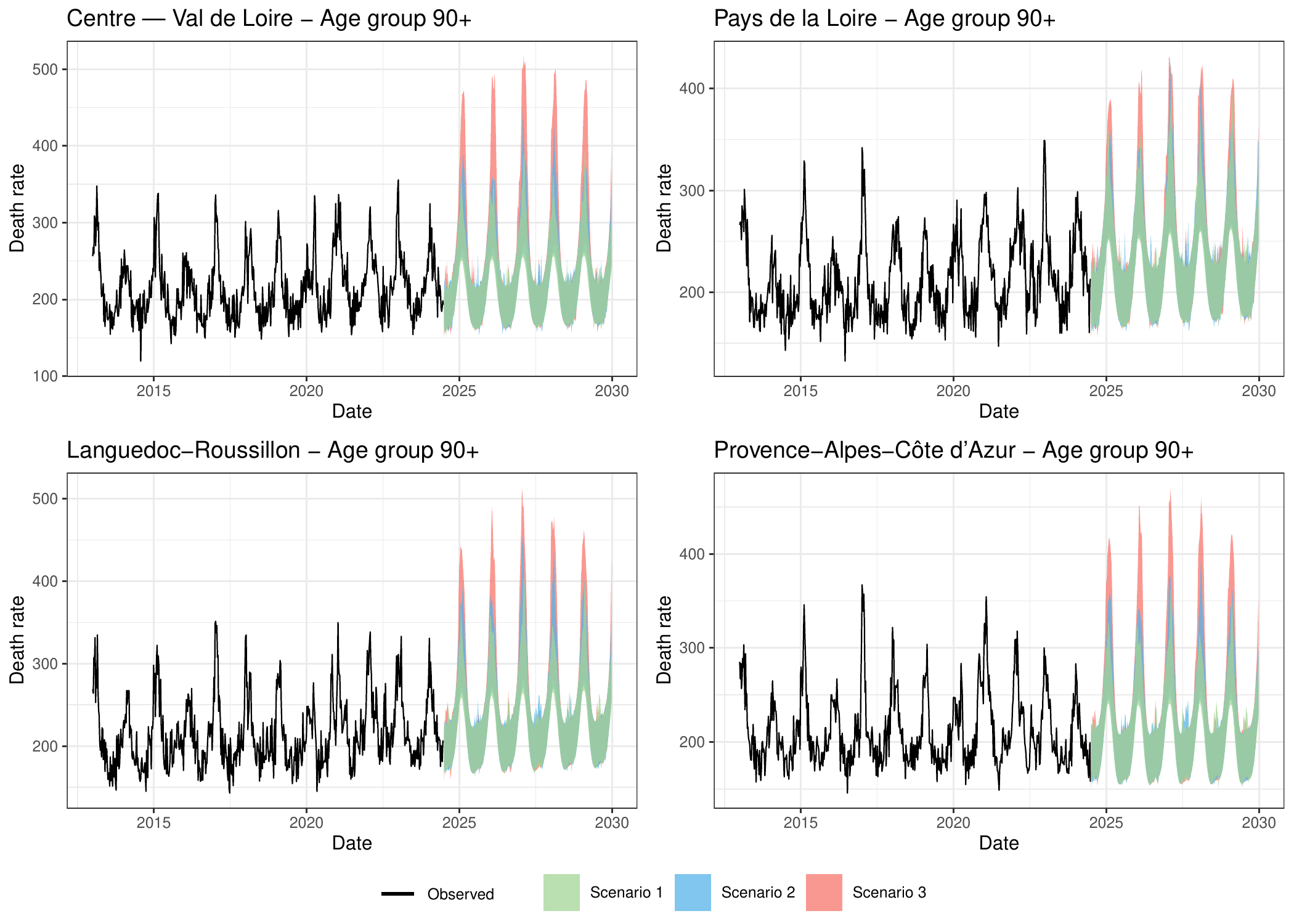}
    \caption{Based on the moderate, high, and severe influenza scenario, we present $95\%$ prediction intervals for the weekly death rates per 1\ 000 person-years in Centre-Val de Loire, Pays de la Loire, Languedoc-Roussillon, and Provence-Alpes-Côte d'Azur for the age group 90+. The predictions range from the 27th ISO week of 2024 to the 52nd ISO week of 2029. The intervals take into account parameter, spatial, state and Poisson uncertainty based on 25\ 000 bootstrap samples. The black line depicts the observed death rates per 1\ 000 person-years, from the first ISO week of 2013 to the 26th ISO week of 2024.}
    \label{fig:forecastscenarios}
\end{figure}

We forecast death rates from the 27th ISO-week of 2024 till the 52nd ISO week of 2029 for all NUTS 2 regions in France. Specifically, we construct the temperature anomalies, extreme temperature indicators, and the influenza activity index (anomalies) based on these temperature and influenza scenarios during the forecasting period. We then apply the forecasting strategy discussed in Section~\ref{sec:quantify.uncertainty} to incorporate parameter, spatial, state, and Poisson uncertainty in the predicted mortality outcomes. We construct three prediction intervals corresponding to three different scenarios: scenario 1 combines the moderate influenza scenario and the RCP 2.6 temperature projection, scenario 2 focuses on the high influenza scenario and the RCP 4.5 temperature projection, and, lastly, scenario 3 combines the severe influenza scenario with the RCP 8.5 temperature projection. Figure~\ref{fig:forecastscenarios} shows the 95$\%$ prediction intervals for the death counts at age group 90+ in the NUTS 2 regions Centre-Val de Loire, Pays de la Loire, Languedoc-Roussillon, and Alpes Côte d'Azur where the intervals in green, blue, and red correspond to scenario 1, 2, and 3, respectively. The prediction intervals for scenarios 1 and 2 only differ slightly, which can be explained by the more severe winter temperatures in the RCP 2.6 temperature projection compared to the RCP 4.5 scenario. Such scenario-based mortality forecasts can offer insurers useful insights into how different climate and influenza scenarios impact future mortality risks. Section F of the Online Appendix contains the prediction intervals for all age groups and NUTS 2 regions considered in this paper.

\begin{figure}[!ht]
    \centering
    \includegraphics[width=0.85\linewidth]{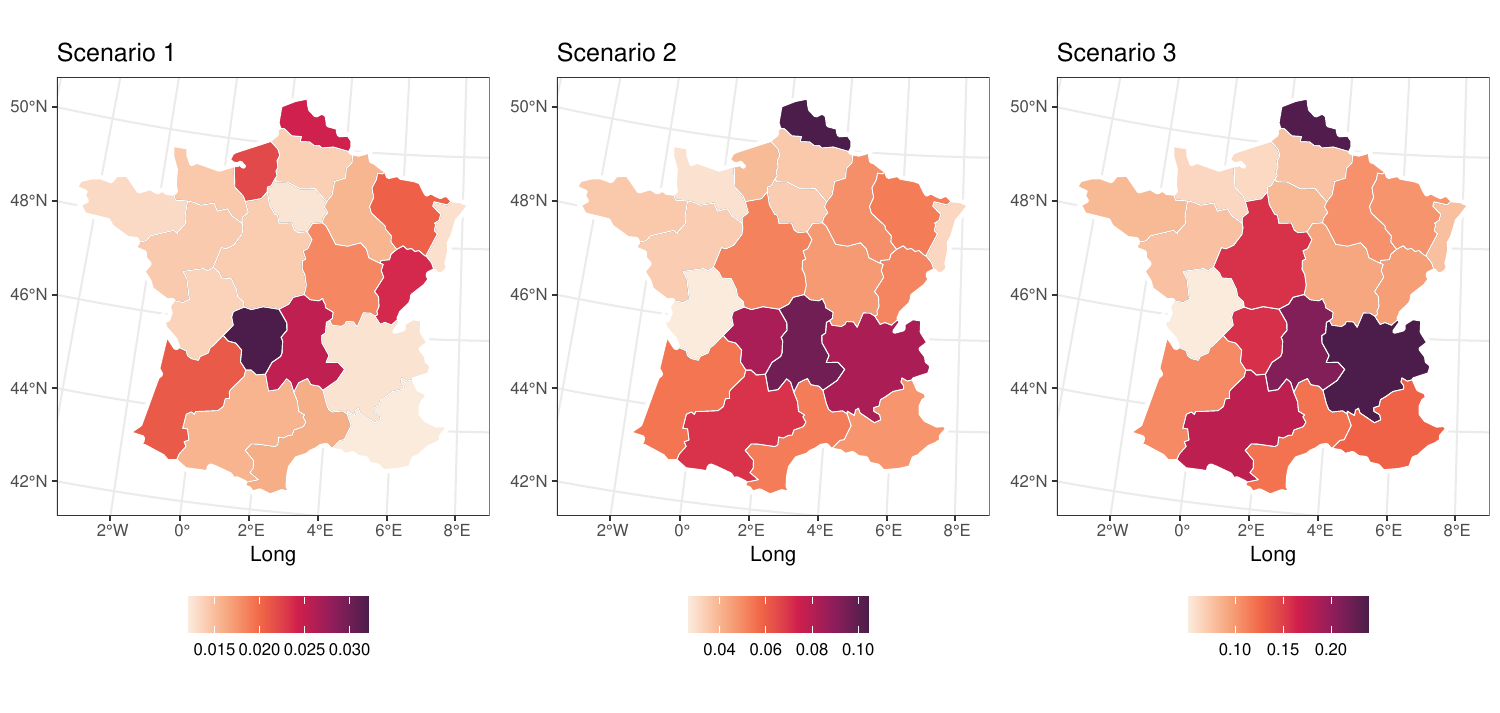}
    \caption{We present the relative excess deaths compared to the baseline mortality model for the age group 90+ in the French NUTS 2 regions under the three presented scenarios.}
    \label{fig:excessdeaths}
\end{figure}

We estimate the excess deaths relative to the baseline mortality model under the three considered scenarios. Hereto, we project the baseline mortality model on the forecasting period, from the 27th ISO week of 2024 until the 52nd ISO week of 2029, and compute the estimated forecasted number of deaths per 1 000 person-years per week. We sum up the results over the entire forecasting period. Next, we calculated the number of deaths (per 1 000 person-years per week) for the median quantile of each scenario. We then determine the relative excess deaths by comparing the scenario-based death counts with the baseline setting, expressed as a relative increase. We do this analysis for each region and age group. Figure~\ref{fig:excessdeaths} shows the results for the 90+ age group, while Section G of the Online Appendix displays the results for the age groups 65-69, 70-74, 75-79, 80-84, and 85-89. The results highlight clear regional differences. Based on the generated scenarios for influenza and temperature, we find that the southern NUTS 2 regions are generally more affected across all considered scenarios. Scenario 1 leads to moderate increases in deaths compared to the baseline, while Scenario 3 shows much sharper rises. Although the scenarios for the involved covariates (influenza, temperature) are purely hypothetical, they provide useful insights into how different conditions can impact regions in France very differently.
 
\section{Conclusion} \label{sec:conclusion}
This paper introduces an intuitive framework for incorporating environmental and epidemic factors from detailed, open-source data into a weekly mortality modeling framework. Specifically, we develop a three-state regime-switching model, which includes a baseline state reflecting seasonal mortality trends, an environmental shock state for periods of heightened mortality due to heatwaves, and a respiratory shock state to account for increased mortality linked to influenza or COVID-19 hospitalizations. 

Based on a case study on 21 regions and six older age groups in France, we detect regime periods of higher mortality associated with heat waves and increased influenza activity or hospital admissions. In the environmental shock state, we estimate that the factor amplifying baseline mortality is highest for the oldest age groups. Furthermore, we find evidence of harvesting effects, reducing mortality in the weeks following a hot week. In the respiratory shock state, the amplification of baseline mortality is associated with increased influenza activity, cold temperatures, and COVID-19-related hospitalizations. The model also captures harvesting effects. Through a back-test, we show that the model provides reliable short-term predictions of weekly mortality, aligning well with observed data.

We emphasize that the main purpose of this paper is to make short-term mortality predictions, and that the model in its current form is not intended for long-term predictions. To do this we must consider future improvements in heat resilience or the mitigating impact of public health adjustments such as increased vaccination rates or expanded cooling centers \citep{bobb2014heat, ferguson2006strategies}. These interventions reduce heat- and respiratory-related mortality \citep{stone2014avoided, simonsen2005impact}. The large-scale implementation of such measures usually takes several years and is therefore especially relevant to consider in long-term projections. However, this is beyond the scope of our short-term view in this paper.

\section*{Data and code availability statement} 
The datasets used in this paper are publicly available. We consult the deaths by week, sex, 5-year age group and NUTS 2 region on Eurostat (\url{https://ec.europa.eu/eurostat/web/products-datasets/-/demo_r_mwk2_05}). We retrieve the daily average temperature from the E-OBS dataset on the Copernicus Climate Data Store (\url{https://cds.climate.copernicus.eu/datasets/insitu-gridded-observations-europe}), the incidences rates for influenza activity from the French Sentinelles network (\url{https://www.sentiweb.fr/france/en/?page=table&maladie=25}), and the hospital data related to the COVID-19 epidemic from the French government's open data platform (\url{https://www.data.gouv.fr/fr/datasets/donnees-hospitalieres-relatives-a-lepidemie-de-covid-19/}). The $\texttt{R}$ code used to implement and analyze the case study presented in this paper is accessible in the GitHub repository: \url{https://github.com/jensrobben/GranMoMoRS}.

\section*{Funding statement} This study is part of the research programme at the Research Centre for Longevity Risk, a joint initiative of NN Group and the University of Amsterdam, with additional funding from the Dutch government’s Public-Private Partnership (PPP) programme. Karim Barigou acknowledges the financial support from the AMF research funds in integrated risk management of financial institutions (Fonds AMF-GIRIF) from Université Laval.

\section*{Conflict of interest disclosure} 
The authors declare no conflict of interest.

{\bibliography{References}}

\newpage 
\begin{center}
    {\huge \bf Supplementary material for “Granular mortality modeling with temperature and epidemic shocks: a three-state regime-switching approach”}
\end{center}

\appendix

\renewcommand\thefigure{\thesection.\arabic{figure}}
\renewcommand\thetable{\thesection.\arabic{table}}  
\setcounter{figure}{0}
\setcounter{table}{0}

\section{Data overview} \label{app:overview}
\paragraph{NUTS 2 regions.} Table~\ref{tab:nuts2overview} lists the names of the NUTS 2 regions in Metropolitan France together with their corresponding NUTS 2 geocode.

\begin{table}[!ht]
\centering
\caption{The NUTS 2 regions in Metropolitan France that constitute the set $\mathcal{R}$. \label{tab:nuts2overview}}
\label{tab:nuts2_regions}

\begin{minipage}[!ht]{0.45\textwidth}
\centering
\begin{small}
\begin{tabular}{ll}
\toprule
\textbf{NUTS 2} & \textbf{Region Name}          \\
\midrule
FR10                & Île-de-France                \\
FRB0                & Centre - Val de Loire        \\
FRC1                & Bourgogne                     \\
FRC2                & Franche-Comté                \\
FRD1                & Basse-Normandie              \\
FRD2                & Haute-Normandie              \\
FRE1                & Nord-Pas de Calais           \\
FRE2                & Picardie                     \\
FRF1                & Alsace                        \\
FRF2                & Champagne-Ardenne            \\
FRF3                & Lorraine                     \\
\bottomrule
\end{tabular}
\end{small}
\end{minipage}%
\hspace{0.5cm}
\begin{minipage}[t]{0.45\textwidth}
\centering
\begin{small}
\begin{tabular}{ll}
\toprule
\textbf{NUTS 2} & \textbf{Region Name}          \\
\midrule
FRG0                & Pays de la Loire             \\
FRH0                & Bretagne                      \\
FRI1                & Aquitaine                     \\
FRI2                & Limousin                     \\
FRI3                & Poitou-Charentes             \\
FRJ1                & Languedoc-Roussillon         \\
FRJ2                & Midi-Pyrénées               \\
FRK1                & Auvergne                      \\
FRK2                & Rhône-Alpes                  \\
FRL0                & Provence-Alpes-Côte d’Azur  \\
\: & \\
\bottomrule
\end{tabular}
\end{small}
\end{minipage}
\end{table}

\begin{figure}[!h]
    \centering
    \includegraphics[width=0.95\linewidth]{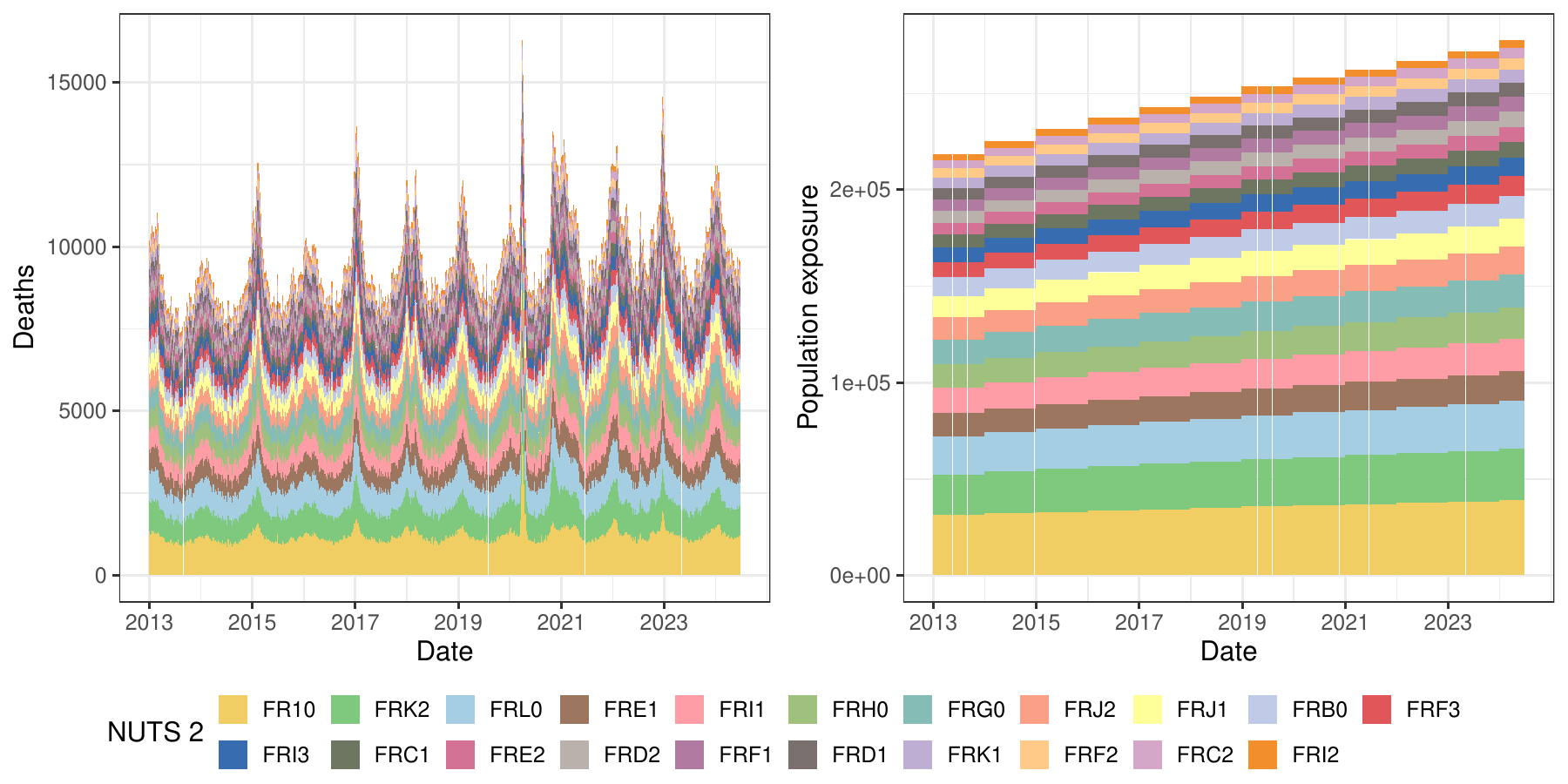}
    \caption{Stacked death counts (left) and exposures (right) for the 21 French NUTS 2 regions from the first ISO week of 2013 till the 26th ISO week of 2024. The death counts per region and time point are aggregated among the different age groups under consideration.}
    \label{figA:overviewDE}
\end{figure}

\begin{figure}[!ht]
    \centering
    \includegraphics[width=0.95\linewidth]{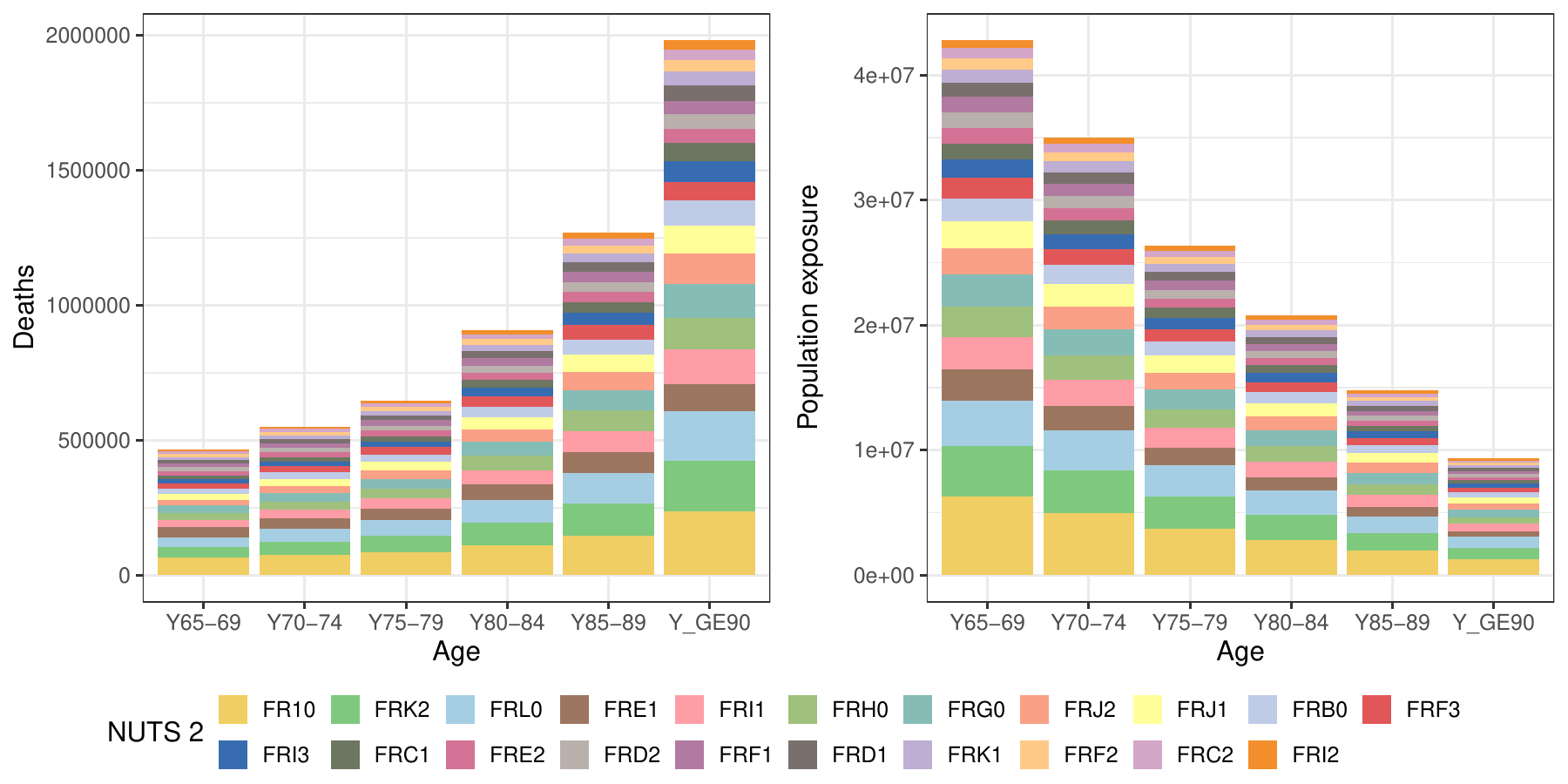}
    \caption{Stacked death counts (left) and exposures (right) for the 21 French NUTS 2 regions by age group. The death counts per region and age group are aggregated among the different ISO weeks under consideration.}
    \label{figA:overviewAge}
\end{figure}

\paragraph{Population exposures.} We obtain the population count of individuals aged 65 and older as of January 1 of each year, disaggregated by sex, age and NUTS 2 region, from Eurostat \citep{Eurostat_DEMO_R_PJANAGGR3}. Let $t$ represent any (ISO-)week within the specified time range $\mathcal{T}$, and let $y(t)$ denote the year corresponding to that week. Then, following the approach outlined by \cite{stmfnote}, we estimate the weekly exposure as:
\begin{equation} \label{eq:popexpo}
E_{x,t}^{(r)} = \frac{P_{x,y(t)}^{(r)} + P_{x,y(t)+1}^{(r)}}{2 \cdot 52.18},
\end{equation}
where \( P_{x,y(t)}^{(r)} \) represents the population count at age $x$ in region \( r \in \mathcal{R} \) on January 1 of year \( y(t) \). The value 52.18 reflects the average number of weeks in a year. Whereas this approach assumes a constant weekly exposure every year, a more refined method would be to apply linear interpolation to ensure a smoother transition of weekly exposures between consecutive years. Figures~\ref{figA:overviewDE} and~\ref{figA:overviewAge} provide a detailed overview of the population count data.

\paragraph{Exploratory plots.} Figure~\ref{figA:overviewDE} illustrates the stacked death counts and population exposures for the 21 French NUTS 2 regions over time, aggregated across age groups. Table~\ref{tab:nuts2_regions} provides the corresponding region names for each NUTS 2 geocode. The left panel shows a sharp spike in death counts at the beginning of 2020, which can be attributed to the COVID-19 pandemic. Additionally, we observe a spike during the winter seasons of 2014/2015 and 2016/2017 which coincides with severe influenza outbreaks. The right panel shows that population exposure remains constant within each year’s ISO weeks, see~\eqref{eq:popexpo}.

\renewcommand\thefigure{\thesection.\arabic{figure}}
\renewcommand\thetable{\thesection.\arabic{table}}  
\setcounter{figure}{0}
\setcounter{table}{0}

Figure~\ref{figA:overviewAge} displays the stacked death counts and population exposures for the 21 French NUTS 2 regions, categorized by age group and aggregated across all considered ISO weeks. The highest number of deaths occurs in the oldest age group (90+), despite having the smallest population exposure. Conversely, the youngest age group (65–69) has the largest population exposure but the lowest number of deaths.

\section{Details on the calibration procedure of the regime-switching model}
\subsection{Notation}
In addition to the notation introduced in Section~\ref{subsubsec:notation}, we denote the collection of age-specific death counts observed up to time point $t \in \mathcal{T}$ as:
\begin{align*}
\mathcal{D}_{t} = \left\{\mathcal{D}_{x,t}^{(r)} \mid r \in \mathcal{R}, x \in \mathcal{X}\right\},
\end{align*}
where $\mathcal{D}_{x,t}^{(r)} = \{D_{x,t'}^{(r)} \mid t' \leq t\}.$ We introduce similar notation for the collection of all states, $\mathcal{S}_t$, and covariate vectors, $\mathcal{Z}_t$, up to time point $t \in \mathcal{T}$:
\begin{align*}
\mathcal{S}_t = \left\{ \mathcal{S}_{t}^{(r)} \mid r \in \mathcal{R} \right\}, \quad \mathcal{Z}_t = \left\{ \mathcal{Z}_{t}^{(r)} \mid r \in \mathcal{R} \right\},
\end{align*}
with $\mathcal{S}_t^{(r)} = \{ S_{t'}^{(r)} \mid t' \leq t\}$ and $\mathcal{Z}_t^{(r)} = \{ \boldsymbol{z}_{t'}^{(r)} \mid t' \leq t\}$.
\subsection{Complete data log-likelihood} \label{app:condloglik}
We first derive the complete-data likelihood of all death counts and regime states, conditional on the spatial effect $\boldsymbol{u}^\ast$ and the covariate information $\mathcal{Z}_T$. We hereby assume that the temporal evolution of death counts and regime states is independent across regions, conditionally on the spatial effect $\boldsymbol{u}^\ast$ and region-specific covariates. Any spatial dependence across regions is assumed to be entirely captured through the spatial effect $\boldsymbol{u}^\ast$. We obtain:
\begin{align*}
&\mathbb{P}\left(\mathcal{D}_T, \mathcal{S}_T\mid \boldsymbol{u}^\ast,\mathcal{Z}_T; \boldsymbol{\theta}\right) \\
&\hspace{1cm}= \displaystyle \prod \limits_{r\in \mathcal{R}} \mathbb{P}\left( \left(\mathcal{D}_{x,T}^{(r)}\right)_{x\in\mathcal{X}},  \mathcal{S}_T^{(r)} \mid u_r^\ast, \mathcal{Z}_T^{(r)} \:;\boldsymbol{\theta}\right) \\
&\hspace{1cm}= \displaystyle \prod \limits_{r\in \mathcal{R}} \Bigg\{ \mathbb{P} \left( \left(D_{x,1}^{(r)}\right)_{x\in\mathcal{X}} = \left(d_{x,1}^{(r)}\right)_{x\in\mathcal{X}}, S_1^{(r)} = s_1^{(r)} \mid u_r^\ast, \mathcal{Z}_1^{(r)} \:; \boldsymbol{\theta}\right) \cdot \\ &\hspace{2cm}\prod_{t=2}^{T} \mathbb{P}\left(\left(D_{x,t}^{(r)}\right)_{x\in\mathcal{X}} = \left(d_{x,t}^{(r)}\right)_{x\in\mathcal{X}}, S_t^{(r)} = s_t^{(r)} \mid u_r^\ast, \mathcal{Z}_t^{(r)} , \left(\mathcal{D}_{x,t-1}^{(r)}\right)_{x\in\mathcal{X}}, \mathcal{S}_{t-1}^{(r)}\: ; \: \boldsymbol{\theta}\right)\Bigg\},
\end{align*}
where we apply the law of conditional probability $T-1$ times to obtain the last equality. Next, within each region $r$, we impose a temporal Markov property on the latent regime state processes, and rewrite the likelihood as:
\begin{align*}
&\mathbb{P}\left(\mathcal{D}_T, \mathcal{S}_T\mid \boldsymbol{u}^\ast,\mathcal{Z}_T; \boldsymbol{\theta}\right) \\
&\hspace{1cm}= \displaystyle \prod \limits_{r\in \mathcal{R}} \Bigg\{ \mathbb{P} \left(\left(d_{x,1}^{(r)}\right)_{x\in\mathcal{X}}, s_1^{(r)} \mid u_r^\ast, \mathcal{Z}_1^{(r)} \:; \boldsymbol{\theta}\right) \cdot \\
&\hspace{3cm} \prod_{t=2}^{T} \mathbb{P}\left(\left(d_{x,t}^{(r)}\right)_{x\in\mathcal{X}}, s_t^{(r)} \mid u_r^\ast, \mathcal{Z}_t^{(r)} , \left(\mathcal{D}_{x,t-1}^{(r)}\right)_{x\in\mathcal{X}}, \mathcal{S}_{t-1}^{(r)}\: ; \: \boldsymbol{\theta}\right)\Bigg\} \\
&\hspace{1cm}= \prod_{r\in\mathcal{R}} \Bigg\{ \mathbb{P}\left(\left(d_{x,1}^{(r)}\right)_{x\in\mathcal{X}} \mid s_1^{(r)}, \boldsymbol{z}_1^{(r)} \:;\: \boldsymbol{\alpha}_{s_1^{(r)}}\right) \cdot \mathbb{P}\left(s_1^{(r)} \mid u_r^\ast, \boldsymbol{z}_1^{(r)} \: ; \: \boldsymbol{\beta}\right) \cdot \\
&\hspace{3cm} \displaystyle \prod_{t=2}^{T}  \mathbb{P}\left(\left(d_{x,t}^{(r)}\right)_{x\in \mathcal{X}} \mid s_t^{(r)}, \boldsymbol{z}_t^{(r)} \:;\: \boldsymbol{\alpha}_{s_t^{(r)}}\right) \cdot \mathbb{P}\left(s_t^{(r)} \mid s_{t-1}^{(r)},  u_r^\ast,\boldsymbol{z}_t^{(r)} \: ; \: \boldsymbol{\beta}\right) \Bigg\}.
\end{align*}

Next, we assume that the death counts random variables $D_{x,t}^{(r)}$ are independent among the different age categories $x \in \mathcal{X}$, conditionally on the regime state $S_t^{(r)}$. Under this assumption, we can rewrite the likelihood to:
\begin{align*}
&\mathbb{P}\left(\mathcal{D}_T, \mathcal{S}_T\mid \boldsymbol{u}^\ast,\mathcal{Z}_T; \boldsymbol{\theta}\right) \\
&\hspace{1cm}= \prod_{r\in\mathcal{R}} \Bigg\{ \left[\: \prod_{x\in\mathcal{X}}\mathbb{P}\left(d_{x,1}^{(r)} \mid s_1^{(r)}, \boldsymbol{z}_1^{(r)} \:;\: \boldsymbol{\alpha}_{s_1^{(r)},x}\right)\right] \cdot \mathbb{P}\left(s_1^{(r)} \mid u_r^\ast, \boldsymbol{z}_1^{(r)} \: ; \: \boldsymbol{\beta}\right) \cdot \\
&\hspace{3cm} \displaystyle \prod_{t=2}^{T}  \left[\: \prod_{x\in\mathcal{X}}\mathbb{P}\left(d_{x,t}^{(r)} \mid s_t^{(r)}, \boldsymbol{z}_t^{(r)} \:;\: \boldsymbol{\alpha}_{s_t^{(r)},x}\right)\right] \cdot \mathbb{P}\left(s_t^{(r)} \mid s_{t-1}^{(r)},  u_r^\ast,\boldsymbol{z}_t^{(r)} \: ; \: \boldsymbol{\beta}\right) \Bigg\}.
\end{align*}

We will treat $\rho_{s_1^{(r)}} := \mathbb{P}(s_1^{(r)} \mid u_r^\ast, \boldsymbol{z}_1^{(r)} \:;\: \boldsymbol{\beta})$ as additonal parameters in the likelihood and obtain:
\begin{align*}
&\mathbb{P}\left(\mathcal{D}_T, \mathcal{S}_T\mid \boldsymbol{u}^\ast,\mathcal{Z}_T; \boldsymbol{\theta}\right) \\
&\hspace{1cm}=\displaystyle \prod \limits_{r\in \mathcal{R}} \scalebox{1.15}{\Bigg\{}
\left[\displaystyle \sum_{j=0}^N \mathbbm{1}\left\{S_1^{(r)} = j\right\} \cdot \left( \: \prod_{x\in\mathcal{X}}\mathbb{P}\left(d_{x,1}^{(r)} \mid S_1^{(r)} = j, \boldsymbol{z}_{1}^{(r)} \:;\: \boldsymbol{\alpha}_{j,x}\right)\right) \cdot \rho_j \right] \cdot  \\ 
&\hspace{2.5cm}\displaystyle \prod_{t=2}^T \scalebox{1.1}{\Bigg[} \displaystyle \sum_{j=0}^N \sum_{i=0}^N \mathbbm{1}\left\{S_{t-1}^{(r)} = i, S_t^{(r)} = j \right\} \cdot \left(\:\prod_{x\in\mathcal{X}}\mathbb{P}\left(d_{x,t}^{(r)} \mid S_t^{(r)} = j, \boldsymbol{z}_{t}^{(r)} \:;\: \boldsymbol{\alpha}_{j,x}\right)\right) \cdot \\
&\hspace{4cm} p_t^{ij}\left(\boldsymbol{z}_t^{(r)} \:;\: u_r^\ast,\boldsymbol{\beta}\right) \scalebox{1.1}{\Bigg]} \scalebox{1.15}{\Bigg\}}.
\end{align*}
Taking the logarithm leads to:
{\allowdisplaybreaks \begin{align*} 
&\log \mathbb{P}\left(\mathcal{D}_T, \mathcal{S}_T\mid \boldsymbol{u}^\ast,\mathcal{Z}_T; \boldsymbol{\theta}\right) \\
&\hspace{1cm}= \displaystyle \sum\limits_{r\in \mathcal{R}} \scalebox{1.15}{\Bigg\{}
\sum_{x\in\mathcal{X}} \sum_{j=0}^N \mathbbm{1}\left\{S_1^{(r)} = j\right\} \cdot \log \mathbb{P}\left(d_{x,1}^{(r)} \mid S_1^{(r)} = j, \boldsymbol{z}_{1}^{(r)} \:;\: \boldsymbol{\alpha}_{j,x}\right) + \\
&\hspace{2.5cm}\displaystyle \sum_{j=0}^N  \mathbbm{1}\left\{S_1^{(r)} = j\right\} \cdot \log \rho_j \: + \stepcounter{equation}\tag{\theequation}{\label{eq:condloglikkk}}\\
&\hspace{2.5cm} 
\displaystyle \sum_{x\in \mathcal{X}} \sum_{t=2}^T  \sum_{j=0}^N \mathbbm{1}\left\{S_t^{(r)} = j \right\} \cdot \log \mathbb{P}\left(d_{x,t}^{(r)} \mid S_t^{(r)} = j, \boldsymbol{z}_{t}^{(r)} \:;\: \boldsymbol{\alpha}_{j,x}\right) + \\
&\hspace{2.5cm} \sum_{t=2}^T \sum_{j=0}^N \sum_{i=0}^N \mathbbm{1}\left\{S_{t-1}^{(r)} = i, S_t^{(r)} = j \right\} \cdot \log p_t^{ij}\left(\boldsymbol{z}_t^{(r)} \:;\: u_r^\ast, \boldsymbol{\beta}\right)\scalebox{1.15}{\Bigg\}},
\end{align*}} 
\hspace{-0.15cm}where~\eqref{eq:trans.prob} specifies the transition probabilities $p_t^{ij}(\boldsymbol{z}_t^{(r)} \:;\: u_r^\ast, \boldsymbol{\beta})$, and $\mathbb{P}(d_{x,t}^{(r)} \mid S_t^{(r)} = j, \boldsymbol{z}_{t}^{(r)} \:;\: \boldsymbol{\alpha}_{j,x})$ represents the probability mass function of the state-dependent Poisson distributions given in~\eqref{eq:cond.dens}.

Following (\ref{eq:objL}) in the main paper, we can then calculate the (unconditional) expected complete-data log-likelihood as:
\begin{gather}
\begin{aligned} 
&\log \mathcal{L}(\boldsymbol{\theta}, \tau \mid \mathcal{D}_T, \mathcal{S}_T, \mathcal{Z}_T) \\
&\hspace{1.75cm}=\displaystyle \sum\limits_{r\in \mathcal{R}} \scalebox{1.15}{\Bigg\{}
\sum_{x\in\mathcal{X}} \sum_{j=0}^N \mathbbm{1}\left\{S_1^{(r)} = j\right\} \cdot \log \mathbb{P}\left(D_{x,1}^{(r)} = d_{x,1}^{(r)} \mid S_1^{(r)} = j, \boldsymbol{z}_{1}^{(r)} \:;\: \boldsymbol{\alpha}_{j,x}\right) + \\
&\hspace{3.3cm}\displaystyle \sum_{j=0}^N  \mathbbm{1}\left\{S_1^{(r)} = j\right\} \cdot \log \rho_j \: + \\
&\hspace{3.3cm} 
\displaystyle \sum_{x\in \mathcal{X}} \sum_{t=2}^T  \sum_{j=0}^N \mathbbm{1}\left\{S_t^{(r)} = j \right\} \cdot \log \mathbb{P}\left(D_{x,t}^{(r)} = d_{x,t}^{(r)} \mid S_t^{(r)} = j, \boldsymbol{z}_{t}^{(r)} \:;\: \boldsymbol{\alpha}_{j,x}\right) + \\
&\hspace{3.3cm} \sum_{t=2}^T \sum_{j=0}^N \sum_{i=0}^N \mathbbm{1}\left\{S_{t-1}^{(r)} = i, S_t^{(r)} = j \right\} \cdot \log p_t^{ij}\left(\boldsymbol{z}_t^{(r)} \:;\: u_r^\ast, \boldsymbol{\beta}\right)\scalebox{1.15}{\Bigg\}} - \\
&\hspace{2.2cm} \frac{1}{2} \log(\det(\boldsymbol{Q}(\tau)^{-1})) - \frac{1}{2} \left(\boldsymbol{u}^\ast(\boldsymbol{\theta})\right)^\top \boldsymbol{Q}(\tau) \boldsymbol{u}^\ast(\boldsymbol{\theta}) - \frac{1}{2} \log(\det(\boldsymbol{S}^{-1}(\boldsymbol{\theta},\boldsymbol{u}^\ast(\boldsymbol{\theta})))),
\end{aligned}
\label{eq:LogLik}
\raisetag{4cm}
\end{gather}
where
\begin{equation}\label{eq:maxU}
\begin{aligned} 
\boldsymbol{u}^\ast &:= \boldsymbol{u}^\ast(\boldsymbol{\theta}) = \argmax_{\boldsymbol{u}} \big\{ \mathbb{E}\left[ \log \mathbb{P}\left(\mathcal{D}_T, \mathcal{S}_T, \boldsymbol{u} \mid \mathcal{Z}_T; \boldsymbol{\theta}, \tau\right)\right]\big\} \\ 
\boldsymbol{S}^{-1} &:= \boldsymbol{S}^{-1}(\boldsymbol{\theta},\boldsymbol{u}^\ast(\boldsymbol{\theta})) = - \frac{\partial^2 \mathbb{E} \left[ \log \mathbb{P}\left(\mathcal{D}_T, \mathcal{S}_T, \boldsymbol{u} \mid \mathcal{Z}_T \:; \boldsymbol{\theta}, \tau \right)\right]}{\partial \boldsymbol{u} \, \partial \boldsymbol{u}^\top} \Big|_{\boldsymbol{u} = \boldsymbol{u}^\ast(\boldsymbol{\theta})}.
\end{aligned}
\end{equation}
Note that we consider $\smash{\rho_j := \mathbb{P}(S_1^{(r)} = j \mid u_r^{\ast}, \boldsymbol{z}_1^{(r)} ; \boldsymbol{\beta})}$ in \eqref{eq:LogLik} as additional region-independent parameters in the model. Appendix~\ref{app:hessian} details the calculation procedure of the Hessian $\boldsymbol{S}^{-1}(\boldsymbol{\theta},\boldsymbol{u}^\ast(\boldsymbol{\theta}))$. 

\subsection{Hessian of the expected complete-data log-likelihood}  \label{app:hessian}
Following~\eqref{eq:maxU}, we  first take the expectation of the (unconditional) complete-data log-likelihood with respect to the regime state process:
\begin{align*}
&\mathbb{E}\left[ \log \mathbb{P}\left(\mathcal{D}_T, \mathcal{S}_T, \boldsymbol{u} \mid \mathcal{Z}_T; \boldsymbol{\theta}, \tau\right)\right] \\
&\hspace{2cm}= \mathbb{E}\left[ \log \mathbb{P}\left(\mathcal{D}_T, \mathcal{S}_T \mid \boldsymbol{u}, \mathcal{Z}_T; \boldsymbol{\theta}\right)\right] + \log f(\boldsymbol{u}\:; \tau) \\
&\hspace{2cm}= \mathbb{E}\left[ \log \mathbb{P}\left(\mathcal{D}_T, \mathcal{S}_T \mid \boldsymbol{u}, \mathcal{Z}_T; \boldsymbol{\theta}\right)\right] - \frac{R}{2}\log(2\pi) - \frac{1}{2} \log(\det(\boldsymbol{Q}(\tau)^{-1})) - \frac{1}{2} \boldsymbol{u}^\top \boldsymbol{Q}(\tau) \boldsymbol{u},
\end{align*}
with $\mathbb{P}(\mathcal{D}_T, \mathcal{S}_T \mid \boldsymbol{u}, \mathcal{Z}_T; \boldsymbol{\theta})$ the complete-data log-likelihood, given the spatial effect $\boldsymbol{u}$ (see~\eqref{eq:condloglikkk}), and $\boldsymbol{Q}(\tau)$ the precision matrix that depends on the precision parameter $\tau$.

We then calculate the expectation of the expression in~\eqref{eq:condloglikkk}, conditionally on the spatial effect $\boldsymbol{u}$, the death counts $\mathcal{D}_T$, and the covariate information $\mathcal{Z}_T$. We obtain:
{\allowdisplaybreaks \begin{align*}
&\mathbb{E}\left[ \log \mathbb{P}\left(\mathcal{D}_T, \mathcal{S}_T \mid \boldsymbol{u}, \mathcal{Z}_T; \boldsymbol{\theta}\right)\right] \\
&\hspace{1cm}= \displaystyle \sum\limits_{r\in \mathcal{R}} \scalebox{1.15}{\Bigg\{}
\sum_{x\in\mathcal{X}} \sum_{j=0}^N \mathbb{P}\left(S_1^{(r)} = j \mid u_r, \boldsymbol{z}_1^{(r)}; \boldsymbol{\theta}\right) \cdot \log \mathbb{P}\left(d_{x,1}^{(r)} \mid S_1^{(r)} = j, \boldsymbol{z}_{1}^{(r)} \:;\: \boldsymbol{\alpha}_{j,x}\right) + \\
&\hspace{2.5cm}\displaystyle \sum_{j=0}^N  \mathbb{P}\left(S_1^{(r)} = j \mid u_r, \boldsymbol{z}_1^{(r)}; \boldsymbol{\theta}\right) \cdot \log \rho_j \: + \stepcounter{equation}\tag{\theequation}\label{eqA:ECDLL}\\
&\hspace{2.5cm} 
\displaystyle \sum_{x\in \mathcal{X}} \sum_{t=2}^T  \sum_{j=0}^N \mathbbm{P}\left(S_t^{(r)} = j \mid u_r, \mathcal{D}_{t}^{(r)}, \mathcal{Z}_t^{(r)}\:; \boldsymbol{\theta}\right) \cdot \log \mathbb{P}\left(d_{x,t}^{(r)} \mid S_t^{(r)} = j, \boldsymbol{z}_{t}^{(r)} \:;\: \boldsymbol{\alpha}_{j,x}\right) + \\
&\hspace{2.5cm} \sum_{t=2}^T \sum_{j=0}^N \sum_{i=0}^N \mathbbm{P}\left(S_{t-1}^{(r)} = i, S_t^{(r)} = j \mid u_r, \mathcal{D}_{t}^{(r)}, \mathcal{Z}_t^{(r)}\:; \boldsymbol{\theta}\right) \cdot \log p_t^{ij}\left(\boldsymbol{z}_t^{(r)} \:;\: u_r, \boldsymbol{\beta}\right)\scalebox{1.15}{\Bigg\}},
\end{align*}}
\hspace{-0.15cm}where $\mathcal{D}_{t}^{(r)} := \{\mathcal{D}_{x,t}^{(r)} \mid x\in\mathcal{X}\}$ is the set of death counts for region $r$ at all ages $x\in\mathcal{X}$ up to time $t$.

In the EM algorithm, we take the expectation of the complete-data log-likelihood conditional on the current best-estimate of the spatial effect $\boldsymbol{u}$ and regime-switching parameter vector $\boldsymbol{\theta}$, see Appendix~\ref{app:modelestimationEM}. As such, the marginal and joint state probabilities in~\eqref{eqA:ECDLL} are considered to be constant when computing the negative Hessian $\boldsymbol{S}^{-1}(\boldsymbol{\theta}, \boldsymbol{u}^\ast(\boldsymbol{\theta}))$ from~\eqref{eq:maxU}. The diagonal entries $(r,r)$ of the negative Hessian equal:
\begin{align*}
- \displaystyle \sum_{t=2}^T \sum_{j=0}^N \sum_{i=0}^N \mathbb{P}\left\{S_{t-1}^{(r)} = i, S_t^{(r)} = j \mid u_r,\mathcal{D}_{t}^{(r)},\mathcal{Z}_t^{(r)} \:; \boldsymbol{\theta} \right\} \cdot \dfrac{\partial^2 \log p_t^{ij}\left(\boldsymbol{z}_t^{(r)} \:; u_r, \boldsymbol{\beta}\right)}{\partial u_r^2} + Q_{rr}(\tau),
\end{align*}
where $Q_{rr}(\tau)$ is entry $(r,r)$ of the precision matrix $\boldsymbol{Q}(\tau)$. Using the expression for the regime state transition probabilities in~\eqref{eq:trans.prob}, we obtain:
\begin{align*}
\frac{\partial^2 \log p_t^{ij}\left(\boldsymbol{z}_t^{(r)} \:; u_r, \boldsymbol{\beta}\right)}{\partial u_r^2} =  
-p_t^{i0}\left(\boldsymbol{z}_t^{(r)} \:; u_r, \boldsymbol{\beta}\right) \cdot \left(1 - p_t^{i0}\left(\boldsymbol{z}_t^{(r)} \:; u_r, \boldsymbol{\beta}\right)\right),
\end{align*}
for $i,j \in \{0,1,...,N\}$ with $(ij) \notin \{(12),(21)\}$.  Entry $(r,s)$ of the Hessian is then equal to:
\begin{equation} \label{eq:hessi}
\begin{aligned} 
&\Bigg(\displaystyle \sum_{t=2}^T \sum_{j=0}^N \sum_{i=0}^N \mathbb{P}\left\{S_{t-1}^{(r)} = i, S_t^{(r)} = j \mid u_r,\mathcal{D}_{t}^{(r)},\mathcal{Z}_t^{(r)} \:; \boldsymbol{\theta} \right\} \cdot p_t^{i0}\left(\boldsymbol{z}_t^{(r)} \:; u_r, \boldsymbol{\beta}\right) \cdot \\
&\hspace{6.75cm}\left(1 - p_t^{i0}\left(\boldsymbol{z}_t^{(r)} \:; u_r, \boldsymbol{\beta}\right)\right)\Bigg) \cdot \mathbbm{1}\left\{r = s\right\} + Q_{rs}(\tau).
\end{aligned}
\end{equation}
We conclude that the negative Hessian $\boldsymbol{S}^{-1}(\boldsymbol{\theta}, \boldsymbol{u}^\ast(\boldsymbol{\theta})$ is the sum of the precision matrix $\boldsymbol{Q}(\tau)$ and a diagonal matrix with as entries the second order partial derivatives of the log transition probabilities, summed over all the states and periods under consideration. 

\subsection{Expectation-maximization algorithm} \label{app:modelestimationEM}
We provide details on the proposed EM algorithm used to optimize the incomplete-data log-likelihood in Section~\ref{subsec:modelestimationEM}. For iteration $k+1$, we apply:

\paragraph{E-step.} We compute the conditional expectation of the complete-data log-likelihood with respect to the regime-state process, given the current best estimates of the regime-switching parameter vector $\boldsymbol{\theta}_k^\ast$ and the spatial effect $\boldsymbol{u}_k^\ast$:
 
{\allowdisplaybreaks
\begin{align*}
&\mathbb{E}\left[\log \mathcal{L}(\boldsymbol{\theta}, \tau \mid \mathcal{D}_T, \mathcal{S}_T, \mathcal{Z}_T) \mid \boldsymbol{u}_{k}^\ast, \mathcal{D}_T, \mathcal{Z}_T; \boldsymbol{\theta}_k^\ast\right] \\
&\hspace{0.5cm}= \displaystyle \sum\limits_{r\in \mathcal{R}} \scalebox{1.15}{\Bigg\{}
\sum_{x\in\mathcal{X}} \sum_{j=0}^N \mathbb{P}\left(S_1^{(r)} = j \mid u_{k,r}^\ast, \boldsymbol{z}_1^{(r)}; \boldsymbol{\theta}_k^\ast\right) \cdot \log \mathbb{P}\left(d_{x,1}^{(r)} \mid S_1^{(r)} = j, \boldsymbol{z}_{1}^{(r)} \:;\: \boldsymbol{\alpha}_{j,x}\right) + \\
&\hspace{2cm}\displaystyle \sum_{j=0}^N  \mathbb{P}\left(S_1^{(r)} = j \mid u_{k,r}^\ast, \boldsymbol{z}_1^{(r)}; \boldsymbol{\theta}^\ast_k\right) \cdot \log \rho_j \: +  \tag{\stepcounter{equation}\theequation}\\
&\hspace{2cm} 
\displaystyle \sum_{x\in \mathcal{X}} \sum_{t=2}^T  \sum_{j=0}^N \mathbb{P}\left(S_t^{(r)} = j \mid u_{k,r}^\ast, \mathcal{D}_{t}^{(r)}, \mathcal{Z}_t^{(r)}\:; \boldsymbol{\theta}_k^\ast\right) \cdot \log \mathbb{P}\left(d_{x,t}^{(r)} \mid S_t^{(r)} = j, \boldsymbol{z}_{t}^{(r)} \:;\: \boldsymbol{\alpha}_{j,x}\right) + \\
&\hspace{2cm} \sum_{t=2}^T \sum_{j=0}^N \sum_{i=0}^N \mathbb{P}\left(S_{t-1}^{(r)} = i, S_t^{(r)} = j \mid u_{k,r}^\ast, \mathcal{D}_{t}^{(r)}, \mathcal{Z}_t^{(r)}\:; \boldsymbol{\theta}_k^\ast\right) \cdot \log p_t^{ij}\left(\boldsymbol{z}_t^{(r)} \:; u_r, \boldsymbol{\beta}\right)\scalebox{1.15}{\Bigg\}} - \\
&\hspace{1.25cm} \frac{1}{2} \log(\det(\boldsymbol{Q}(\tau)^{-1})) - \frac{1}{2} \left(\boldsymbol{u}^\ast(\boldsymbol{\theta})\right)^\top \boldsymbol{Q}(\tau) \boldsymbol{u}^\ast(\boldsymbol{\theta}) - \frac{1}{2} \log(\det(\boldsymbol{S}^{-1}(\boldsymbol{\theta},\boldsymbol{u}^\ast(\boldsymbol{\theta})))),
\end{align*}

where $u_{k,r}^\ast$ is the $r$-th component of the spatial effect $\boldsymbol{u}_k^\ast$, and $\boldsymbol{\beta}^\ast_k$ is a subset of the parameter vector $\boldsymbol{\theta}^\ast_k$, related to the parameters in the transition probabilities. We refer to the involved state probabilities, i.e., $\mathbb{P}(S_t^{(r)} = j \mid u_{k,r}^\ast, \mathcal{D}_{t}^{(r)}, \mathcal{Z}_t^{(r)}\:; \boldsymbol{\theta}_k^\ast)$ and $\mathbb{P}(S_{t-1}^{(r)} = i, S_t^{(r)} = j \mid u_{k,r}^\ast, \mathcal{D}_{t}^{(r)}, \mathcal{Z}_t^{(r)}\:; \boldsymbol{\theta}_k^\ast)$, as the filtered marginal and joint state probabilities, respectively. The term filtered indicates that these probabilities are conditioned only on information available up to and including time $t$. The computation of these marginal and joint state probabilities follows the iterative procedure described below.
\begin{enumerate}
\item[\textbf{1.}] Calculate the state-dependent Poisson death count and transition probabilities from~\eqref{eq:cond.dens} and~\eqref{eq:trans.prob}, using the current best estimates of the regime-switching parameter vector $\boldsymbol{\theta}_k^\ast$ and the spatial effect $\boldsymbol{u}_k^\ast$.
\item[\textbf{2.}] For every region $r\in \mathcal{R}$, we calculate the filtered joint state probabilities by iterating over the following steps for $t = 2,3,\ldots, T$:
 \begin{enumerate}
 \item[\textbf{(a)}] For $t=2$, we compute:
 \begin{align*}
 &\mathbb{P}\left(S_1^{(r)} = i,S_2^{(r)} = j,\left(D_{x,2}^{(r)}\right)_{x\in\mathcal{X}} = \left(d_{x,2}^{(r)}\right)_{x\in\mathcal{X}} \mid u^\ast_{k,r}, \mathcal{D}_1^{(r)}, \mathcal{Z}_2^{(r)} \: ; \boldsymbol{\theta}_k^\ast\right) \\ &\hspace{0.75cm}= 
\rho_{i,k}^{\ast} \cdot p_2^{ij}\left(\boldsymbol{z}_2^{(r)} \:; u^\ast_{k,r}, \boldsymbol{\beta}_{k}^\ast \right) \cdot \prod_{x\in\mathcal{X}}\mathbb{P}\left(D_{x,2}^{(r)} = d_{x,2}^{(r)} \mid S_2^{(r)} = j, \boldsymbol{z}_2^{(r)} \: ; \: \boldsymbol{\alpha}_{j,x,k}^{\ast}\right),
 \end{align*}
 with $i,j \in \{0,1,...,N\}$. For $t\geq 2$, we compute the probability:
 \begin{align*}
  &\mathbb{P}\left(S_{t-1}^{(r)} = i,S_{t}^{(r)} = j,\left(D_{x,t}^{(r)}\right)_{x\in\mathcal{X}} = \left(d_{x,t}^{(r)}\right)_{x\in\mathcal{X}} \mid u_{k,r}^\ast, \mathcal{D}_{t-1}^{(r)}, \mathcal{Z}_t^{(r)} \:;  \boldsymbol{\theta}^\ast_k\right)  \\ &\hspace{0.9cm}=  
  \sum_{h = 0}^N \mathbb{P}\left(S_{t-2}^{(r)} = h,S_{t-1}^{(r)} = i \mid u_{k,r}^\ast,\mathcal{D}_{t-1}^{(r)}, \mathcal{Z}_{t-1}^{(r)}\:;  \boldsymbol{\theta}^\ast_k\right) \cdot p_t^{ij}\left(\boldsymbol{z}_t^{(r)} \:; u_{k,r}^\ast, \boldsymbol{\beta}_{k}^\ast \right) \cdot 
  \\ &\hspace{2.1cm} \prod_{x\in\mathcal{X}}\mathbb{P}\left(D_{x,t}^{(r)} = d_{x,t}^{(r)} \mid S_{t}^{(r)} = j, \boldsymbol{z}_t^{(r)} \: ; \: \boldsymbol{\alpha}_{j,x,k}^{\ast}\right),
 \end{align*}
 with $i,j \in \{0,1,...,N\}$. In the above calculation, we relied on the assumption that the death counts are conditionally independent, given the regime state the system is in. The second and third terms are the transition probabilities and the state-dependent Poisson probabilities from step 1 and the first term is obtained from step 2.c of the previous iteration at time $t-1$. 
 \item[\textbf{(b)}] Using step 2.a, we calculate:
 \begin{align*}
 &\mathbb{P}\left(\left(D_{x,t}^{(r)}\right)_{x\in\mathcal{X}} = \left(d_{x,t}^{(r)}\right)_{x\in\mathcal{X}} \mid u_{k,r}^\ast, \mathcal{D}_{t-1}^{(r)}, \mathcal{Z}_t^{(r)} \: ; \boldsymbol{\theta}_k^\ast\right) \\ &\hspace{0.5cm}= \displaystyle \sum_{i = 0}^N \sum_{j = 0}^N \mathbb{P}\left(S_{t-1}^{(r)} = i,S_{t}^{(r)} = j,\left(D_{x,t}^{(r)}\right)_{x\in\mathcal{X}} = \left(d_{x,t}^{(r)}\right)_{x\in\mathcal{X}} \mid u_{k,r}^\ast, \mathcal{D}_{t-1}^{(r)}, \mathcal{Z}_t^{(r)} \: ;  \boldsymbol{\theta}^\ast_b\right).
 \end{align*} 
 \item[\textbf{(c)}] From step 2.a and 2.b, we obtain the time $t$ filtered joint state probabilities for $i,j \in \{0,1,..,N\}$:
 \begin{align*}
 &\mathbb{P}\left(S_{t-1}^{(r)} = i,S_{t}^{(r)} = j \mid u_{k,r}^\ast, \mathcal{D}_{t}^{(r)}, \mathcal{Z}_t^{(r)} \: ; \boldsymbol{\theta}_k^\ast\right) \\ &\hspace{1.5cm}= \dfrac{\mathbb{P}\left(S_{t-1}^{(r)} = i,S_{t}^{(r)} = j,\left(D_{x,t}^{(r)}\right)_{x\in\mathcal{X}} = \left(d_{x,t}^{(r)}\right)_{x\in\mathcal{X}} \mid u_{k,r}^\ast, \mathcal{D}_{t-1}^{(r)}, \mathcal{Z}_t^{(r)} \: ; \boldsymbol{\theta}_k^\ast\right)}{\mathbb{P}\left(\left(D_{x,t}^{(r)}\right)_{x\in\mathcal{X}} = \left(d_{x,t}^{(r)}\right)_{x\in\mathcal{X}} \mid u_{k,r}^\ast, \mathcal{D}_{t-1}^{(r)}, \mathcal{Z}_t^{(r)} \: ; \boldsymbol{\theta}_k^\ast\right)}.
 \end{align*}
 \end{enumerate}
 After iterating over these steps for $t=2,3,...,T$, we obtain all filtered joint state probabilities. We obtain the filtered marginal state probabilities as:
 \begin{align*}
\mathbb{P}\left(S_{t}^{(r)} = j \mid u_{k,r}^\ast, \mathcal{D}_{t}^{(r)}, \mathcal{Z}_t^{(r)} \: ; \: \boldsymbol{\theta}_k^\ast\right)  = \displaystyle \sum_{i = 0}^N \mathbb{P}\left(S_{t-1}^{(r)} = i,S_{t}^{(r)} = j \mid u_{k,r}^\ast, \mathcal{D}_{t}^{(r)}, \mathcal{Z}_t^{(r)} \: ; \boldsymbol{\theta}_k^\ast\right),
\end{align*}
for $j=0,1,...,N$ and $t=2,3,...,T$.
\end{enumerate}

\paragraph{M-step.} We update the regime-switching parameter vector $\boldsymbol{\theta}$ by solving:
\begin{align*}
\boldsymbol{\theta}_{k+1}^\ast = \argmax_{\boldsymbol{\theta}} \mathbb{E}\left[ \log \mathcal{L}(\boldsymbol{\theta}, \tau \mid \mathcal{D}_T, \mathcal{S}_T, \mathcal{Z}_T) \mid \boldsymbol{u}_{k}^\ast, \mathcal{D}_T, \mathcal{Z}_T; \boldsymbol{\theta}_k^\ast\right].
\end{align*}
For computational convenience, we can split this expected complete-data log-likelihood into a part that only depends on the parameter vectors $\boldsymbol{\alpha}_{j} := (\boldsymbol{\alpha}_{j,x})_{x\in\mathcal{X}}$, for each $j\in \{1,2\}$, and a part that only depends on the parameters $\boldsymbol{\beta}_{01}, \boldsymbol{\beta}_{02}, \boldsymbol{\beta}_{11}$, and $\boldsymbol{\beta}_{22}$. As such, we split the maximization problem into three separate maximization problems.

\paragraph{Update spatial effect.} We update the mode $\boldsymbol{u}^\ast$ from~\eqref{eq:maxU} under the current optimal parameter vector $\boldsymbol{\theta}^\ast_{k+1}$:
\begin{align*}
\boldsymbol{u}^\ast_{k+1} := \boldsymbol{u}^\ast (\boldsymbol{\theta}^\ast_{k+1}) = \argmax_{\boldsymbol{u}}  \mathbb{E}\left[ \log \mathcal{L}\left(\boldsymbol{\theta}^\ast_{k+1}, \tau \mid \mathcal{D}_T, \mathcal{S}_T, \mathcal{Z}_T) \right) \mid \boldsymbol{u}_{k}^\ast, \mathcal{D}_T, \mathcal{Z}_T; \boldsymbol{\theta}_k^\ast\right] .
\end{align*}

\subsection{Optimizing the precision parameter}  \label{subsubsec:precisionparam}
The precision parameter $\tau$ appears in the ICAR prior distribution applied to the spatial effect $\boldsymbol{U}$ and controls the degree of spatial autocorrelation in the transition probabilities of the different regions (see~(\ref{eq:icardens}) in the paper). We estimate $\tau$ by maximizing the so-called profile log-likelihood. Let $\mathcal{G}$ be a grid of candidate values for $\tau$. For each $\tau \in \mathcal{G}$, we obtain the optimal parameter vector $\boldsymbol{\theta}^\ast(\tau)$ in the three-state regime-switching process by maximizing the incomplete-data log-likelihood using the EM-algorithm summarized in Section~\ref{subsec:modelestimationEM} of the paper:
\begin{align*}
\boldsymbol{\theta}^{\ast}(\tau) = \argmax_{\boldsymbol{\theta}} \log \tilde{\mathcal{L}}(\boldsymbol{\theta},\tau \mid \mathcal{D}_T, \mathcal{Z}_T).
\end{align*}
The optimal precision parameter $\tau$ is then the value that maximizes the profile log-likelihood:
\begin{align*}
\tau^{\ast} = \argmax_{\tau \in \mathcal{G}} \log \tilde{\mathcal{L}}(\boldsymbol{\theta}^{\ast}(\tau), \tau \mid \mathcal{D}_T, \mathcal{Z}_T).
\end{align*}

\subsection{Parameter uncertainty in the regime-switching model} \label{app:paramuncertainty}
We describe the procedure to calculate the Fisher Information Matrix (FIM) of the unknown parameter vector $\boldsymbol{\theta}$ in the regime-switching model according to the approach outlined in \cite{meng2017efficient}.
\begin{itemize}
\item[1.] We apply the EM-algorithm to maximize the incomplete-data log-likelihood using the observed death counts dataset, $\mathcal{D}_T$, and the optimal precision parameter $\tau^\ast$, as determined in Section~\ref{subsubsec:precisionparam}. This maximization results in the optimal parameter vector $\smash{\boldsymbol{\theta}^{\ast}\left(\tau^{\ast}\right)}$ after reaching an absolute convergence in the EM-algorithm, see Section~\ref{subsec:modelestimationEM}.
\item[2.] We define the expectation of the complete-data log-likelihood with respect to the conditional distribution of the regime state process $\mathcal{S}_T$, given the observed death counts dataset $\mathcal{D}_T$, the covariate information $\mathcal{Z}_T$, and the optimal parameter vector $\boldsymbol{\theta}^\ast$ and spatial effect $\boldsymbol{u}^\ast$, obtained from the EM-algorithm:
\begin{align*}
Q\left(\boldsymbol{\theta} \mid \mathcal{D}_T, \mathcal{Z}_T, \boldsymbol{\theta}^{\ast}\right) =   \mathbb{E}\left[\log \mathcal{L}\left(\boldsymbol{\theta}, \tau^{\ast} \mid \mathcal{D}_T, \mathcal{S}_T, \mathcal{Z}_T\right) \mid \boldsymbol{u}^\ast, \mathcal{D}_T, \mathcal{Z}_T\:; \boldsymbol{\theta}^{\ast} \right],
\end{align*}
where $\tau^\ast$ is the value of the precision parameter $\tau$ that maximizes the profile log-likelihood, see Section~\ref{subsubsec:precisionparam}. Subsequently, we calculate the derivative of the expected complete-data log-likelihood with respect to the parameter vector $\boldsymbol{\theta}$, evaluated in $\boldsymbol{\theta}^\ast$. We define:
\begin{align} \label{eq:Diff.vector}
\boldsymbol{G}(\boldsymbol{\theta}^\ast \mid \mathcal{D}_T, \mathcal{Z}_T) &= \dfrac{\partial Q(\boldsymbol{\theta} \mid \mathcal{D}_T, \mathcal{Z}_T,\boldsymbol{\theta}^{\ast})}{\partial \boldsymbol{\theta}} \Bigg|_{\boldsymbol{\theta} = \boldsymbol{\theta}^\ast}.
\end{align}
Since the complete-data log-likelihood can be expressed as a sum of separate terms that depend on the parameter groups $\boldsymbol{\alpha}_1 := (\boldsymbol{\alpha}_{1,x})_{x\in\mathcal{X}}$, $\boldsymbol{\alpha}_2 :=  (\boldsymbol{\alpha}_{2,x})_{x\in\mathcal{X}}$, and $\boldsymbol{\beta} := (\boldsymbol{\beta}_0, \boldsymbol{\beta}_1, \boldsymbol{\beta}_2)$ individually, we calculate the derivatives with respect to each parameter group separately.

\paragraph{Derivative w.r.t. $\boldsymbol{\alpha}_1$.} We obtain: \vspace{-0.8cm}
\begin{center}
\resizebox{0.94\textwidth}{!}{
\begin{minipage}{\textwidth}
\begin{equation}\label{eq:Diff.alpha.1}
\begin{aligned} 
\boldsymbol{G}(\boldsymbol{\alpha}_{1,x} \mid \mathcal{D}_T, \mathcal{Z}_T) &= \displaystyle \sum_{r \in \mathcal{R}} \sum_{t = 1}^T P\left(S_t^{(r)} = 1 \mid u_r^\ast, \mathcal{D}_{t}^{(r)}, \mathcal{Z}_{t}^{(r)} ;  \boldsymbol{\theta}^\ast\right) \left(d_{x,t}^{(r)} - \hat{b}_{x,t}^{(r)} e^{\left(\boldsymbol{z}_t^{(r)}\right)^\top \boldsymbol{\alpha}_{1,x}} \right) \boldsymbol{z}_t^{(1,r)},
\end{aligned}
\end{equation}
\end{minipage}}
\end{center}
for every $x \in \mathcal{X}$, where $\boldsymbol{z}_t^{(1,r)}$ is the subset of the covariate vector $\boldsymbol{z}_t^{(r)}$ for which the parameter coefficients in $\boldsymbol{\alpha}_{1,x}$ are not set to zero.

\paragraph{Derivative w.r.t. $\boldsymbol{\alpha}_2$.} In a similar way, we obtain: \vspace{-0.8cm}
\begin{center}
\resizebox{0.94\textwidth}{!}{
\begin{minipage}{\textwidth}
\begin{equation} \label{eq:Diff.alpha.2}
\begin{aligned}
\boldsymbol{G}(\boldsymbol{\alpha}_{2,x} \mid \mathcal{D}_T, \mathcal{Z}_T) &= \displaystyle \sum_{r \in \mathcal{R}} \sum_{t = 1}^T P\left(S_t^{(r)} = 2 \mid u_r^\ast, \mathcal{D}_{t}^{(r)}, \mathcal{Z}_{t}^{(r)} ;  \boldsymbol{\theta}^\ast \right) \left(d_{x,t}^{(r)} - \hat{b}_{x,t}^{(r)} e^{\left(\boldsymbol{z}_t^{(r)}\right)^\top \boldsymbol{\alpha}_{2,x}} \right) \boldsymbol{z}_t^{(2,r)},
\end{aligned}
\end{equation}
\end{minipage}}
\end{center}
for every $x \in \mathcal{X}$, where $\boldsymbol{z}_t^{(2,r)}$ is the subset of the covariate vector $\boldsymbol{z}_t^{(r)}$ for which the parameter coefficients in $\boldsymbol{\alpha}_{2,x}$ are not set to zero.

\paragraph{Derivative w.r.t. $\boldsymbol{\beta}$.}
We calculate the derivative of the expected complete-data log-likelihood with respect to $\boldsymbol{\beta}_{ij}$, where $ij \in \{01, 02, 11, 22\}$ (see~\eqref{eq:trans.prob}):
\vspace{-0.8cm}
\begin{center}
\resizebox{0.95\textwidth}{!}{
\begin{minipage}{\textwidth}
\begin{equation}\label{eq:Diff.beta.ij}
\begin{aligned} 
&\boldsymbol{G}(\boldsymbol{\beta}_{ij} \mid \mathcal{D}_T, \mathcal{Z}_T) = \displaystyle \sum_{r\in\mathcal{R}} \sum_{t= 2}^T \sum_{j' = 0}^N \mathbb{P}\left(S_{t-1}^{(r)} = i, S_t^{(r)} = j' \mid u_r^\ast, \mathcal{D}_t^{(r)}, \mathcal{Z}_t^{(r)}; \boldsymbol{\theta}^\ast\right) \dfrac{\partial \log p_t^{ij'}\left(\boldsymbol{z}_t^{(r)}; u_r^\ast, \boldsymbol{\beta}\right)}{\partial \boldsymbol{\beta}_{ij}}  \\
&\hspace{3.5cm}  - \frac{1}{2} \dfrac{\partial \log \left( \det \left(\boldsymbol{S}^{-1}\left(\boldsymbol{\theta}, \boldsymbol{u}^\ast\right)\right)\right)}{\boldsymbol{\beta}_{ij}}.
\end{aligned}
\end{equation}
\end{minipage}}
\end{center}
We first calculate the derivative of the log transition probabilities with respect to $\boldsymbol{\beta}_{ij}$. A straightforward calculation shows that:
\begin{align} \label{eq:score1}
\dfrac{\partial \log p_t^{ij'}\left(\boldsymbol{z}_t^{(r)}; u_r^\ast, \boldsymbol{\beta}\right)}{\partial \boldsymbol{\beta}_{ij}} = 
\begin{cases}
-p_t^{ij}\left(\boldsymbol{z}_t^{(r)}; u_r^\ast, \boldsymbol{\beta}\right) \boldsymbol{z}_t^{(ij,r)}  & \:\: j\neq j' \vspace{0.1cm}\\
\left(1 - p_t^{ij}\left(\boldsymbol{z}_t^{(r)}; u_r^\ast, \boldsymbol{\beta}\right)\right) \boldsymbol{z}_t^{(ij,r)} &  \;\: j = j',
\end{cases}
\end{align}
where $\boldsymbol{z}_t^{(ij,r)}$ is the subset of the covariate vector $\boldsymbol{z}_t^{(r)}$ for which the parameter coefficients in $\boldsymbol{\beta}_{ij}$ are not set to zero.

Next, we calculate the derivative of the logarithm of the determinant of the Hessian $\smash{\boldsymbol{S}^{-1}(\boldsymbol{\theta}, \boldsymbol{u}^{\ast})}$, see~\eqref{eq:hessi}, with respect to the $l$-th entry of the parameter vector $\boldsymbol{\beta}_{ij}$, i.e., $\beta_{ijl}$. We apply Jacobi's formula and obtain:
\begin{align*}
\dfrac{\partial \log \left( \det \boldsymbol{S}^{-1}\left(\boldsymbol{\theta}, \boldsymbol{u}^\ast\right)\right)}{\beta_{ijl}} = \text{tr}\left(\boldsymbol{S}\left(\boldsymbol{\theta}, \boldsymbol{u}^\ast\right) \dfrac{\partial \boldsymbol{S}^{-1}\left(\boldsymbol{\theta}, \boldsymbol{u}^\ast\right)}{\partial \beta_{ijl}}\right).
\end{align*}
Note that the derivative of the Hessian with respect to $\beta_{ijl}$ is a diagonal matrix with entries:
\begin{equation*}
\begin{aligned}
&\left(\dfrac{\boldsymbol{S}^{-1}\left(\boldsymbol{\theta}, \boldsymbol{u}^\ast\right)}{\partial \beta_{ijl}}\right)_{rr} = 
\displaystyle \sum_{t=2}^T \sum_{j'=0}^N \mathbb{P}\left(S_{t-1}^{(r)} = i, S_t^{(r)} = j' \mid u_r^\ast, \mathcal{D}_t^{(r)}, \mathcal{Z}_t^{(r)}; \boldsymbol{\theta}^\ast\right) \cdot \\ &\hspace{3.75cm} p_t^{i0}\left(\boldsymbol{z}_t^{(r)}; u_r^\ast, \boldsymbol{\beta}\right) \cdot\left(2p_t^{i0}\left(\boldsymbol{z}_t^{(r)}; u_r^\ast, \boldsymbol{\beta}\right) - 1 \right) \cdot p_t^{ij}\left(\boldsymbol{z}_t^{(r)}; u_r^\ast, \boldsymbol{\beta}\right) \cdot  z_{t,l}^{(ij,r)},
\end{aligned}
\end{equation*}
for any $r\in \mathcal{R}$. This then leads to:
\begin{equation} \label{eq:score2}
\begin{aligned}
&\dfrac{\partial \log \left( \det \boldsymbol{S}^{-1}\left(\boldsymbol{\theta}, \boldsymbol{u}^\ast\right)\right)}{\boldsymbol{\beta}_{ij}} \\
&\hspace{1cm}= \displaystyle \sum_{r\in \mathcal{R}} \left(\boldsymbol{S}\left(\boldsymbol{\theta}, \boldsymbol{u}^\ast\right)\right)_{rr} \Bigg[\displaystyle \sum_{t=2}^T \sum_{j'=0}^N \mathbb{P}\left(S_{t-1}^{(r)} = i, S_t^{(r)} = j' \mid u_r^\ast, \mathcal{D}_t^{(r)}, \mathcal{Z}_t^{(r)}; \boldsymbol{\theta}^\ast \right) \cdot \\ &\hspace{3cm} p_t^{i0}\left(\boldsymbol{z}_t^{(r)}; u_r^\ast, \boldsymbol{\beta}\right) \cdot\left(2p_t^{i0}\left(\boldsymbol{z}_t^{(r)}; u_r^\ast, \boldsymbol{\beta}\right) - 1 \right) \cdot p_t^{ij}\left(\boldsymbol{z}_t^{(r)}; u_r^\ast, \boldsymbol{\beta}\right) \cdot  \boldsymbol{z}_t^{(ij,r)}\Bigg].
\end{aligned}
\end{equation}
We substitute the results from~\eqref{eq:score1} and~\eqref{eq:score2} into~\eqref{eq:Diff.beta.ij} to derive the gradient of the expected complete-data log-likelihood with respect to $\boldsymbol{\beta}_{ij}$, for $ij \in \{01,02,11,22\}$.

Using the derivatives from~\eqref{eq:Diff.alpha.1}, \eqref{eq:Diff.alpha.2}, and \eqref{eq:Diff.beta.ij}, the derivative of the expected complete-data log-likelihood in~\eqref{eq:Diff.vector} results in:
\begin{align*}
\boldsymbol{G}\left( \boldsymbol{\theta}^\ast \mid \mathcal{D}_T, \mathcal{Z}_T \right) = \begin{pmatrix}
\left(\boldsymbol{G} \left(\boldsymbol{\alpha}_{1,x}^\ast \mid \mathcal{D_T}, \mathcal{Z}_T \right)\right)_{x\in \mathcal{X}} \\
\left(\boldsymbol{G}\left(\boldsymbol{\alpha}_{2,x}^\ast \mid \mathcal{D_T}, \mathcal{Z}_T \right)\right)_{x\in \mathcal{X}} \\
\boldsymbol{G}\left( \boldsymbol{\beta}_{01}^\ast \mid \mathcal{D}_T, \mathcal{Z}_T \right) \\
\boldsymbol{G}\left( \boldsymbol{\beta}_{02}^\ast \mid \mathcal{D}_T, \mathcal{Z}_T \right) \\
\boldsymbol{G}\left( \boldsymbol{\beta}_{11}^\ast \mid \mathcal{D}_T, \mathcal{Z}_T \right) \\
\boldsymbol{G}\left( \boldsymbol{\beta}_{22}^\ast \mid \mathcal{D}_T, \mathcal{Z}_T \right) 
\end{pmatrix} \in \mathbb{R}^q.
\end{align*}
\item[3.] We generate $M$ synthetic datasets of death counts by simulating from the state-dependent distribution function in~\eqref{eq:cond.dens}, using the optimal parameter vector $\smash{\boldsymbol{\theta}^{\ast}\left(\tau^{\ast}\right)}$. This involves two steps: (i) we first generate $M$ sample paths for the state process $\mathcal{S}_T$ of the underlying Markov chain, based on the estimated transition probabilities in~\eqref{eq:trans.prob} using $\smash{\boldsymbol{\theta}^{\ast}\left(\tau^{\ast}\right)}$, and (ii) we subsequently sample from the Poisson distribution conditioned on the simulated states. During this sampling process, we keep the covariates $\smash{\boldsymbol{z}_t^{(r)}}$, for $j=1,2$, fixed. We denote the generated $M$ synthetic death counts datasets as $\mathcal{D}_{T,1}'$, $\mathcal{D}_{T,2}'$, ..., $\mathcal{D}_{T,M}'$ and the corresponding underlying simulated Markov chains as $\mathcal{S}_{T,1}'$, $\mathcal{S}_{T,2}'$, ..., $\mathcal{S}_{T,M}'$.
\item[4.] We generate $M$ perturbation vectors $\boldsymbol{\Delta}_k \in \mathbb{R}^q$ for $k = 1, 2, \ldots, M$, where each $\Delta_{kj}$ is independently drawn across different $k$ and $j$. Each $\Delta_{kj}$ follows a scaled Rademacher distribution with a scale factor $c$, meaning it takes the value $+c$ with a probability of 0.5 and $-c$ with a probability of 0.5.
\item[5.] For each $k=1,..,M$, we define the first-order differences:
\begin{align*}
    \delta \boldsymbol{G}_k = \boldsymbol{G}\left(\boldsymbol{\theta}^\ast + \boldsymbol{\Delta}_k \mid \mathcal{D}_{T,k}', \mathcal{Z}_t\right) - \boldsymbol{G}\left(\boldsymbol{\theta}^\ast - \boldsymbol{\Delta}_k | \mathcal{D}_{T,k}', \mathcal{Z}_t\right).
\end{align*}
According to \cite{meng2017efficient}, we can then approximate the Hessian of the incomplete-data log-likelihood for death counts dataset $\mathcal{D}_{T,k}'$ as:
\begin{align}
    \hat{\boldsymbol{H}}_k \left(\boldsymbol{\theta}^\ast \mid \mathcal{D}_{T,k}', \mathcal{Z}_t \right) = 0.5 \cdot \left(\dfrac{\delta \boldsymbol{G}_k}{2} \left[\Delta_{k1}^{-1}, \Delta_{k2}^{-1}, ..., \Delta_{kq}^{-1}\right] + \left(\dfrac{\delta \boldsymbol{G}_k}{2} \left[\Delta_{k1}^{-1}, \Delta_{k2}^{-1}, ..., \Delta_{kq}^{-1}\right] \right)^\top \right)
\end{align}
Lastly, we approximate the Fisher information matrix by taking the negative average of all $\hat{\boldsymbol{H}}_k \left(\boldsymbol{\theta}^\ast \mid \mathcal{D}_{T,k}', \mathcal{Z}_t \right)$. The diagonal entries of the inverse Fisher information matrix serve as measures of parameter uncertainty, enabling the construction of confidence intervals around the estimated parameter vector $\boldsymbol{\theta}^\ast$.
\end{itemize}

\section{Constructing temperature and influenza scenarios for short-term mortality forecasting} \label{apps:scenariooooos}

\subsection{Temperature scenario} \label{subsec:tempscen}
We use temperature projections based on the RCP 2.6, RCP 4.5, and RCP 8.5 scenarios from the Copernicus Climate Change Service \citep{cds2}. These scenarios represent a low-emission scenario aiming to limit warming (RCP 2.6), a medium stabilization scenario (RCP 4.5), and a high-emission, worst-case scenario that leads to significant global warming (RCP 8.5). Since we focus on short-term mortality forecasting (up to 2030), differences between the RCP scenarios remain small because the climate system only responds slowly to changes in emissions. Over this short-term period, natural variations, such as ocean cycles or weather patterns, cause temperatures under the RCP 4.5 or RCP 2.6 scenario to occasionally exceed those under the RCP 8.5 scenario. However, our goal is to provide plausible temperature projections for the near future, rather than assessing long-term differences.

We extract bias-adjusted daily average temperature projections with a 5 km resolution from the 27th ISO week of 2024 to January 1, 2030. The bias adjustment reduces bias in climate model projections by comparing them with observed reference data. Using the population-weighted approach, we transform the daily temperature projections to regional-level (NUTS 2) projections for each RCP scenario. We then compute daily anomalies and extreme indicators from these regional temperature levels and average them weekly (see Section~\ref{subsec:data} of the paper). Figure~\ref{fig:rcpsirs} shows the scenario projections for the daily average temperature in Pays de la Loire and Provence-Alpes-Côte d'Azur. Section E of the Online Appendix contains the results for the other regions.

\begin{figure}[!htb]
    \centering
    \includegraphics[width=0.95\linewidth]{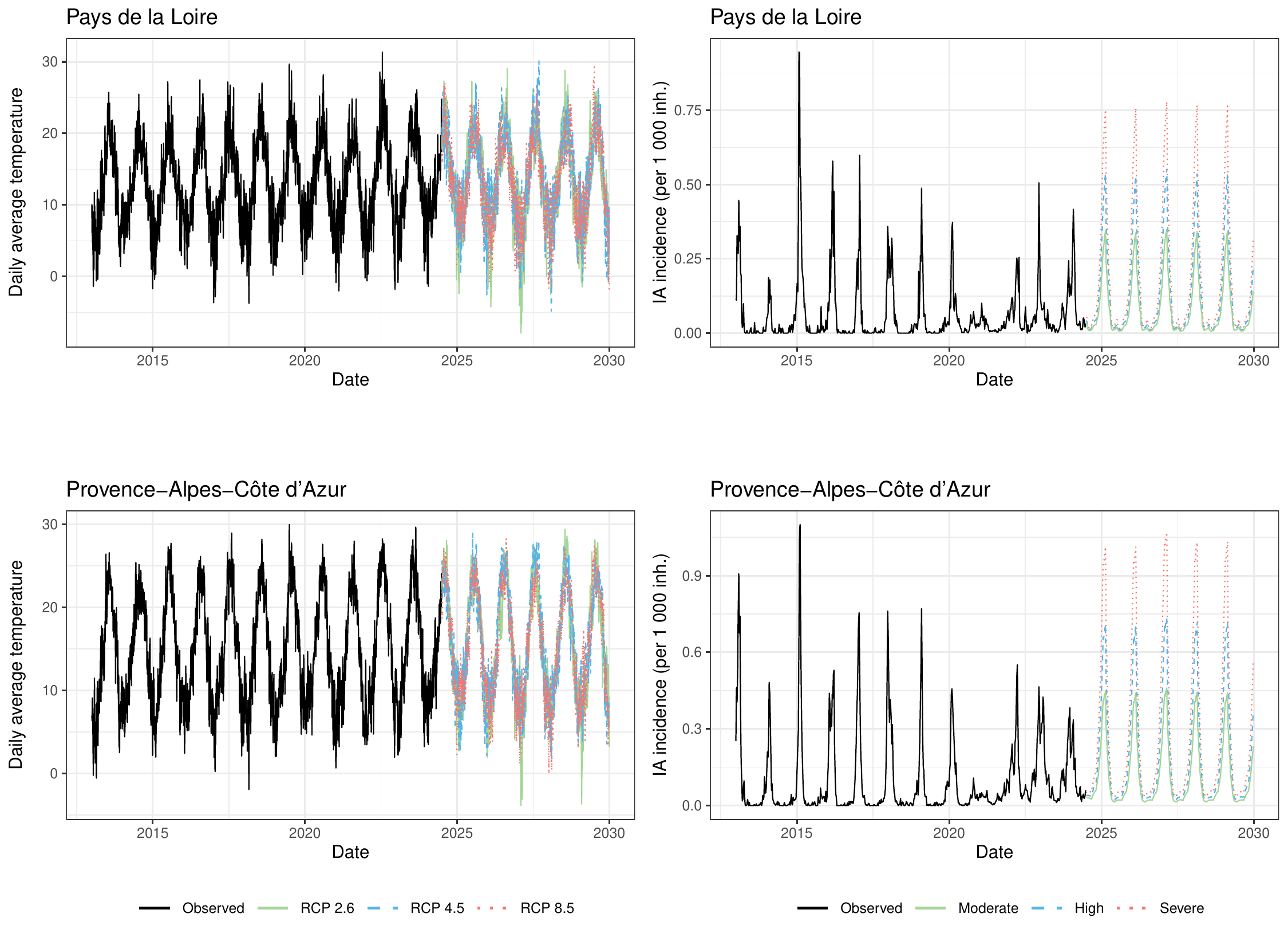}
    \caption{Projections for the daily average temperature based on the RCP 2.6, RCP 4.5, and RCP 8.5 scenarios (left panel), and for the weekly influenza activity per 1 000 inhabitants based on the proposed Bayesian stochastic SIRS model (right panel). We provide a moderate, high, and severe incidence scenario for influenza activity using the point-wise $50\%$, $80\%$, and $95\%$ quantiles of the posterior predictive samples, respectively. The top panels show the result for Pays de la Loire, while the bottom panels depict the results for Provences-Alpes-Côte d'Azur.}
    \label{fig:rcpsirs}
\end{figure}

\subsection{Influenza scenario} \label{subsec:respscen}
To generate plausible scenarios for the weekly number of new influenza incidences, we apply the Bayesian stochastic SIRS model based on the framework introduced by \cite{bucyibaruta2023discrete}. Unlike traditional SIRS models that track the total number of currently infected individuals, our approach focuses explicitly on modeling the weekly number of new infections. Let $I_t$ denote the number of new influenza cases per 1 000 inhabitants in week $t$, as reported by the French Sentinelles network. Additionally, let $S_t$ represent the number of susceptible individuals per 1 000 inhabitants at week $t$, and let $R_t$ denote the individuals who are not susceptible to influenza infection in week $t$. This latter group includes those who are either still infected or have acquired immunity from a prior infection.

Given the number of susceptible and newly infected individuals at week $t-1$, we model the number of new infections at week $t$ as:
\begin{equation*}
\begin{aligned}
I_t &\sim \text{POI}(\lambda_t), \\
\lambda_t &= \frac{S_{t-1}}{N} \cdot (I_{t-1} + I_{t-2})^\kappa \cdot \exp(r_t + \epsilon_t),
\end{aligned}  
\end{equation*}
where $\lambda_t$ is the expected number of new infections, which depends on the susceptible population at week $t-1$, the number of infections in the two preceding weeks, and the rate $r_t$ incorporates seasonal variation. The parameter $\kappa$ allows for non-linear transmission effects, and the error terms $\epsilon_t \sim \mathcal{U}[0,\zeta]$ allow to capture unexplained random variations that are not accounted for by the deterministic model structure. We update the susceptible and non-susceptible compartments recursively:
\begin{equation*}
\begin{aligned}
S_t &= S_{t-1} - \frac{S_{t-1}}{N} \cdot (I_{t-1} + I_{t-2})^\kappa \cdot \exp(r_t + \epsilon_t) + \psi R_{t-1}, \\
R_t &= (1 - \psi) R_{t-1} + I_{t-1}.
\end{aligned}  
\end{equation*}
Here, $\psi$ represents the rate at which individuals transition back to the susceptible compartment. The number of non-susceptible individuals at week $t$, i.e., $R_t$, consists of those who were infected in the previous week as well as a fraction of those who were not susceptible a week before. Lastly, we update the susceptible population $S_t$ based on the previous week's susceptible population, the expected infections at week $t$, and the fraction of individuals re-entering susceptibility.

We impose the following structure for $r_t$ to capture the seasonal dependence in the influenza infections \citep{bucyibaruta2023discrete}:
\begin{align*}
    r_t &= \varphi_{q(t)}  \vspace{0.75cm}\\
    q(t) &= \begin{cases}
        \min\left(\text{floor}\left(\frac{w(t)}{4}\right) + 1, 13\right) &\hspace{0.5cm}\text{if } w(t) \text{ mod }4 \neq 0 \\
         \text{floor}\left(\frac{w(t)}{4}\right)  &\hspace{0.5cm}\text{if } w(t) \text{ mod } 4 = 0,
    \end{cases}
\end{align*}
where $w(t)$ is the ISO-week number in the year $y(t)$ of time point $t$. Since some years contain 53 ISO-weeks, we restrict the maximum number of period effects to 13. 

We apply Bayesian inference through Markov Chain Monte Carlo (MCMC) methods. Hereto, we use the JAGS (Just Another Gibbs Sampler) implementation to estimate the posterior distribution of the weekly new infections $I_t$. Furthermore, we assign independent normal priors with zero mean and unit variance to the parameters $\phi_j$, for $j=1,2,..13$, and uniform priors to the parameters $\psi$, $\kappa$, and $\zeta$. The model structure supports forecasting by treating future infections as missing data. More specifically, we extend the observed time series of infections by appending NA values for the weeks we aim to forecast. JAGS interprets these NA entries as parameters to be sampled, which enables the MCMC algorithm to produce posterior predictive samples for both the model parameters and the unobserved future infections. We repeat this procedure for every NUTS 2 region within France.

For scenario-based forecasting, each posterior predictive sample represents a complete trajectory of weekly new infections. We define a moderate, high, and severe influenza scenario by taking the point-wise $50\%$, $80\%$, and $95\%$ quantile of the posterior predictive samples. Figure~\ref{fig:rcpsirs} illustrates these three trajectories for weekly influenza in Pays de la Loire and Provence-Alpes-Côte d'Azur. The influenza scenarios for the remaining NUTS 2 regions can be found in Section E of the Online Appendix. Lastly, we set COVID-19 hospitalizations to zero in the forecasting period. However, users can create alternative scenarios to explore the potential impact of new respiratory infections such as COVID-19 on mortality, if desired.


\end{document}